\documentclass[12pt,a4paper,table]{article}
\usepackage{jheppub}
\usepackage{amsmath}
\usepackage{amsfonts}
\usepackage{tikz-cd,tikz}
\usepackage{physics}
\usepackage{float}
\usepackage{bbold}
\usepackage{caption}
\usepackage{hyperref}

\usepackage{graphicx} % Required for inserting images

\def\[{\left[}
\def\]{\right]}
\def\({\left(}
\def\){\right)}

\newcommand\beqa{\begin{eqnarray}}

\newcommand{\mycomment}[1]{}
\newcommand\eeqa{\end{eqnarray}}
\newcommand{\beq}{\begin{eqnarray}}
\newcommand{\eeq}{\end{eqnarray}}
\newcommand{\la}[1]{\label{#1}}
\newcommand{\eq}[1]{(\ref{#1})}

\newcommand{\bR}{{\bf R}}
\newcommand{\bB}{{\bf B}}

\newcommand{\ep}{\varepsilon}
\newcommand{\si}{\sigma}
\newcommand{\kp}{\kappa}

\newcommand{\ii}{i}
\newcommand{\betheQ}{\mathbb{Q}}
\newcommand{\bfB}{\mathbf{B}}
\newcommand{\bfR}{\mathbf{R}}

\newcommand{\algosp}{\mathfrak{osp}}
\newcommand{\algsu}{\mathfrak{su}}
\newcommand{\algso}{\mathfrak{so}}
\newcommand{\algsp}{\mathfrak{sp}}

\newcommand{\sigmaBES}{\sigma_{\text{BES}}}
\newcommand{\fd}{f}
\newcommand{\fu}{{f^{*}}}
\newcommand{\mcF}{\mathcal{F}}

\newcommand{\bBullet}{\bB_{\bullet}}
\newcommand{\bBulletP}{\bB_{\bullet,(+)}}
\newcommand{\bBulletM}{\bB_{\bullet,(-)}}
\newcommand{\bBulletBar}{\bB_{\bar{\bullet}}}
\newcommand{\bBulletBarP}{\bB_{\bar{\bullet},(+)}}
\newcommand{\bBulletBarM}{\bB_{\bar{\bullet},(-)}}

\newcommand{\rBullet}{\bR_{\bullet}}
\newcommand{\rBulletP}{\bR_{\bullet,(+)}}
\newcommand{\rBulletM}{\bR_{\bullet,(-)}}

\newcommand{\rBulletBarP}{\bR_{\bar{\bullet},(+)}}
\newcommand{\rBulletBarM}{\bR_{\bar{\bullet},(-)}}

\newcommand{\mcN}{\mathcal{N}}
\newcommand{\hmcN}{\hat{\mathcal{N}}}

\newcommand{\mcS}{\mathcal{S}}

\newcommand{\mcA}{\mathcal{A}}
\newcommand{\mcB}{\mathcal{B}}

\newcommand{\bP}{\mathbf{P}}

\newcommand{\bQ}{\mathbf{Q}}
\newcommand{\fQ}{\mathcal{Q}}

\newcommand{\algsl}{\mathfrak{sl}}

\newcommand{\alggl}{\mathfrak{gl}}

\title{Gluing Quantum Spectral Curves:\\
A Two-Copy $\mathfrak{osp}(4|2)$ Construction}
\author{Filipp Chernikov${}^\bullet$, Simon Ekhammar${}^{\bullet \circ}$, Nikolay Gromov${}^{\bullet}$ and Benjamin Smith${}^{\bullet}$}
\affiliation{
    $\bullet$ Mathematics Department, King's College London,
    The Strand, London WC2R 2LS, UK\\
    $\circ$ Department of Physics and Astronomy, Uppsala University, Box 516, SE-751 20 \, \, Uppsala, Sweden}
    \emailAdd{filipp.chernikov@kcl.ac.uk}
    \emailAdd{simon.ekhammar@kcl.ac.uk}
    \emailAdd{nikolay.gromov@kcl.ac.uk}
 \emailAdd{benjamin.g.smith@kcl.ac.uk}
 
\date{February 2024}

\abstract{
We propose a Quantum Spectral Curve  for planar string theory on AdS$_3 \times$S$^{3} \times$S$^{3} \times$S$^{1}$ supported by pure Ramond-Ramond flux. Our proposal is built on symmetry considerations and integrability-based functional relations. To test our construction, we consider the large volume limit and successfully reproduce the crossing equations and the correct structure of the Bethe equations found in the literature. In a symmetric subsector, we find agreement with previously known results and furthermore extend the Asymptotic Bethe Ansatz to include massless modes. Beyond this sector, we identify an interesting puzzle regarding the compatibility of crossing equations with braiding unitarity for individual dressing phases, which warrants further investigation and may require additional physical insights or novel structures not previously encountered in related systems. As we expect the QSC to be exact in the planar limit, our proposal may open the way for non-perturbative analysis of this holographic system.
}

\begin{document}

\maketitle
\newpage
\section{Introduction} 

The AdS/CFT correspondence has established a profound link between string/M-theory in Anti-de Sitter (AdS) space and conformal field theory (CFT) \cite{Maldacena:1997re, Witten:1998qj,Gubser:1998bc}. 
The AdS/CFT correspondence is a so-called strong/weak coupling duality; it requires non-perturbative methods to unleash its true potential. Excitingly, in the planar limit, and usually for highly supersymmetric configurations, integrability has emerged in some key examples such as AdS$_5 \times$S$^5$, enabling exact computations \cite{Metsaev:1998it, Minahan:2002ve,Beisert:2003jj,Beisert:2005fw,Gromov:2009tv,Bombardelli:2009ns,Arutyunov:2009ur}. In this theory, integrability has made it possible to study the system from the classical string regime all the way to the weakly coupled CFT regime. For reviews, see \cite{Beisert:2010jr,Gromov:2017blm,Levkovich-Maslyuk:2019awk,Kazakov:2018ugh}. 

In this paper, we take steps to extend exact non-perturbative planar integrability to the case of AdS$_3 \times $S$^3 \times $S$^3\times $S$^1$ when the radii of the two S$^{3}$ are equal. There is an extensive literature on superstring theory on AdS$_3\times $S$^3\times$$\mathcal{M}^4$, with $\mathcal{M}^4$ some compact 4-dimensional manifold \cite{Gukov:2004ym, Elitzur:1998mm, Dolan2002,deBoer:1998ip, Giveon:1998ns, Giveon:2001up, Giveon:2003ku}, a setting in which integrability has long been studied \cite{Babichenko:2009dk,OhlssonSax:2011ms,Zarembo:2010sg,Zarembo:2010yz,Cagnazzo:2012se}. Historically, the main focus of integrability has been the case $\mathcal{M}^4 = T^4$, and significant effort has been made in studying these theories using various techniques \cite{Borsato:2013qpa,Borsato:2014hja,Borsato:2016kbm,Ekhammar:2021pys,Cavaglia:2021eqr,Frolov:2021bwp,Cavaglia:2022xld}. In contrast, the case of $\mathcal{M}_3 =  \text{S}^3 \times $S$^1$ is still relatively unexplored, although important steps have been taken \cite{OhlssonSax:2011ms,Borsato:2012ud,Borsato:2012ss,Borsato:2015mma}. 

The primary technical tool of this paper is the Quantum Spectral Curve (QSC), the most advanced integrability framework available to tackle the spectral problem \cite{Gromov:2013pga}. In particular, the QSC has enjoyed great success in planar $\mathcal{N}=4$ SYM and ABJM theory \cite{Gromov:2014caa, Cavaglia:2014exa, Bombardelli:2017vhk}, where it provides an efficient method to calculate the anomalous dimensions of local single trace operators \cite{Gromov:2015wca, LevkovichMaslyuk:2011ty} as well as quantities such as the slope function and exact correlators \cite{Giombi_2018, Bianchi:2014laa, Gromov:2014eha, Gromov:2014bva}. More recently, the QSC has integrated with conformal bootstrap techniques, usually referred to as ``Bootstrability", to study integrated correlators on a 2-dimensional defect in $\mathcal{N}=4$ \cite{Cavaglia:2022yvv, Cavaglia:2022qpg}
as well as for local operators in \cite{Caron-Huot:2022sdy,Caron-Huot:2024tzr}. Finally, the Q-function of the QSC serves as one of the key building blocks for the modern integrability constructions of correlation functions~\cite{Bercini:2022jxo,Basso:2022nny,Basso:2025mca}.

Due to its success in higher dimensions, the QSC was recently conjectured for AdS$_{3} \times $S$^{3} \times $T$^{4}$ supported by pure Ramond-Ramond flux \cite{Ekhammar:2021pys,Cavaglia:2021eqr}. The proposed QSC was ``bootstrapped'' from symmetry assumptions, bypassing the standard route of deriving it from the so-called thermodynamic Bethe Ansatz (TBA). These equations were only subsequently proposed for AdS$_3 \times $S$^3 \times$T$^4$ \cite{Frolov:2021bwp}. The main purpose of this paper is to extend this construction to the case of AdS$_3 \times $S$^3 \times $S$^3 \times $S$^1$ by leveraging its underlying $\algosp(4|2)$ symmetry.

An important motivation to extend the frameworks proposed for T$^4$ to S$^{3} \times$S$^1$ is to provide another example for studying the many additional nuances that appear in AdS$_3$ integrability compared to higher-dimensional examples. In particular, the presence of gapless modes on the string worldsheet introduces a significant technical challenge to the underlying integrable structures; these excitations were only recently identified within the QSC framework \cite{Ekhammar:2024kzp}. Furthermore, there is also an increased interest in this particular background from orthogonal points of view~\cite{witten2024instantonslargen4algebra}. 

The remainder of this paper is structured as follows. In Section~\ref{sec:ABJMQSC}, we review the QSC construction for ABJM.
In Section~\ref{sec:OSPQSC} we propose a QSC for AdS$_3 \times$S$^{3}\times$S$^3\times$S$^{1}$. In~\ref{Sect:consequences} we present two different presentations of the QSC, the so-called $\bQ\tau$-and $\bP\nu$-systems. We also identify a particular simple subsector of solutions, known as the symmetric sector.
Finally, in Section~\ref{Sect:ABA}, we test our proposal by deriving a large volume limit and comparing the results with the ABA equations available in the literature. 
Finally, we conclude in Section~\ref{Sec:conclusions}.

\section{Quantum Spectral Curve and its AdS$_4$/CFT$_3$ Realisation} \label{sec:ABJMQSC}

Our goal is to construct a proposal for QSC for free strings on AdS$_3\times $S$^3\times $S$^3\times $S$^1$. The construction is guided primarily by symmetry considerations. In the coset construction of \cite{Babichenko:2009dk}, the key starting point is that the (subspace) AdS$_3\times $S$^3\times $S$^3$ has super-isometry algebra $\algosp(4|2)^{\oplus 2}$. In analogy with all other known examples of the QSC, we will assume that this symmetry survives on a quantum level, and we therefore require that our QSC exhibits $\algosp(4|2)$-symmetry. 

The QSC is constructed from a collection of Q-functions, which depend on a complex variable $u$, the spectral parameter. These Q-functions satisfy various finite difference equations among themselves, forming a so-called QQ-system. The algebraic structure of the QQ-system is generally believed to be fully determined by the underlying symmetry algebra, in our case $\mathfrak{osp}(4|2)$. Altough QQ-systems have been studied intensively for $\alggl({m|n})$ \cite{Chernyak:2020lgw, Ryan:2020rfk, Kazakov:2015efa, Marboe_2017}, much less is known about other superalgebras. For recent progress on explicit operator constructions of $Q$-functions for $\algosp$, see \cite{Frassek:2023tka, Ferrando:2020vzk, ekhammar2021extendedsystemsbaxterqfunctions}. In particular, to the best of the authors' knowledge, the only case of an $\algosp$ QQ-system studied with typical QSC notation in the literature 
is that of $\algosp({6|4})$ of ABJM \cite{Bombardelli:2017vhk}; see \cite{Tsuboi:2023sfs} for related extensions. In the first part of this section, we review the ABJM QSC \cite{Bombardelli:2017vhk}, which we use to motivate our proposal for the AdS$_3\times $S$^3\times $S$^3\times $S$^1$ QSC in the next section.

The quasiclassical limit has been a key aspect of previously studied cases of the QSC \cite{Gromov:2014caa,Bombardelli:2017vhk}. The semi-classical limit of type IIB string theory on AdS$_3\times $S$^3\times $S$^3\times $S$^1$ has been previously studied in \cite{Abbott:2012dd}. In the second half of this section, we review these results with a view to their application to our proposed QSC. 

\subsection{The ABJM QSC: A Constructive Example}
The ABJM QSC is based on the superalgebra $\algosp(6|4)$ and
was derived from the Thermodynamic Bethe Ansatz (TBA) in \cite{Bombardelli_2010,Bombardelli:2017vhk}. Let us remind the reader of the key result; see also \cite{Brizio:2024nso} for a comprehensive review.

\paragraph{Algebraic construction of Q-functions.}
The basic building blocks of the $\algosp(6|4)$ QQ-system are two matrices: $Q_{a|i}(u)$ and $Q^{a}{}_{|i}(u)$. Here, the subscript $a=1,\dots,4$ indexes the $\algso_{6}\simeq \algsu_{4}$ representation $\mathbf{4}$, while the superscript $a$ is for $\bar{\mathbf{4}}$. Finally, $i=1,\dots 4$ labels the $\algsp_4$ representation $\mathbf{4}$ \footnote{All notation introduced in this paragraph is specific to the $\algosp({6|4})$ case and does not hold for the rest of the paper.}. The two matrices can be contracted to produce the main ``bosonic'' Q-functions (traditionally denoted as $\bP$ and $\bQ$)
\begin{equation}\label{PQABJMspin_1}
    \bP_{ab}=Q_{b|i}^-Q_{a|j}^+\kp^{ij}\,,
    \quad
    \bP^{ab}=-Q^{b}{}_{|i}^-Q^{a}{}_{|j}^+\kp^{ij}\,,
    \quad
     \bQ_{ij}=Q^{a}{}_{|i}^+Q_{a|j}^-\;.
\end{equation}
Here $\kappa_{ij}$ is the invariant $\algsp({4})$ inner form, given as $\kappa_{ij} = (-1)^{i+1}\delta_{i+j,5}$. In \eqref{PQABJMspin_1}, we have introduced standard notation for shifting the spectral parameter
\begin{equation}
    f^{\pm}\equiv f\left(u\pm\tfrac{\ii}{2}\right)\,,
    \quad
    f^{[k]}\equiv f\left(u+\tfrac{\ii k}{2}\right)\,.
\end{equation}
The antisymmetric tensors $\bP_{ab}(u)$ and $\bP^{ab}(u)$ 
are not independent but are inverse matrices of each other, up to a rescaling; more precisely, they are 
related as follows:
\begin{equation}\label{PPrel}
    \bP_{ab}\bP^{ba}=4\;\;,\;\;\mathrm{Pf}(\bP_{ab})=1\;,
\end{equation}
where $\mathrm{Pf}(\bP_{ab})$ is the Pfaffian of the matrix $\bP_{ab}$.
Similarly, the determinants of $Q_{a|i}$ and $Q^{a}{}_{|i}$ must be constant, which, in our convention, becomes
\begin{equation}\label{ABJMdet}
    \det(Q_{a|i})=\det(Q^{a}{}_{|i})=-1\,.
\end{equation}
The matrix $Q^{a}{}_{|i}$ is related to the inverse transpose of $Q_{a|i}$
\begin{equation}\label{ABJMQinv}
    Q^{a}{}_{|i}=Q^{a|j}\kp_{ji}\,,
    \quad 
    Q_{a|i}Q^{a|j}=\delta_i^{j}\,,
    \quad
    Q_{a|i}Q^{b|i}=\delta_a^{b}\,.
\end{equation}
Using \eqref{ABJMQinv} in \eqref{PQABJMspin_1}, we find the following finite difference equations
\begin{equation}
    Q_{a|i}^+=\bP_{ab}(Q^{b}{}_{|i})^{-}\,,
    \quad
    (Q^{a}{}_{|i})^{+}=\bP^{ab}Q_{b|i}^-\,,
\end{equation}
which we can combine to obtain a matrix-Baxter equation for $Q_{a|i}$, i.e., a linear finite difference equation
\begin{equation}\label{ABJMBaxter}
    Q_{a|i}^+=\bP_{ab}(\bP^{bc})^{[-2]}Q_{c|i}^{[-3]}\,.
\end{equation}
Provided that the functions $\bP$ are known, $Q_{a|i}$ can then be found by solving (\ref{ABJMBaxter}). This is the standard approach to numerically or analytically solving the QSC.

\paragraph{Analytic properties of $\bP$ and $\bQ$.} 

Historically, the analytic properties of Q-functions were understood by decoding them from the TBA. Here, we recap the final output, which appears to be universal for all iterations of the QSC relevant to AdS/CFT integrability. 

The first requirement is that $\bP$ and $\bQ$ both have powerlike asymptotics fixed by the charges of the state under consideration.

The functions $\bP_{ab}(u)$ have a single quadratic branch cut on the first Riemann sheet stretching from $-2g$ to $2g$ along the real line and can, hence, be written as a Laurent series expansion in the Zhukovsky variable $x(u)$:
\begin{equation}\label{P_laurent}
    \bP_{ab}\propto \sum_{n} \dfrac{c_{A,n}}{x^n}~~,~~x+\frac{1}{x}=\frac{u}{g}\,,
\end{equation}
which is absolutely convergent on the first sheet, $\abs{x}>1$, and a bit further under the cut.

\begin{figure}
    \centering
    \includegraphics[width=0.7\linewidth]{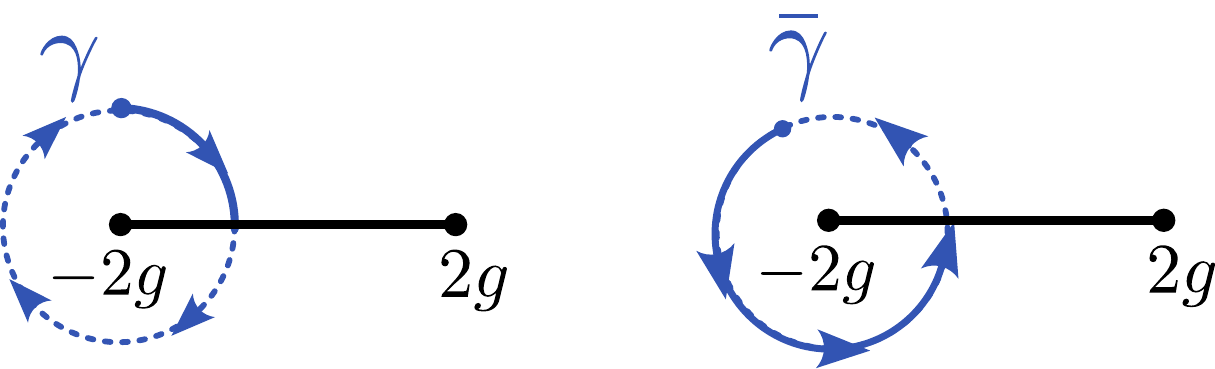}
     \caption{Two contours we use for analytic continuation. For the ABJM case, there is no difference between the two, since all cuts are quadratic there. For AdS$_3$, however, one has to distinguish between the two.}
    \label{fig:gammacontour}
\end{figure}

The cut structure of $\bQ$ is more complicated. We require that $\bQ$ is analytic in the upper half-plane. Due to the QQ-relations \eqref{PQABJMspin_1} and the analytic properties of $\bP$, $\bQ$ must feature an infinite ladder of cuts in the lower-half-plane. However, a semblance of democracy between $\bP$ and $\bQ$ is restored by demanding that $\bQ$ has only one cut when considered as a function with long cuts. In practice, this means that upon traversing the branch points $\pm 2g$, one should find a function that is analytic in the \emph{lower half-plane}. Let us introduce notation $f^{\gamma}$ for the analytic continuation of a function $f$ around $-2g$ clockwise, and $f^{\bar{\gamma}}$ for going anti-clockwise. We depict this action in Figure~\ref{fig:gammacontour}.
With this notation at hand, we can state the additional analytic property we impose on $\bQ$, usually referred to as the \emph{gluing condition}. We require
\begin{equation}\label{ABJMglue}
    \bQ_{ij}^{\gamma}=F_i^{~l}\kappa_{lm}(\bQ^{mn})^{*} \kappa_{nk} F_{j}^{~k}\,,
\end{equation}
where $\bQ^{*}$ is the complex conjugate of $\bQ$ and $\bQ_{ij}\bQ^{jk}=\delta^k_i$.  As argued in \cite{Bombardelli:2017vhk}, the matrix $F_{i}^{~l}$ is constant. Furthermore, many of the components of $F$ are, in fact, zero, as was shown in \cite{Bombardelli:2017vhk}. Schematically, it looks as follows:
\begin{equation}
    F_{i}^{~l}=\begin{pmatrix}
        \star &0&0&\star\\0&\star&0&0\\
        0&0&\star&0\\
        \star&0&0&\star
    \end{pmatrix}\,,
\end{equation}
where $\star$ is generally a non-zero number.

\subsection{Quasiclassics in AdS$_3 \times$S$^3 \times$S$^3 \times$S$^{1}$}\label{subsec:Quasi}
Another source of information is the classical integrable structure of the worldsheet sigma model, known as the classical spectral curve~\cite{Kazakov:2004qf,SchaferNameki:2010jy}. Using the classical curve, one can construct infinitely many conserved integrals of motion, often conveniently encoded in the so-called quasi-momenta $p$, which are related to the eigenvalues of the classical monodromy matrix.  

As the QSC must know about the classical worldsheet in the appropriate limit, the Q-functions should encode the quasi-momenta. The exact relation between the Q-functions and the quasi-momenta is similar to the relationship between a wave function and its quasiclassical approximation in terms of the classical momenta. The relation is well established in $\mathcal{N}=4$ and ABJM \cite{Gromov:2014caa, Gromov:2008bz}, schematically it takes the form
\begin{align}\label{PQpq}
    \bP \sim e^{-g \int \hat{p} \, du}\,,
    \quad
    \bQ \sim e^{ \, g \int \tilde{p} \, du}\,,
\end{align}
which should still be decorated with appropriate indices to be an exact equation. We will explain how to do this in AdS$_3$ in the next section.

The classical curve for AdS$_3 \times$S$^3 \times$S$^{3}$ has been studied in \cite{Abbott:2012dd}. The algebraic charges are the conformal scaling dimension $\Delta$, the spin $S$, and the charges from the two spheres, which we denote as $J_{1},\;J_2$ and $K_{1},\;K_2$. They are encoded in $6+6$ quasi-momenta as
\begin{align} \label{Eqn:quasiclass}
    \begin{pmatrix}
        \tilde{p}_{1} \\
        \tilde{p}_{2}\\
        \hat{p}_{1}\\
        \hat{p}_{2} \\
        \hat{p}_{3} \\
        \hat{p}_{4}
    \end{pmatrix}
    \simeq 
    \frac{1}{2u} \begin{pmatrix}
        +\Delta - S\\
        -\Delta + S\\
        +J_1-K_1\\
        +J_2-K_2\\
        -J_2+ K_2 \\
        -J_1 + K_1
    \end{pmatrix}
    \,,
    \quad
    \begin{pmatrix}
        \tilde{\bar{p}}_{1} \\
        \tilde{\bar{p}}_{2}\\
        \hat{\bar{p}}_{1}\\
        \hat{\bar{p}}_{2} \\
        \hat{\bar{p}}_{3} \\
        \hat{\bar{p}}_{4}
    \end{pmatrix}
    \simeq 
    \frac{1}{2u} \begin{pmatrix}
        +\Delta + S\\
        -\Delta - S\\
        +J_1+K_1\\
        +J_2+K_2\\
        -J_2- K_2 \\
        -J_1- K_1
    \end{pmatrix}\;.
\end{align}
This determines, through \eqref{PQpq}, the asymptotic behaviour of the Q-functions when all quantum numbers are large. This information can be used to determine how the large $u$ asymptotic of the Q-functions is related to the quantum numbers of the global symmetry.

\section{The AdS${}_3\times$S${}^3\times$S${}^3\times$S${}^1$ QSC Proposal}\label{sec:OSPQSC}
In this section, we formulate our proposal for a $\algosp({4|2})^{\oplus 2}$ QSC. At its heart, the QSC construction is a system of functional relations between Q-functions -- the QQ-system -- which is expected to be universal among different integrable systems with the same symmetry. In an ideal world, the QQ-system corresponding to an integrable model would be derived from first principles, for example, by starting from the underlying Yangian when considering rational spin chains. 

As stated above, since we currently lack a thermodynamic Bethe ansatz or a similar framework to describe free strings on AdS$_3 \times$S$^3 \times$S$^3 \times$S$^1$, we cannot derive the QSC directly by following the standard approach. We will instead rely on intuition derived from systems with similar symmetries and general features of integrable models, such as the ABJM QQ-system described in the previous section. Since $\algosp({4|2})$ is a subalgebra of $\algosp({6|4})$, part of the strategy to recover the $\algosp({4|2})$ QSC is to truncate the ABJM QSC reviewed in the previous section. This truncation is depicted schematically in Figure~\ref{DyDiagABJM}.

\begin{center}
\begin{minipage}{0.05\textwidth}
~~
\end{minipage}
\begin{minipage}{0.4 \textwidth}
    \begin{tikzpicture}[cross/.style={path picture={ 
  \draw[black]
(path picture bounding box.south east) -- (path picture bounding box.north west) (path picture bounding box.south west) -- (path picture bounding box.north east);
}}]
        \node at (-3,1) {$\algosp({6|4})$};
        \draw[] (-2.5,0)--(-1.25,0){};
        \draw[] (-1.25,0)--(0,0){};
        \draw[] (0,0)--({1.5/sqrt(2)},{1/sqrt(2)}){};
        \draw[] (0,0)--({1.5/sqrt(2)},{-1/sqrt(2)})
        {};
        \draw [ dashed] (0.65,0) ellipse (1.3 and 1.75);
        \draw[fill=white,thick] ({1.5/sqrt(2)},{1/sqrt(2)}) circle (0.2) node[anchor=south,yshift=4] {$Q_{a|i}$};
        \draw[fill=gray!20,cross,thick] (0,0) circle (0.2) node[anchor=south,yshift=4] {$\fQ_{A|IJ}$};
        \draw[fill=white,thick] ({1.5/sqrt(2)},-{1/sqrt(2)}) circle (0.2) node[anchor=north,yshift=-4]{$Q^{a}{}_{|i}$};
        \draw[fill=white,thick] ({-1.25},{0}) circle (0.2) node[anchor=south,yshift=4] {$\fQ_{A|I}$};
        \draw[fill=gray!20,cross,thick] (-2.5,0) circle (0.2) node[anchor=south,yshift=4] {$\bP_A$};
    \end{tikzpicture}
\end{minipage}
\begin{minipage}{0.05\textwidth}
~~
\end{minipage}
\begin{minipage}{0.4 \textwidth}
    \begin{tikzpicture}[cross/.style={path picture={ 
  \draw[black]
(path picture bounding box.south east) -- (path picture bounding box.north west) (path picture bounding box.south west) -- (path picture bounding box.north east);
}}]
        \node at (-1,1) {$\algosp({4|2})$};
        \draw[] (0,0)--({1/sqrt(2)},{1/sqrt(2)}){};
        \draw[] (0,0)--({1/sqrt(2)},{-1/sqrt(2)}){};
        \draw[fill=white,thick] ({1/sqrt(2)},{1/sqrt(2)}) circle (0.2) node[anchor=south,yshift=4] {};
        \draw[fill=gray!20,cross,thick] (0,0) circle (0.2) node[anchor=south,yshift=4] {};
        \draw[fill=white,thick] ({1/sqrt(2)},-{1/sqrt(2)}) circle (0.2) node[anchor=north,yshift=-4]{};
    \end{tikzpicture}
\end{minipage}

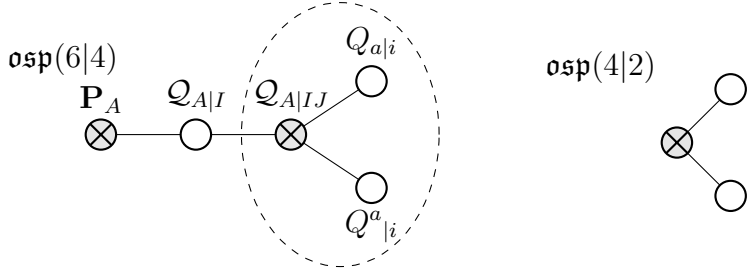
\captionof{figure}{The Dynkin diagram for $\algosp({6|4})$ (on the left) and its subdiagram $\algosp({4|2})$ (on the right). Each node has a Q-function associated with it. The diagram encodes relations between the functions, forming a closed $QQ$-system.}
\label{DyDiagABJM}
\end{center}

Restricting $\algosp({6|4})$ to $\algosp({4|2})$ suggests that we should replace the two basic matrices $Q_{a|k}$ and $Q^{a}{}_{|k}$ in the ABJM case with $Q_{a|k}$ and $Q_{\dot{a}|k}$, where $a,\dot{a}= 1,2$ are the two spinor representations of $\algso({4})$ while $k=1,2$ is associated with the fundamental representation of $\algsl({2})$. All indices $a,\dot{a},k$ can be raised and lowered by a fully antisymmetric tensor $\ep$
\begin{equation}
\ep^{\dot{a}\dot{b}}=\ep_{\dot{a}\dot{b}}=\ep_{ab}=\ep^{ab}=\ep^{ij}=\ep_{ij}=\begin{pmatrix}
    0&1\\-1&0
\end{pmatrix}\;,
\end{equation}
and we will use conventions
\begin{equation}
    Q^{a}{}_{|k} = \epsilon^{ab}Q_{b|k}\;,
    \quad
    Q_{a|k} = -\epsilon_{ab}Q^{b}{}_{|k}\;.
\end{equation}
We  also use vector indices $A=1\dots 4$ and $I=1,2,3$ which are raised and lowered by the metrics $\eta_{AB}$ and $\rho_{IJ}$
\begin{equation}
    \eta^{AB}=\eta_{AB}=\begin{pmatrix}
0&0&0&1\\0&0&1&0\\0&1&0&0\\1&0&0&0
\end{pmatrix}\;,
\quad
\rho^{IJ}=\rho_{IJ}=\begin{pmatrix}
    0&0&1\\0&1&0\\1&0&0
\end{pmatrix}\;.
\end{equation}
These are appropriately reduced (and rescaled) versions of those encountered in ABJM.

Vector and spinor representations are connected through $\sigma$-matrices. For $\algso(4)$, we will use two sets of $2\times2$ matrices $(\sigma_A)^{\dot{a}b}$ and $(\sigma^A)^{a\dot{b}}$. The $\algsl_2$ representation is built using three Pauli matrices $(\Sigma_I)^{ij}$. Further explicit expressions for $\sigma,\Sigma$-matrices are given in Appendix~\ref{app:conventions}. 

Let us outline the structure of this section. In section~\ref{subsec:AlgebraicRelations}, we will define the algebraic properties of a single $\algosp({4|2})$ QQ-system, and subsequently, in section~\ref{subsec:AnalyticStructure} and section~\ref{subsec:Gluing}, we will postulate the analytic structure of the Q-functions. Finally, in section~\ref{subsec:AsymptoticSection} we specify asymptotics for the two copies of $\algosp({4|2})$ Q-systems in terms of the quantum numbers of the state under consideration.

\subsection{The $\mathfrak{osp}({4|2})$ QQ-system}\label{subsec:AlgebraicRelations}
As explained in Section~\ref{sec:ABJMQSC}, key players in the ABJM QSC are the Q-functions $\bP$ and $\bQ$, we will now construct them for $\mathfrak{osp}(4|2)$ starting from $Q_{a|i},Q_{\dot{a}|i}$.

In analogy with \eqref{PQABJMspin_1} we define
\begin{equation}\label{eq:DefP}
     \bP_{A} \equiv  \epsilon^{kl}\sigma_{A}^{\dot{a}b} Q^-_{\dot{a}|k}Q^+_{b|l}\,.
\end{equation}
Due to the finite shifts of Q-functions in \eqref{eq:DefP} this definition does not treat the two spinor representations of $\algso({4})$ in a democratic way (as $Q_{\dot a|i}$ is shifted by $-i/2$, whereas $Q_{b|l}$ is shifted in the opposite direction). It has previously been observed in Q-systems for $\algso({2n})$ that one can form a consistent system by demanding that one finds the same $\bP_{a}$ upon exchanging the two representations \cite{ekhammar2021extendedsystemsbaxterqfunctions,Ekhammar:2021myw,Ferrando:2020vzk}, the same phenomenon is also true in ABJM. We follow this prescription and require
\begin{equation}\label{eq:bPDemocracy}
\epsilon^{kl}
Q^-_{\dot{a}|k}Q^+_{b|l} = \epsilon^{kl}
Q^+_{\dot{a}|k}Q^-_{b|l}\,.
\end{equation}
By taking the determinant of \eq{eq:bPDemocracy}, we get
\beq
\frac{\det Q_{a|k}^+}{\det Q_{\dot a|k}^+} = 
\frac{\det Q_{a|k}^-}{\det Q_{\dot a|k}^-}\;,
\eeq
implying that the ratio $\frac{\det Q_{a|k}}{\det Q_{\dot a|k}}$ is a periodic function. As usual, we will require our Q-functions to be upper-half-plane analytic and with a finite number of zeros; hence, this periodic function should be a constant. 
If, in addition, we set $\bP_A \bP^A$ to be a constant, we also conclude that  
each individual determinant is a constant. By fixing the normalisation, we set for convenience
\begin{equation}\label{detcond}
    \det Q_{a|k} = \det Q_{\dot{a}|k} = -1
    \,,
\end{equation}
which then also implies that
\begin{equation}\label{detcond2}
\bP_A \,\bP^{A} = -2\;.
\end{equation}

Just as for $\bP_A$, we will build $\bQ_I$ from the same basic building blocks. However, from group theory, it is clear that we need only $Q_{a|i}$ to construct a vector, $\mathbf{3}$, of $\algsp({2})$ and the singlet $\bQ_{\circ}$. We define
\begin{equation}\label{eq:bQDef}
     \bQ_{K} = +\Sigma^{ij}_{K} \epsilon^{ab}Q^{-}_{a|i}Q^{+}_{b|j}\,,
     \quad
     \bQ_{\circ}=Q_{a|k}^-Q_{b|l}^+\ep^{ab}\ep^{kl}\,,
\end{equation}
and find from \eqref{eq:bPDemocracy}, and using \eqref{detcond}, that $\bQ_K,\bQ_{\circ}$ can also be constructed using $Q_{\dot{a}|k}$ as
\begin{equation}\label{eq:bQDef2}
     \bQ_{K} = -\Sigma^{ij}_{K}\epsilon^{\dot{a}\dot{b}}Q_{\dot{a}|i}^- Q_{\dot{b}|j}^+\,,
     \quad
     \bQ_{\circ} =
     Q_{\dot{a}|k}^-Q_{\dot{b}|l}^+\ep^{\dot{a}\dot{b}}\ep^{kl}\,.
\end{equation}
Note that whereas $\bQ_{\circ}$ is not sensitive to interchanging the dotted and undotted indices, the vector representation $\bQ_{K}$ changes the sign.

Using the QQ-relations presented above, one can derive all further relations used in this paper. We emphasise that the above set presents the complete defining set for the QQ-system, which is a major part of the QSC construction.

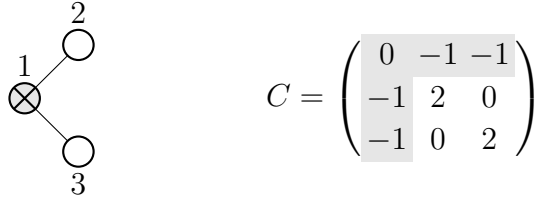
\begin{figure}
\begin{center}
\begin{tikzpicture}[cross/.style={path picture={ 
\draw[black]
(path picture bounding box.south east) -- (path picture bounding box.north west) (path picture bounding box.south west) -- (path picture bounding box.north east);
}}]
  % your two other lines
  \draw[] (0,0)--({1/sqrt(2)},{1/sqrt(2)});
  \draw[] (0,0)--({1/sqrt(2)},{-1/sqrt(2)});

  % circle at (1/√2,1/√2)
  \draw[fill=white,thick]
       ({1/sqrt(2)},{1/sqrt(2)}) circle (0.2) 
       node[anchor=south,yshift=4] {2};

  % crossed circle at (0,0), now with gray background
  \draw[fill=gray!20,cross,thick]
       (0,0) circle (0.2) 
       node[anchor=south,yshift=4] {1};

  % circle at (1/√2,–1/√2)
  \draw[fill=white,thick]
       ({1/sqrt(2)},-{1/sqrt(2)}) circle (0.2) 
       node[anchor=north,yshift=-4]{3};
\node[] (Cartan) at (5,0) {$C = \left(
  \begin{array}{>{\columncolor{gray!20}}c c c}                \rowcolor{gray!20}            % first row gray
   0  & -1 & -1  \\
  -1  &  2 &  0  \\
  -1  &  0 &  2
  \end{array}
\right)$};
    \end{tikzpicture}
\captionof{figure}{Dynkin diagrams for $\algosp({4|2})$ and its Cartan matrix, encoding the structure of the Bethe equations.}
\label{DyDiagGrade1}
\end{center}
\end{figure}

\paragraph{Exact Bethe equations from QQ-relations.}
To motivate the definition \eqref{eq:bPDemocracy}, we now show how to reproduce the exact Bethe equations from our QQ-system. A standard way to deduce Bethe equations is to consider the Cartan-matrix, $C_{AB}$, of the underlying (super-)algebra and use it to define Bethe equations as $\prod_{B=1}^{3}\frac{\fQ^{[C_{AB}]}_{B}}{\fQ^{[-C_{AB}]}_{B}} = (-1)^{\frac{C_{AA}}{2}}$ at $\fQ_{A}=0$. There is no unique Cartan matrix in a superalgebra, since there exist different inequivalent choices of positive roots. Below, we consider the specific choice depicted in figure~\ref{DyDiagGrade1}. For this particular choice, the Bethe equations are

\begin{equation}\label{eq:BetheEquationsGrading1}
\colorbox{gray!20}{$\displaystyle
\dfrac{\fQ_{2}^+\fQ_{3}^+}{\fQ_{2}^-\fQ_{3}^-}\Biggl|_{\fQ_1=0}=1$}\;,
\quad
\dfrac{\fQ_{2}^{++}\fQ_1^-}{\fQ_{2}^{--}\fQ_1^+}\Biggl|_{\fQ_{2}=0}=-1\;,
\quad
\dfrac{\fQ_{3}^{++}\fQ_1^-}{\fQ_{3}^{--}\fQ_1^+}\Biggl|_{\fQ_{3}=0}=-1\;.
\end{equation}
Let us now show how to reproduce these equations and identify $\fQ_{A}$ in our formalism. First, we simply write out the definitions of $\bQ_1$ using both \eqref{eq:bQDef} and \eqref{eq:bQDef2}. For the first component of $\bQ_{1}$, we find
\begin{equation}\label{eq:bQ1QQ}
    -Q_{1|1}^+ Q_{2|1}^- + Q_{1|1}^- Q_{2|1}^+ = \bQ_{1}\;,
    \quad
    Q_{\dot{1}|1}^+ Q_{\dot{2}|1}^- - Q_{\dot{1}|1}^- Q_{\dot{2}|1}^+ = \bQ_{1}\;,
\end{equation}
and the third identity we need is
\begin{equation}\label{eq:bQbP1QQ}
    \bP_1\bQ_1 = Q^+_{1|1}Q^+_{\dot{1}|1}-Q^-_{1|1}Q^-_{\dot{1}|1} \,,
\end{equation}
which is a direct consequence of the definitions \eq{eq:DefP}, \eq{eq:bQDef} and \eq{detcond} (for detailed derivation see Appendix~\ref{App:QQ}).

We now shift $u$ in \eqref{eq:bQ1QQ} as $u\rightarrow u \pm \frac{\ii}{2}$ and thereafter evaluate at $\fQ_{1|1}=0$ and $\fQ_{\dot{1}|1} = 0$. In \eqref{eq:bQbP1QQ} we simply evaluate at $\bQ_{1}=0$, this gives the equations
\begin{equation}\label{eq:OurBAE}
    \frac{Q^+_{1|1}Q^+_{\dot{1}|1}}{Q^-_{1|1}Q^-_{\dot{1}|1}}\bigg|_{\bQ_1=0} = 1\,,
    \quad
    \frac{Q^{++}_{1|1}}{Q^{--}_{1|1}}\frac{\bQ^-_1}{\bQ^+_1}\bigg|_{Q_{1|1} = 0} = -1\,,
    \quad
    \frac{Q^{++}_{\dot{1}|1}}{Q^{--}_{\dot{1}|1}}\frac{\bQ^-_1}{\bQ^+_1}\bigg|_{Q_{\dot{1}|1} = 0} = -1\,,
\end{equation}
and upon identifying $\bQ_{1} = \fQ_{1}, \fQ_{2} = Q_{1|1},\fQ_{3} = Q_{\dot{1}|1},$ we reproduce \eqref{eq:BetheEquationsGrading1}. 

\paragraph{Spinor notation and Baxter equations.}
In the previous paragraph, we defined $\bP$ as a vector of $\algso(4)$. However, it is often much more convenient to use spinor indices. We define $\bP_{\dot{a}b}$ as
\begin{equation}
    \bP_{\dot{a} b} \equiv \bP^{A}(\sigma_{A})_{\dot{a}b}\;.
\end{equation}
Substituting this into \eqref{eq:DefP} and \eqref{eq:bPDemocracy} gives
\begin{equation}\label{eq:SpinorPQRel}
    \bP_{\dot{a}b}  = -Q^{+}_{\dot{a}|p} \epsilon^{pk}Q^{-}_{b|k} =
    -Q^-_{\dot{a}|k} \epsilon^{kp} Q^{+}_{b|p}\,,
\end{equation}
which is reminiscent of the ABJM results \eqref{PQABJMspin_1}.
For the next step, it is convenient to introduce the Q-functions with raised indexes, using $\epsilon$
\begin{equation}    Q^{a|k}=\ep^{ab}\ep^{kp}Q_{b|p}~~,~~Q^{\dot{a}|k}=\ep^{\dot{a}\dot{b}}\ep^{kp}Q_{\dot{b}|p}~~,~~\bP^{\dot{a}c}=\ep^{\dot{a}\dot{b}}\ep^{cd}\bP_{\dot{b}d}\,,
\end{equation}
which due to the normalisation \eqref{detcond} are minus inverse of $Q_{a|p}$ with lower indexes
$
Q^{a|p}Q_{a|q}=-\delta^{p}_{q}$ and $Q^{\dot{a}|p}Q_{\dot{a}|q}=-\delta^{p}_{q}$.
This allows us to linearise the equation for $Q$ \eq{eq:SpinorPQRel}
\begin{equation}\label{SpinorPQ1}
    Q_{a|k}^+=(Q^{\dot{b}|l})^-\bP_{\dot{b}a}\ep_{lk}\,,
    \quad
    Q_{a|k}^-=(Q^{\dot{b}|l})^+\bP_{\dot{b}a }\ep_{lk}\,.
\end{equation}
By applying this relation twice, we can get a finite difference equation for $Q_{a|i}$
\begin{equation}\label{eq:BaxterQai}
    Q^+_{a|k} = \bP_{\dot{b}a}\left(\bP^{\dot{b} c} \right)^{[-2]}Q^{[-3]}_{c|k}\,,
    \quad
    Q^+_{\dot{a}|k} = \bP_{\dot{a}b}\left(\bP^{\dot{c}b} \right)^{[-2]}Q^{[-3]}_{\dot{c}|k}\;,
\end{equation}
which we will henceforth refer to as a matrix Baxter equation.

Analogously, one can express $\bQ$ using spinor indices as
\begin{equation}\label{eq:Qspindef}
    \bQ_{kl}=\bQ^I(\Sigma_I)_{kl}\,.
\end{equation}
It turns out to be even more convenient to define an object from which we can readily recover not only $\bQ_{kl}$ but also $\bQ_{\circ}$. We define
\begin{equation}\label{hatQ}
    \hat{\bQ}_{kl}\equiv \bQ_{kl}+\dfrac{1}{2}\bQ_{\circ}\ep_{kl}=\ep^{ab}Q_{a|k}^+Q_{b|l}^-=\ep^{\dot{a}\dot{b}}Q_{\dot{a}|k}^-Q_{\dot{b}|l}^+\,,
\end{equation}
where in order to find the last two equalities, one uses the completeness relations for $\Sigma_I$, we have listed them in \eqref{eq:Fierz}. We also find
\begin{equation}
    \bQ_{I} = -\Sigma^{kl}_{I} \hat{\bQ}_{kl}\,,
    \quad
    \bQ_{\circ} = \epsilon^{kl}\hat{\bQ}_{kl}\,.
\end{equation}

Finally, let us mention the following relation, mapping $\bQ$ to $\bP$
\beq\label{PtoQ}
\bP_{\dot a b}=-Q_{\dot a| i}^+Q_{b| j}^+ \hat\bQ^{ij}=
+Q_{\dot a| j}^-Q_{b| i}^- \hat\bQ^{ij}
\eeq
where we used $\epsilon$ to raise the index in $\hat\bQ_{ij}$.
\subsection{Analytic Structure of Q-Functions}\label{subsec:AnalyticStructure}

In the previous section, we proposed an algebraic formulation of an $\algosp({4|2})$ Q-system. In order to obtain a discrete spectrum from this system, it must be supplemented with analytic constraints on the Q-functions. In this section, in analogy with ABJM and AdS$_3\times$S$^3\times $T$^4$, we conjecture what analytic structure to impose on the the Q-functions in order to reproduce the spectrum of free strings on AdS$_3 \times$S$^{3} \times$S$^3 \times$S$^1$.

The main feature that distinguishes the Q-functions that appear in AdS/CFT integrability, in comparison to other integrable models, is that they feature branch cuts. We will assume the same to be true for our Q-functions and allow for branch points situated at $[-2g+in,2g+in]\,,\, n\in \mathbb{Z}$ (for some Q-functions, like $Q_{a|i}$, the branch points can be shifted by $i/2$). 

The size of the cut is parametrised by parameter $g$. For ${\cal N}=4$ SYM, it is known that $g=\frac{\sqrt{\lambda}}{4\pi}$ with $\lambda$ the 't Hoof coupling, while for ABJM this relation is more complicated and was conjectured in \cite{Gromov:2014eha}. In the current case, we do not have an exactly known dual CFT, but we expect that when $g$ becomes large, the dual classical string theory becomes the right description for operators with large quantum numbers and hence $g$ should be related to the string tension.

\begin{figure}[h]
\begin{center}
    \begin{minipage}{0.3\textwidth}
    \begin{tikzpicture}<2->
    \node[] (a) at (-1.3,1.3) {$\bP$};
    \draw[thick] (-1,-1) rectangle (1,1);
    \node[circle,inner sep=1pt, fill=black] (p1) at (0.5,0) {};
    \node[circle,inner sep=1pt, fill=black] (p1) at (-0.5,0) {};
    \draw[blue] (-0.5,0)--(0.5,0);
    \end{tikzpicture}  
    \end{minipage}
    \begin{minipage}{0.3\textwidth}
    \begin{tikzpicture}<2->
    \node[] (a) at (-1.3,1.3) {$\bQ^{\downarrow}$};
    \draw[thick] (-1,-1) rectangle (1,1);
    \node[circle,inner sep=1pt, fill=black] (p1) at (0.5,0) {};
    \node[circle,inner sep=1pt, fill=black] (p1) at (-0.5,0) {};
    \draw[blue] (-0.5,0)--(0.5,0);
    \node[circle,inner sep=1pt, fill=black] (p1) at (0.5,-0.5) {};
    \node[circle,inner sep=1pt, fill=black] (p1) at (-0.5,-0.5) {};
    \draw[blue] (-0.5,-0.5)--(0.5,-0.5);
    \end{tikzpicture}  
    \end{minipage}
    \begin{minipage}{0.3\textwidth}
    \begin{tikzpicture}<2->
    \node[] (a) at (-1.3,1.3) {$\bQ^{\uparrow}$};
    \draw[thick] (-1,-1) rectangle (1,1);
    \node[circle,inner sep=1pt, fill=black] (p1) at (0.5,0) {};
    \node[circle,inner sep=1pt, fill=black] (p1) at (-0.5,0) {};
    \draw[blue] (-0.5,0)--(0.5,0);
    \node[circle,inner sep=1pt, fill=black] (p1) at (0.5,0.5) {};
    \node[circle,inner sep=1pt, fill=black] (p1) at (-0.5,0.5) {};
    \draw[blue] (-0.5,0.5)--(0.5,0.5);
    \end{tikzpicture}  
    \end{minipage}
\end{center}
    \caption{The analytic structure of the functions $\bP$ and $\bQ^{\downarrow/\uparrow}$.}
    \label{fig:CutStructure}
\end{figure}
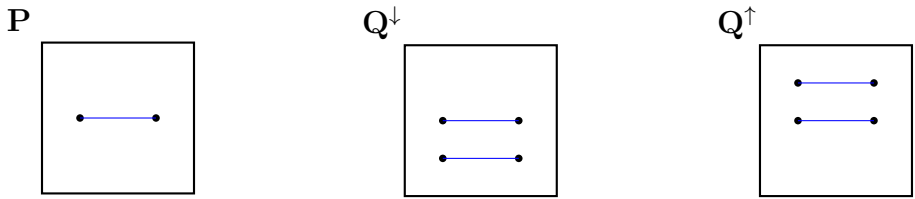

We demand that $\bP_A$ is an analytic function away from a single branch cut on the real axis located at $[-2g,2g]$. This simple structure also dictates the cut structure of $Q_{a|k}$ and $Q_{\dot{a}|k}$ through the Baxter equation (\ref{eq:BaxterQai}). Indeed, the Baxter equation has two types of solutions 
\begin{equation}\nonumber
    Q_{a|i}^{\downarrow} ~-~\text{Upper half-plane analytic (UHPA)}\,,
    \quad
    Q_{a|i}^{\uparrow} ~-~\text{Lower half-plane analytic (LHPA)}\;,
\end{equation}
each one having an infinite ladder of branch cuts.
For real-valued parameters, the two solutions can be simply related by complex conjugation, or in the case of parity symmetric solutions, one can also use $u\to -u$ to generate one set from the other.
The functions $Q^{\downarrow}_{a|i}$ 
in general have cuts at $(-2g-\frac{\ii}{2}-\ii n,2g-\frac{\ii}{2}-\ii n), \, n \in \mathbb{Z}_{\geq0}$ 
while $Q^{\uparrow}_{a|i}$ have cuts at $(-2g+\frac{\ii}{2}+\ii n,2g+\frac{\ii}{2}+\ii n), \, n \in \mathbb{Z}_{\geq0}$. 
These ladders of cuts further propagate into $\bQ_{ij}$ through \eqref{eq:bQDef}. This ensures that $\bQ^{\downarrow}_{ij}$ is analytic in the upper half-plane and $\bQ^{\uparrow}_{ij}$ is analytic in the lower half-plane. We summarise the analytic structure of $\bP$ and $\bQ$ in Figure~\ref{fig:CutStructure}.

The four solutions, $Q^{\uparrow/\downarrow}_{a|i},\;i=1,2$, cannot be independent of each other as they form a set of $4$ solutions of a second-order finite difference Baxter equation~\eq{eq:BaxterQai}. Indeed, they are related to each other by a matrix $\Omega^j_{~i}$
\begin{equation}\label{ud}
Q_{a|i}^{\uparrow}=Q_{a|j}^{\downarrow}(\Omega^j_{~i})^+~,~Q_{\dot{a}|i}^{\uparrow}=Q_{\dot{a}|j}^{\downarrow}(\dot{\Omega}^j_{~i})^+\,,
\quad
\Omega^{[4]} = \Omega\,,
\quad
\dot{\Omega}^{[4]} = \dot{\Omega}\,,
\end{equation}
where the $2\ii$ periodicity follows from the fact that the Baxter equation \eqref{eq:BaxterQai} relates values with a shift by $2\ii$ only, unlike in ${\cal N}=4$ SYM. Notice that $\dot{\Omega}$ and $\Omega$ are related as
\beq\label{WWd}
\dot{\Omega}^j_{~i}=(\Omega^j_{~i})^{[+2]}\;\;,\;\;
{\Omega}^j_{~i}=(\dot\Omega^j_{~i})^{[+2]}
\;,
\eeq
which is a consequence of \eqref{SpinorPQ1} and the fact that we can express $\bP$ in terms of both $Q^{\uparrow/\downarrow}$.

\subsection{Gluing Conditions}\label{subsec:Gluing}
To complete the construction of the QSC, we must also specify the monodromies of $\bQ$ around its branch-points on the real axis. In analogy with all other known iterations of the QSC, the idea is to require that $\bQ$ is a function with a single long cut $(-\infty,-2g)\cup(2g,+\infty)$ and is consistent with QQ-relations. 
This means that, when we analytically continue $\bQ$ through the shortcut on the real axis, we find a new sheet without any cuts in the lower half-plane. There are many ways to accomplish this, but following the AdS$_3\times $S$^3\times$T$^4$ analysis \cite{Cavaglia:2021eqr,Ekhammar:2021pys} we will propose that the answer is to connect two hitherto unrelated $\algosp(4|2)$ QQ-systems upon analytic continuation.

We thus postulate that there exist two copies of so-far independent $\algosp(4|2)$ QQ-systems. We will distinguish these systems using barred indices, i.e $\bQ_{kl}$ is from the first QQ-system and $\bQ_{\bar{k}\bar{l}}$ belongs to the second system. When we can be cavalier about indices, we will allow ourselves to simply write $\bQ,\bar{\bQ}$. We then glue along the short branch-cut according to
\begin{equation}\label{Qdown}
   \hat{\bQ}_{ij}^{\downarrow}(u+\ii0)=F_i^{~\bar{k}}F_{j}^{~\bar{l}}\hat{\bQ}_{\bar{k}\bar{l}}^{\uparrow}(u-\ii0)\,,
   \quad
   \hat{\bQ}_{\bar{i}\bar{j}}^{\downarrow}(u+\ii0)=F_{\bar{i}}^{~k}F_{\bar{j}}^{~l}\hat{\bQ}_{kl}^{\uparrow}(u-\ii0)\,.
\end{equation}
for $u \in (-2g,2g)$. Here $\hat{\bQ}$ was previously defined as \eqref{hatQ}. Alternatively, using our convention for the contour $\gamma$ (see  fig.\ref{fig:gammacontour}), we rewrite (\ref{Qdown}) as
\begin{equation}\label{glue}
(\hat{\bQ}_{ij}^{\downarrow})^{\gamma}=F_i^{~\bar{k}}F_{j}^{~\bar{l}}\hat{\bQ}_{\bar{k}\bar{l}}^{\uparrow}~~,~~(\hat{\bQ}_{\bar{i}\bar{j}}^{\downarrow})^{\gamma}=F_{\bar{i}}^{~k}F_{\bar{j}}^{~l}\hat{\bQ}_{kl}^{\uparrow}\,.
\end{equation}
Note that the way we perform gluing is by transforming separately the two indices. This should be the case as the $F$-matrix must be a symmetry of the QQ-system to ensure consistency with all QQ-relations introduced previously.

\begin{figure}[h]
\begin{center}
    \begin{minipage}{0.3\textwidth}
    \begin{tikzpicture}<2->
    \node[] (a) at (-1.3,1.3) {$\bQ^{\downarrow}$};
    \draw[thick] (-1,-1) rectangle (1,1);
    \node[circle,inner sep=1pt, fill=black] (p1) at (0.5,0) {};
    \node[circle,inner sep=1pt, fill=black] (p1) at (-0.5,0) {};
    \draw[blue] (-0.5,0)--(0.5,0);
    \node[circle,inner sep=1pt, fill=black] (p1) at (0.5,-0.5) {};
    \node[circle,inner sep=1pt, fill=black] (p1) at (-0.5,-0.5) {};
    \draw[blue] (-0.5,-0.5)--(0.5,-0.5);
    \draw[red, thick,dashed,->] ({3/4},0) arc[start angle=0, end angle=180, radius=10pt];
    \end{tikzpicture}  
    \end{minipage}
    \begin{minipage}{0.3\textwidth}
    \begin{tikzpicture}<2->
    \node[] (a) at (-1.3,1.3) {$\bar{\bQ}^{\uparrow}\propto \bQ^{\gamma}$};
    \draw[thick] (-1,-1) rectangle (1,1);
    \node[circle,inner sep=1pt, fill=black] (p1) at (0.5,0) {};
    \node[circle,inner sep=1pt, fill=black] (p1) at (-0.5,0) {};
    \draw[blue] (-0.5,0)--(0.5,0);
    \node[circle,inner sep=1pt, fill=black] (p1) at (0.5,0.5) {};
    \node[circle,inner sep=1pt, fill=black] (p1) at (-0.5,0.5) {};
    \draw[blue] (-0.5,0.5)--(0.5,0.5);
    \draw[red, thick,dashed,->] ({0},0) arc[start angle=180, end angle=360, radius=10pt];
    \end{tikzpicture}  
    \end{minipage}
    \begin{minipage}{0.3\textwidth}
    \begin{tikzpicture}<2->
    \node[] (a) at (-1.5,1.3) {$\bQ^{\downarrow}$};
    \node[] (a) at (-1.5,-0.9) {$\bar{\bQ}^{\uparrow}$};
    \draw[thick] (-1,-1) rectangle (1,1);
    \node[circle,inner sep=1pt, fill=black] (p1) at (0.5,0) {};
    \node[circle,inner sep=1pt, fill=black] (p1) at (-0.5,0) {};
    \draw[blue] (-0.5,0)--(-1,0);
    \draw[blue] (0.5,0)--(1,0);
    \draw[red, thick,dashed,->] ({3/4},{1/10}) arc[start angle=20, end angle=340, radius=10pt];
    \end{tikzpicture}  
    \end{minipage}
\end{center}
    \caption{The analytic structure of the functions $\bP_{a},\bP^{a},\bQ^{\downarrow/\uparrow}_i$ and $(\bQ^{\downarrow/\uparrow})^i$}
    \label{fig:ACQ}
\end{figure}
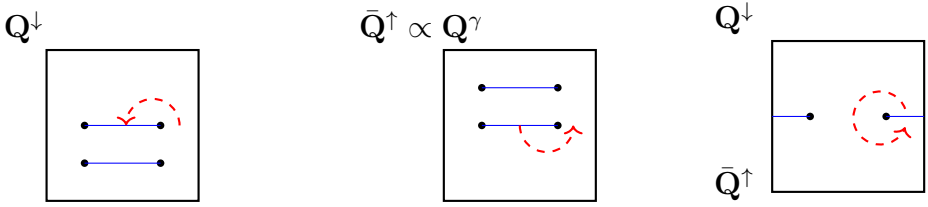

Using \eq{glue} one can combine $\hat \bQ^{\downarrow}_{ij}$, in the upper half plane,
with $\hat \bQ^{\uparrow}_{\bar i\bar j}$ in the lower half plane to create a single function analytic everywhere on the complex plane, except for a long cut $(-\infty,-2g)\cup(2g,+\infty)$ on the real axis. We depict this in Figure~\ref{fig:ACQ}.

As the matrices $F_i^{~\bar{k}}$ and $F_{\bar{i}}^{~k}$ are two constant independent matrices, we find using unimodularity, see \eqref{detcond}, that
 \begin{equation}\label{Fdet}
\det(F_i^{~\bar{k}})^2=\det(F_{\bar{i}}^{~k})^2=1\,.
\end{equation}

\subsection{Asymptotic Behaviour} \label{subsec:AsymptoticSection}
The asymptotic behaviour of Q-functions encodes the quantum number of the state under consideration, which in particular includes $\Delta$ -- the conformal dimension. Motivated by the quasiclassical behaviour of the Q-functions in section~\ref{sec:ABJMQSC}, and taking into account small shifts due to QQ-relations, we find
\begin{equation}\label{asymp}
    \bP_{A} \simeq \mathcal{A}_{A} \, u^{-M_{A}}\,,
    \quad
    \bQ_{K} \simeq \mcB_I \, u^{\hat{M}_I-1}\,,
    \quad
    \bQ_{\circ} \simeq u^{0}\,,
\end{equation}
where the powers are determined by the $\mathfrak{osp}({4|2})$ charges:
\begin{equation}\label{MaMab}
\begin{split}
    M_{A} &= \{(J_{1}-K_1+1),\;(J_2-K_2),\;-(J_2-K_2),\;-(J_1-K_1+1)\}\,,
    \\
    M_{\bar{A}} &= \{(J_1+K_1),\;\;\;\;\;\;\;(J_2+K_2),\;-(J_2+K_2),\;-(J_1+K_1)\},
\end{split}
\end{equation}
and
\begin{equation}\label{MaMab2}
    \hat{M}_{I} = \{+(\Delta - S+1),\; 0,\; - (\Delta - S+1) \}\,,
    \quad
    \hat{M}_{\bar{I}} = \{(\Delta + S),\; 0,\; -( \Delta + S) \}\,.
\end{equation}
Here $\Delta$ and $S$ are the AdS$_{3}$ charges, representing the conformal dimensions and spin of the gauge theory operators, they enter the asymptotics of $\bQ$. The charges corresponding to the two copies of $S^{3}$ enter through the asymptotic of $\bP$. 

The coefficients $\mcA_{A},\mcB_{K}$ are constrained by the algebraic equations of the QQ-system. As we show in Appendix~\ref{app:Vector} the prefactors $\mcA$ and $\mcB$ are constrained by
\begin{align}
    &\mcA_{A}\mcA^{A} = -2\frac{\prod_{I=1}^{3}\left(M_{A}-\hat{M}_{I}\right)}{\prod_{B\neq A}^{4}\left(M_{A}-M_{B}\right)}\,,
    \quad
    \mcB_{K} \mcB^{K} = \frac{1}{2} \frac{\prod_{A=1}^{4}\left( \hat{M}_K - M_A\right)}{\prod_{K \neq I}^{3}\left( \hat{M}_K- \hat{M}_I\right)}.
\end{align}
Let us also include the corresponding asymptotics of the spinorial-index Q-functions
\begin{equation}\label{Pabasymp}
    \bP_{\dot{a} b} \sim u^{\mcN_{\dot{a}}+\mcN_{b}}\,,
    \quad
    Q_{a|k} \sim u^{\mcN_a+\hmcN_k}\,,
    \quad
    Q_{\dot{a}|k} \sim u^{\mcN_{\dot{a}}+\hmcN_{k}}\,,
\end{equation}
where
\begin{equation}
\begin{split}
    \mcN_{a} = \frac{M_{1}+M_{2}}{2}\,\{-1,1\}\,,
    \quad
    \mcN_{\dot{a}} = \frac{M_{1}- M_{2}}{2}\,\{-1,1\}\,,
    \quad
    \hat{\mathcal{N}}_i &=  \frac{\hat{M}_1}{2}\,\{1,-1\}\,,
\end{split}
\end{equation}
and the same applies to the barred system.

This completes our QSC proposal. In the next section, we derive some additional relations that follow from the proposal. We then subsequently test our proposal by taking the large volume limit in Section~\ref{Sect:ABA} and comparing the result with the ABA equations from the literature. 

\section{Deriving $\bQ\tau$ and $\bP\nu$ Systems}\label{Sect:consequences}

In the previous section, we proposed a QSC by providing a QQ-system supplemented with  analytic conditions. In this section, we will recast the gluing conditions into so-called $\bQ \nu$- and $\bP \tau$-systems. These systems will play a key role when we consider the large volume limit in Section~\ref{Sect:ABA}.

\subsection{Q$\tau$-System}
In this section, we derive the $\bQ\tau$-system from the gluing condition (\ref{glue}). Our method largely follows \cite{Bombardelli:2017vhk}. We start by recasting the gluing conditions as analytic continuation and use the fact that we can map UHPA Q-functions to LHPA Q-functions using the matrix $\Omega$ defined in \eqref{ud}. This gives the relations
\begin{equation}\label{qtau}
\left(\hat{\bQ}_{kl}\right)^{\gamma}=\left(\tau_{k}^{~\bar{m}}\right)^{[+2]}\tau_{l}^{~\bar{n}}\hat{\bQ}_{\bar{m}\bar{n}}\;,
\quad
\left(\hat{\bQ}_{\bar{k}\bar{l}}\right)^{\gamma}=\left(\tau_{\bar{k}}^{~m}\right)^{[+2]}\tau_{\bar{l}}^{~n}\hat{\bQ}_{mn}\;,
 \end{equation}
where we recall that $\gamma$ denotes clockwise analytic continuation around the branch points $-2g$ as depicted in Figure~\ref{fig:gammacontour}, and $\tau$ is defined as
\begin{equation}\label{taudef}
\tau_{k}{}^{\bar{l}}=F_k{}^{\bar{m}}(\Omega^{\bar{l}}{}_{\bar{m}})^{+}\,,
\quad
\tau_{\bar{k}}{}^{l}=F_{\bar{k}}{}^{m}(\Omega^{l}{}_{m})^{+}\,.
\end{equation}
Note that $\tau$ inherits $2\ii$ periodicity from $\Omega$ and that its determinant is fixed from \eqref{Fdet}, to summarize
\begin{equation}\label{Taudet}
(\tau_k^{~\bar{l}})^{[+4]}=\tau_k^{~\bar{l}}\,,
\quad
\det(\tau_k^{~\bar{l}})=\pm1\,.
\end{equation}
Of course the same properties hold for $\tau_{\bar{k}}^{~l}$. From \eq{WWd} it is also natural to introduce the following notation 
\beq\la{taudotdef}
\dot\tau\equiv \tau^{[+2]}\,,
\quad
\dot\tau^{[+2]}= \tau\;.
\eeq
similar to the one used in the ABJM QSC.

The function $\tau$ has an infinite number of cuts, separated by $i$ going to infinity in both upper and lower half-planes. The analytic continuation under the cut on the real axis along $\bar\gamma$ can be extracted from the QQ-relations (\ref{eq:SpinorPQRel}) and (\ref{hatQ}). We derive in Appendix \ref{App:QQ} the following relation

\begin{equation}\label{tautau}(\dot\tau_{k}{}^{\bar{m}}\tau_{l}^{~\bar{n}})^{\bar{\gamma}}-\dot\tau_{k}{}^{\bar{m}}\tau_{l}{}^{\bar{n}}=\hat{\bQ}_{kl}(\hat{\bQ}^{\bar{m}\bar{n}})^{\bar{\gamma}}-(\hat{\bQ}_{kl})^{\gamma}\hat{\bQ}^{\bar{m}\bar{n}}.
\end{equation}
The equations \eqref{qtau} and \eqref{tautau} can be used to express  $\bQ$ and $\tau$ on any sheet in terms of their values on the first Riemann sheet, forming a closed system of equations. To make this even more explicit, we will derive relations connecting $\bQ$ with its analytic continuation along $\gamma$ performed consecutively $n$ times.
By adopting a trick from \cite{Cavaglia:2021eqr,Ekhammar:2021pys} we introduce
\begin{equation}
    U_{kl}{}^{\bar{m}\bar{n}}=(U^{-1})_{kl}{}^{\bar{m}\bar{n}}=-\dot\tau_k{}^{\bar{p}}\tau_l{}^{\bar{q}}(\delta^{\bar{m}}_{\bar{p}}\delta^{\bar{n}}_{\bar{q}}-\hat{\bQ}^{\gamma}_{\bar{p}\bar{q}}\hat{\bQ}^{\bar{m}\bar{n}})\;,
\end{equation}
in terms of which \eqref{tautau} reads
\begin{equation}
    \left(U_{kl}{}^{\bar{m}\bar{n}}\right)^{\bar{\gamma}} = U_{kl}{}^{\bar{m}\bar{n}}\;,
\end{equation}
so we learn that $U$ does not have a cut on the real axis. Now, let us note that we can write \eqref{qtau} as
\begin{equation}\label{qgammabar}
\hat{\bQ}_{kl}{}^{\bar{m}\bar{n}}=U_{kl}{}^{\bar{m}\bar{n}}\hat{\bQ}^{\bar{\gamma}}_{\bar{m}\bar{n}}\,.
\end{equation}
Using that $U$ does not have a cut on the real axis, we deduce
\begin{equation}
\hat{\bQ}^{\gamma^n}_{kl}=U_{kl}{}^{\bar{m}\bar{n}}\hat{\bQ}^{\gamma^{n-1}}_{\bar{m}\bar{n}}\;,
\quad
\hat{\bQ}^{\bar{\gamma}^n}_{kl}=(U^{-1})_{kl}{}^{\bar{m}\bar{n}}\hat{\bQ}^{\bar{\gamma}^{n-1}}_{\bar{m}\bar{n}}\;.
\end{equation}
Since there is no reason for the equality $U_{kl}{}^{\bar{p}\bar{q}}U_{\bar{p}\bar{q}}{}^{mn}=\delta^m_l\delta^n_l$ (or, in other words, that it would be over-constraining to impose one), we have to assume that our model, in general, should feature non-quadratic cuts. This makes our proposed QSC significantly different from ABJM, where all Q-functions have quadratic cuts. Of course, this non-quadratic cut behaviour is expected since it has been observed before in the AdS$_3\times$S$^3\times$T$^4$ QSC and TBA \cite{Cavaglia:2021eqr,Ekhammar:2021pys,Frolov:2021bwp,Frolov:2021fmj}.

\subsection{P$\nu$-System}
The $\bP \nu$-system is the analogue of the $\bQ \tau$-system for $\bP$-functions. Using \eqref{PtoQ} we can map $\bQ$ to $\bP$ in the gluing equation (\ref{qtau}). We find
\begin{equation}\label{Pgamma}
    \bP_{\dot{c}a}^{\gamma}=Q_{\dot{c}|l}^{+}Q_{a|k}^{+}Q^{\dot{\bar{e}}|\bar{l}+}Q^{\bar{f}|\bar{k}+}\dot\tau_{~\bar{k}}^{k}\tau_{~\bar{l}}^{l}\bP_{\dot{\bar{e}}|\bar{f}}\;,
\end{equation}
where we have also used that $Q^{+}_{a|k}$ does not have a cut on the real axis. Let us define two matrices $\nu_a^{~\bar{f}}$ and $\nu_{\dot{c}}^{~\dot{\bar{e}}}$ as
\begin{equation}\label{nudef}
\nu_a^{~~\bar{f}}=
Q_{a|k}^{-}Q^{\bar{f}|\bar{k}-}\tau_{~\bar{k}}^{k}~~,~~\nu_{\dot{c}}^{~~\dot{\bar{e}}}=
Q_{\dot{c}|l}^{-}Q^{\dot{\bar{e}}|\bar{l}-}\dot \tau_{~\bar{l}}^{l}\;.
\end{equation}
Since $Q_{a|i}$ and $\tau$ both have a unit determinant, we have
\begin{equation}\label{nudett}
\det(\nu_a^{~\bar{f}})=\pm 1\,,
\quad
\det(\nu_{\dot{c}}^{~\dot{\bar{e}}})=\pm1\,.
\end{equation}
With the definition \eqref{nudef} at hand, \eq{Pgamma} reads
\begin{equation}\label{pmu1}
\bP^{\gamma}_{\dot{c}a}=(\nu_a^{~\bar{f}}\nu_{\dot{c}}^{~\dot{\bar{e}}})^{[+2]}\bP_{\dot{\bar{e}}\bar{f}}\;.
\end{equation}
Importantly, $\nu_a^{~\bar{f}}\nu_{\dot{c}}^{~\dot{\bar{e}}}$ is \textit{mirror-periodic} (i.e. periodic up to an analytic continuation under the cut):
\begin{equation}\label{mupp}
(\nu_a^{~\bar{f}}\nu_{\dot{c}}^{~\dot{\bar{e}}})^{[+2]}=(\nu_a^{~\bar{f}}\nu_{\dot{c}}^{~\dot{\bar{e}}})^{\gamma}\,
\end{equation}
as we show in detail in Appendix \ref{QwPm Appendix}.  
Thus, the $\bP\nu$-system can be cast into a more natural form using the $\bar\gamma$ contour, just as in AdS$_3 \times$S$^3\times$T$^4$ \cite{Cavaglia:2021eqr,Ekhammar:2021pys}. We find
\beq
\bP^{\bar\gamma}_{{\dot{e}}{f}}=
\nu^{\dot{\bar{c}}}_{~{\dot{e}}}
\nu^{\bar a}_{~{f}}\bP_{\dot{\bar{c}}\bar a}\;,
\eeq
where $\nu^*_{~*}$ are obtained by raising and lowering the indices with $\epsilon$ from $\nu_*^{~*}$ e.g.
\beq
\nu^{a}_{~\bar b}\equiv -\epsilon^{a c}\epsilon_{\bar b{\bar d}}\nu_{c}^{~\bar d}\;.
\eeq

Furthermore, let us show that the functions $\nu$ independently satisfy mirror-periodicity, much like in the ABJM case. To see this, we invert $(\nu_{\dot{c}}^{~\dot{\bar{e}}})^{[+2]}$ on the l.h.s of (\ref{mupp}) to get
\begin{equation}(\nu_a^{~\bar{f}})^{[+2]}=\left(-\frac{(\nu^{\dot{c}}_{~\dot{\bar{e}}})^{[+2]}(\nu_{\dot{c}}^{~\dot{\bar{e}}})^{\gamma}}{2\det(\dot{\nu})}\right)(\nu_a^{~\bar{f}})^{\gamma}\;.\end{equation}
It will be convenient to denote the term in the brackets here as $e^{\ii \mathcal{P}}$. Then, from the above equation, along with (\ref{mupp}), it follows that 
\begin{equation}
\label{nugamma}(\nu_a^{~\bar{f}})^{[+2]}=e^{\ii\mathcal{P}}(\nu_a^{~\bar{f}})^{\gamma}~~~,~~~(\nu_{\dot{c}}^{~\dot{\bar{e}}})^{[+2]}=e^{-\ii\mathcal{P}}(\nu_{\dot{c}}^{~\dot{\bar{e}}})^{\gamma}\,.
\end{equation}
By taking determinant of \eqref{nugamma} and keeping \eqref{nudett} in mind we get $e^{\ii\mathcal{P}}=\pm1$.
This is similar to the ABJM case, constrained to a symmetric sector (relating upper and lower indexes), as in our case, they are related by definition with the $\epsilon$ tensor.

There exists a discontinuity relation also for $\nu$; it reads
\begin{equation}\label{nunu}(\nu_{\dot{c}}^{~\dot{\bar{e}}}\nu_a^{~\bar{f}})^{\gamma}-\nu_{\dot{c}}^{~\dot{\bar{e}}}\nu_a^{~\bar{f}}=(\bP_{\dot{c}a})^{\gamma}\bP^{\dot{\bar{e}}\bar{f}}-\bP_{\dot{c}a}(\bP^{\dot{\bar{e}}\bar{f}})^{\bar{\gamma}}\;.
\end{equation}
To derive this equation, we simply start from the discontinuity equation for $\tau$ and apply $Q_{a|k}$ to map $\tau$ to $\nu$.

Finally, let us emphasise that the $\bP\nu$-system allows us to express $\bP$ and $\nu$ on any Riemann sheet in terms of their values on the main sheet. Following the same method as for the $\bQ \tau$ system, we find the following recursive relations
\begin{equation}
\bP_{\dot{c}a}^{\gamma^n}=W_{\dot{c}a}^{~~\dot{\bar{e}}\bar{f}}\bP_{\dot{\bar{e}}\bar{f}}^{\gamma^{n-1}}~~,~~\bP_{\dot{c}a}^{\bar{\gamma}^n}=(W^{-1})_{\dot{c}a}^{~~\dot{\bar{e}}\bar{f}}\bP_{\dot{\bar{e}}\bar{f}}^{\bar{\gamma}^{n-1}}\,,
\end{equation}
where
\begin{equation}
    W_{\dot{c}a}^{~~\dot{\bar{e}}\bar{f}}=(W^{-1})_{\dot{c}a}^{~~\dot{\bar{e}}\bar{f}}=-(\delta^{\dot{d}}_{\dot{c}}\delta^b_a-\bP_{\dot{c}a}\bP^{\dot{d}b})\nu_{b}{}^{\bar{f}}\nu_{\dot{d}}{}^{\dot{\bar{e}}}\,.
\end{equation}

\subsection{The Symmetric Sector}\label{sec:SymmetricSector}
In this section, we discuss a large subsector of our proposed QSC, wherein we identify Q-functions with dotted indices alongside their undotted counterparts. We will call this sector the \textit{symmetric sector}. The symmetric sector is characterised by the equation
\begin{equation}\label{symsecP}
    \bP_{\dot{a}b} = \bP_{a\dot{b}}\,.
\end{equation}
By definition, $\bP_{a\dot{b}}\equiv \bP_{\dot{b}a}$, and hence \eqref{symsecP}, reduce to $\bP_3=\bP_2$. This, in particular, implies that $M_2=0$ is in the symmetric sector.

Using \eqref{symsecP}, it follows that $Q_{a|k}$ and $Q_{\dot{a}|k}$ satisfy the same Baxter equation. They furthermore have the same asymptotics and are both analytic in the upper half-plane. Hence, they must be related by a constant linear transformation. We can write
\begin{equation}
    Q_{\dot{a}|k}=K_{\dot{a}}^{~a}Q_{a|l}G_{k}^{~l}\;,
\end{equation}
where both matrices $K$ and $G$ are constant, diagonal, and obey $\det(K)\det(G)=1$. They can further be restricted by (\ref{symsecP}), since $\bP$ is related to $Q_{a|i}$ by (\ref{eq:SpinorPQRel}). We find
\begin{equation}
\label{QdQrel}
\begin{split}
    K_{\dot{1}}^{~1}G_{1}^{~1} &=-K_{\dot{2}}^{~2}G_{2}^{~2}\,,
    \quad 
    K_{\dot{2}}^{~2}G_{1}^{~1}=-K_{\dot{1}}^{~1}G_{2}^{~2}\,,
\\
    K_{\dot{1}}^{~1}G_{1}^{~1} &=-K_{\dot{1}}^{~1}G_{2}^{~2}\,,
    \quad
    K_{\dot{2}}^{~2}G_{1}^{~1}=-K_{\dot{2}}^{~2}G_{2}^{~2}\;,
\end{split}
\end{equation}
which can be reduced to
\begin{equation}\label{QdQrel2}
K_{\dot{1}}^{~1}=K_{\dot{2}}^{~2}\,,
\quad
G_{1}^{~1}=-G_{2}^{~2}\,.
\end{equation}
Without loss of generality, we choose $K_{\dot{1}}^{~1}=K_{\dot{2}}^{~2}=1$ and arrive at
\begin{equation}
    K_{\dot{a}}^{~a}=\begin{pmatrix}
        1&0\\0&1
    \end{pmatrix}~~~,~~~G_{k}^{~l}=\begin{pmatrix}
        -i&0\\0&i
    \end{pmatrix}\;,
\end{equation}
resulting in the following simple relation between $Q_{a|k}$ and $Q_{\dot{a}|k}$
\begin{equation}\label{symsecQai}
    Q_{\dot{a}|k}=\begin{pmatrix}
        -i\, Q_{1|1}&i\, Q_{1|2}\\-i \,Q_{2|1}&i\, Q_{2|2}
    \end{pmatrix}\;.
\end{equation}
With the matrices $K$ and $G$ fixed, there is one extra Q-function constraint left to discuss. We recall that $Q_{a|k}$ and $Q_{\dot{a}|k}$ were not independent of each other to begin with. Indeed, there are two ways of defining $\bQ_{ij}$ (\ref{hatQ}), which can be viewed as an additional restriction on the Q-system. It is inherited in the symmetric sector and looks as follows
\begin{equation}
    \hat{\bQ}_{kl}= - G_{k}^{~m}G_{l}^{~n}\hat{\bQ}_{nm}
    \implies \hat{\bQ}_{12}=-\hat{\bQ}_{21}\,,\bQ_{2}=0\,.
\end{equation}
Lastly, let us address the $\bP\nu$ and $\bQ\tau$ systems. We first note that the matrices $\tau$ and $\dot{\tau}$ are related to each other. Indeed, from \eqref{ud} and substituting \eqref{symsecQai}, we find
\begin{equation}
    \dot{\tau}_{k}^{~\bar{l}}=(-1)^{k+\bar{l}}\tau_{k}^{~\bar{l}}\;.
\end{equation}
Recalling that $\dot{\tau}=\tau^{[+2]}$, we see that components of $\tau$ are now either $i$-periodic or $i$-anti-periodic. Finally, since $K$ is simply the identity matrix, we have
\begin{equation}\label{nusymsec}
    \nu_{\dot{c}}^{~\dot{\bar{e}}}=\begin{pmatrix}
        \nu_{1}^{~\bar{1}}&\nu_{1}^{~\bar{2}}\\\nu_{2}^{~\bar{1}}&\nu_{2}^{~\bar{2}}
    \end{pmatrix}\;.
\end{equation}

\section{The Large Volume Limit and the Asymptotic Bethe Ansatz}\label{Sect:ABA}

To test our proposal, we now solve the QSC in the so-called large volume limit and compare our results to the ABA in the literature \cite{OhlssonSax:2011ms,Borsato:2012ss,Borsato:2016xns}. The large volume limit of the QSC has been considered before in AdS$_5 \times$S$^5$, AdS$_4 \times \mathbb{CP}^3$, and AdS$_3\times$S$^3\times$T$^{4}$  \cite{Gromov:2014caa,Bombardelli:2017vhk,Cavaglia:2021eqr,Ekhammar:2021pys,Ekhammar:2024kzp} where the expected ABA equations were recovered. We will follow the procedure outlined in these papers quite closely.

In the large volume limit, the various dressing phases that play a crucial role in the ABA can be recovered from the QSC. This is particularly non-trivial in AdS$_3\times$S$^3 \times$T$^4$ since this background features several distinct dressing phases between particles in different wings (dotted and undotted) and also with massless particles~\cite{Ekhammar:2024kzp}. Finding the exact form of the dressing phases in AdS$_3\times$S$^3 \times$T$^4$ is, in general, tricky, especially in light of the lack of reliable comparisons from gauge theory. Several iterations of these phases have been proposed using non-QSC methods \cite{Borsato:2013hoa,Frolov:2021fmj,Frolov:2025ozz}, and remarkably, the most up-to-date version, which appeared after \cite{Ekhammar:2024kzp}, perfectly agrees with the QSC calculation presented in \cite{Ekhammar:2024kzp}. 

In AdS$_3\times$S$^3\times$S$^3\times$S$^1$,  there are currently no proposals for the dressing phases; only the crossing equations they are expected to obey have been written down \cite{Borsato:2012ud,Borsato:2015mma}. Thus, we will only be able to compare crossing equations, not solutions. 

\subsection{The Large Volume Limit}

The large volume limit is characterised by some of the quantum numbers becoming large. To be precise, we will assume that $\Delta$ and $J_1$ are of order $L\gg1$. 
We will assume that $\bQ$ and $\bP$ scale according to their large $u$ behaviour, which we will keep track of using the small parameter $\ep \simeq u^{-L}$. From the asymptotics \eqref{asymp} we find the following scalings
\begin{equation} \label{pqlarge}
\begin{split}
&\bP_1\sim\ep\,,
\, \bP_2\sim1\,,\,
\bP_3\sim1\,,\,
\bP_4\sim1/\ep\,,\,
\bQ_1\sim1/\ep\,,\,
\bQ_2\sim1\,,\,
\bQ_3\sim\ep\,, \\
&Q_{a|i}\sim Q_{\dot{a}|k}\sim\begin{pmatrix}
    1&\ep\\1/\ep&1
\end{pmatrix}\,,
\quad
Q^{a|k}\sim Q^{\dot{a}|k}\sim\begin{pmatrix}1&1/\ep\\\ep&1\end{pmatrix}\,.
\end{split}
\end{equation}
Here ``$\sim$'' denotes the leading-order dependence on the asymptotically small parameter $\ep$. The $Q$-functions in the second $\algosp({4|2})$ copy have the same asymptotic behaviour. Due to its $2\ii$ periodicity, $\tau$ cannot scale with $\epsilon$ and must behave as $\tau_{k}^{~\bar{l}}\sim1$. On the other hand, $\nu$ does scale non-trivially; we find 
\begin{equation}\label{nudefaba}
    \nu_{1}^{~\bar{2}}\simeq Q_{1|1}^{-}Q_{\bar{1}|\bar{1}}^{-}\tau^{1}_{~\bar{2}}\,,
    \quad
    \nu_{\dot{1}}^{~\bar{\dot{2}}}\simeq Q_{\dot{1}|1}^{-}Q_{\dot{\bar{1}}|\bar{1}}^{-}\dot\tau^{1}_{~\bar{2}}\,,
\end{equation}
as follows from the scaling of $Q_{a|k}$ and $Q_{\dot{a}|k}$ in \eqref{pqlarge}.

The scaling of $\nu$ also implies that the analytic continuation of $\bP$ simplifies significantly. In particular, upon using \eqref{nudefaba} in \eqref{pmu1} for $\bP_{\dot{1}1} = \bP_{1}$ we find
\begin{equation}\label{pmu1aba}
(\bP_{1})^{\gamma}\simeq Q_{1|1}^{+}Q_{\dot{1}|1}^+
\dot\tau^1_{~\bar{2}}\;
\tau^{1}_{~\bar{2}}\;\bQ_{\bar{1}}\;.
\end{equation}

Also, the matrix $\Omega$ that maps UHPA Q-functions to LHPA Q-functions simplifies considerably in the large volume limit. Namely, it must be diagonal. This follows from the fact that $\Omega$ is periodic and hence cannot scale with $\epsilon$. Yet, from its definition in terms of $Q_{a|k}$, the off-diagonal terms naively scale with $\epsilon$, which allows us to conclude that these components must be $0$. Thus, in the ABA limit
\begin{equation}\label{udaba}
    Q_{1|1}^{\uparrow}\propto Q_{1|1}^{\downarrow}(\Omega^1_{~1})^+~,~Q_{\dot{1}|1}^{\uparrow}\propto Q_{\dot{1}|1}^{\downarrow}(\Omega^1_{~1})^{[+3]}\,.
\end{equation}
From the property of $\Omega$ it then follows from \eqref{udaba} that
\begin{equation}\label{udaba1}
    Q_{1|1}^{\uparrow}\propto Q_{1|1}^{\downarrow}(\tau_{\bar{2}}^{~1})^+~,~Q_{\dot{1}|1}^{\uparrow}\propto Q_{\dot{1}|1}^{\downarrow}(\tau_{\bar{2}}^{~1})^{[+3]}\; .
\end{equation}
We will now use these simplifications to construct all the key QSC quantities entering into the ABA equations.

\subsection{Finding $\mu$ and $\omega$} 
We will start by finding explicit expressions for $\nu$ and $\tau$ in the large volume limit. To do so, it is convenient to focus on the following bilinear combinations
\begin{equation}\label{mununu}
    \mu\equiv \nu_{1}{}^{\bar{2}}\nu_{\dot{1}}{}^{\dot{\bar{2}}}\,,
    \quad 
    \omega\equiv \tau^{1}{}_{\bar 2}\;\dot\tau^{1}{}_{\bar 2}\,.
\end{equation}
From \eq{nugamma} and \eq{taudotdef}, it follows that
\beq\la{mugamma}
\mu^{[+2]}=\mu^\gamma\,,
\quad
\omega^{[+2]}=\omega\,.
\eeq
Furthermore, from \eq{nudefaba} we obtain the large volume limit
\beq\la{muQQQQomega}
\mu \simeq Q^-_{1|1}Q^-_{\bar 1|\bar1}Q^-_{\dot 1|1}
Q^-_{\bar{\dot 1}|1}\;\omega\;.
\eeq
As recently understood in the context of AdS$_3\times$S$^3\times$T$^4$ \cite{Ekhammar:2024neh,Ekhammar:2025vig}, with important input from the BFKL regime of $\mathcal{N}=4$ \cite{Ekhammar:2024kzp}, to find massive Bethe equations, one should assume that $\mu$ has a finite number of zeros $u_i$ at a finite distance from the cut on the defining Riemann sheet (when $g\to 0$),
whereas the massless modes correspond to the roots sitting right on the cut on the real axis. For simplicity, we consider only massive excitations in the main text and present the generalisation to massless modes in Appendix~\ref{MasslessABA}.

We will introduce a polynomial $\mathbb{Q}$ that collects all massive roots of $\mu^+$
\begin{equation}
\mathbb{Q}(u)=\prod_{i=1}^M(u-u_i)\;\;,\;\;
\left.\mu^+
\right|_{u=u_i}=0
\,,
\end{equation}
so that the combination $\mu^+/\mathbb{Q}$ does not have zeros or poles on the main sheet.
Now, let us consider the following function
\begin{equation}\label{abaf}
F^2\equiv\dfrac{\mu}{\mu^{[+2]}}\dfrac{\mathbb{Q}^+} {\mathbb{Q}^-}
=
\frac{\mu}{\mu^{\gamma}}\dfrac{\mathbb{Q}^+} {\mathbb{Q}^-}
\,,\end{equation}
which is constructed in such a way that it has no zeros or poles on the first sheet, and we used \eq{mugamma} in the second equality. Furthermore, because of \eq{muQQQQomega} and the periodicity of $\omega$ \eq{mugamma}, we also have
\beq\la{FinQQ}
F^2=\frac{Q^-_{1|1}Q^-_{\bar 1|\bar1}Q^-_{\dot 1|1}
Q^-_{\bar{\dot 1}|1}}{Q_{1|1}^{+}Q_{\bar 1|\bar1}^{+}Q_{\dot 1|1}^{+}
Q_{\bar{\dot 1}|1}^{+}}\frac{\betheQ^+}{\betheQ^-}\;,  
\eeq  
which immediately tells us that $F^2$ is analytic in the upper half plane. From \eq{udaba}, we immediately conclude that $F^2$ does not have any cuts in the lower half plane either. Finally, \eq{FinQQ} implies that $F^2\to 1$, as $u\to\infty$, due to the power-like asymptotic behaviour of $Q_{a|i}$.
To fix the sign of $F$, we require $F\to 1$.

Following the previous analysis of the AdS$_3\times$S$^3\times$T$^4$ QSC \cite{Cavaglia:2021eqr,Ekhammar:2021pys,Ekhammar:2024kzp}, we will make an additional assumption: we assume that $\mu$ has square-root branch cuts in the large volume limit. The square-root cut assumption for $\mu$ implies that $F$ is also a function with square roots. Assuming this is the case, from the last equality in \eq{abaf}, it follows that 
\begin{equation}\label{FF}
FF^{\gamma}=\pm\dfrac{\mathbb{Q^+}}{\mathbb{Q}^-}\,.\end{equation}
To fix the sign, we can evaluate this expression right at the branch point, where $F^\gamma=F$ and $\mu=\mu^\gamma$ are located, implying a plus above, which sets $FF^{\gamma}=\dfrac{\mathbb{Q^+}}{\mathbb{Q}^-}$.

We know that $F$ has a cut with a single cut and no zeros or poles on the main sheet. From \eq{FF}, we learn that on the second sheet, there is still only one cut; however, there are also some zeros and poles. This implies that $F$ is a rational function of the Zhukovsky variable $x$ with no zeros or poles outside the unit circle (corresponding to the upper sheet in $u$); thus, we have the following solution of \eqref{FF}
\begin{equation}\label{Fsol}
   F= \(\prod_{i=1}^M\frac{\sqrt {x_i^-}}{\sqrt {x_i^+}}\) \dfrac{B_{(+)}}{B_{(-)}}\;,
\end{equation}
where
\begin{equation}\label{Bpm}
    B_{(\pm)}(u)=\prod_{i=1}^M\sqrt{\frac{g}{x_i^{\mp}}}\left(\frac{1}{x(u)}-x_i^{\mp}\right)
\;\;,\;\;R_{(\pm)}(x)\equiv B_{(\pm)}(\tfrac1x)\;\;,\;\;
    R_{(\pm)}B_{(\pm)}=(-1)^{M}\mathbb{Q}^{\pm}\,,
\end{equation}
and the factor in \eq{Fsol} is fixed to ensure $F\to 1$. Furthermore, from \eq{Bpm} we see that
\beq
F F^\gamma = 
\(\prod_{i=1}^M\frac{ {x_i^-}}{ {x_i^+}}\) \dfrac{
{\mathbb Q}^+
}{
{\mathbb Q}^-
}\;,
\eeq
which, together with \eq{FF}, implies the momentum quantisation condition
\begin{equation}\label{mom}
    \prod_{i=1}^M\frac{x_i^+}{x_i^-}= 1\;. 
\end{equation}

\paragraph{Finding $\mu$.} 
Having found $F$, we can proceed with $\nu$ and $\tau$. We start by plugging \eqref{Fsol} into \eqref{abaf} which together with \eqref{mom} results in a finite-difference equation 
\begin{equation}\label{eq:nunuFiniteDifference}
\dfrac{\mu}{\mu^{[+2]}}=\dfrac{\mu}{\mu^{\gamma}}=\(\dfrac{B_{(+)}}{B_{(-)}}\)^2\;.
\end{equation}
To solve \eqref{eq:nunuFiniteDifference}, we introduce an upper half-plane analytic function $\fd$ and its complex conjugate $f^*$\footnote{the infinite products in the definition of $f$ are divergent, but the divergence can be absorbed into $u$ independent constant. In other words, in the products we assume the regularisation where we first compute the derivative  $d_u\log f$ as a convergent sum and then integrate in $u$.}
\begin{equation}\label{fud}
    \dfrac{\fd}{\fd^{[+2]}} = \dfrac{{\bf B}_{(+)}}{{\bf B}_{(-)}}\;\;,\;\;
    f \propto \prod_{n=0}^{\infty}\frac{{\bf B}^{[+2n]}_{(+)}}{{\bf B}^{[+2n]}_{(-)}}\;\;,\;\;f^* \propto \prod_{n=0}^{\infty}\frac{{\bf B}^{[-2n]}_{(-)}}{{\bf B}^{[-2n]}_{(+)}}\,,
\end{equation}
where
\beq\label{BRdef}
\mathbf{B}_{( \pm)}=\prod_{k=1}^M\left(1-\frac{1}{x x_k^{\mp}}\right), \quad \mathbf{R}_{( \pm)}=\prod_{i=1}^M\left(x-x_k^{\mp}\right), \quad \mathbf{B}_{( \pm)} \mathbf{R}_{( \pm)}=h^{-M} \mathbb{Q}^{ \pm}\;.
\eeq
Using these new functions, we solve \eqref{eq:nunuFiniteDifference} to find
\begin{equation}\label{wmuaba}
    \mu \propto\mathbb{Q}^-\fu^{--}\fd\,.
\end{equation}

\paragraph{Finding $\omega$.}
In order to find $\omega$, we first use \eq{FinQQ}
to determine
\begin{equation}\label{4q}
Q_{1|1}Q_{\bar{1}|\bar{1}}Q_{\dot{1}|1}Q_{\dot{\bar{1}}|\bar{1}}\propto\mathbb{Q}\;\fd^2\;,
\end{equation} 
as follows from UHP analyticity of the l.s.h. After that, from \eq{muQQQQomega} we immediately determine $\omega$
\begin{equation}\label{wmuaba_2}
\omega \propto\dfrac{f^{*--}}{f}\,.
\end{equation}
The analysis presented above closely follows that of the AdS$_3 \times$S$^3\times$T$^4$ case. However, from this point on, we will find significant differences.

\paragraph{The dispersion relation.}
From \eqref{mununu} together with \eqref{nudefaba} we find the asymptotic of $\mu$ to be
\begin{equation}\label{eq:AsymptoticMu}
    \mu \sim u^{2\Delta-2J_1}\,.
\end{equation}
We can now compare this to our exact expression. The only non-trivial asymptotic behaviour is found in $f$, which behaves as
\begin{equation}
\begin{split}
    \log\(\dfrac{f}{f^{[2]}}\)&=\log\(\dfrac{\mathbf{B}_{(+)}}{\mathbf{B}_{(-)}}\)\sim\dfrac{\gamma}{iu}~\Rightarrow~f\sim u^{\gamma}\;,
    \\
    \gamma &= \ii g \sum_{\alpha\in \{1|1,\dot{1}|1,\bar{1}|\bar{1},\dot{\bar{1}}|\bar{1}\}}\sum_{k=1}^{K_{\alpha}}\left(\frac{1}{x^+_{\alpha,k}}-\frac{1}{x^-_{\alpha,k}}\right)\,.\label{eq:gammaRoots}
\end{split}
\end{equation}
Comparing then \eqref{wmuaba} with \eqref{eq:AsymptoticMu} we find
\begin{equation}
    \Delta = J_1+\frac{M}{2}+ \gamma\,,
\end{equation}
reproducing nicely the expected anomalous part of $\Delta$ in terms of momentum-carrying roots \cite{Borsato:2012ss}.

\subsection{Finding $Q_{a|k}$}\label{subsec:FindingQak}
In this section, we find $Q_{\alpha}\,, \alpha\in\{1|1,\,\bar{1}|\bar{1},\,\dot{1}|1,\,\dot{\bar{1}}|\bar{1}\},$ in the ABA limit. We start from \eq{4q}.
We see that all zeros of the l.h.s. are contained in the polynomial $\mathbb{Q}$, and $f$ is also defined as a product overt the roots of $\mathbb{Q}$. We thus split
\begin{equation}\label{fnote}
    \mathbb{Q}=\prod_{\alpha}\mathbb{Q_\alpha}\;,
    \quad
    \mathbb{Q}_{\alpha} = \prod^{K_{\alpha}}_{i=1}(u-u_{\alpha,i})\;,
    \quad
    \fd=\prod_\alpha \fd_\alpha\;,
    \quad
    \alpha\in\{1|1,\,\bar{1}|\bar{1},\,\dot{1}|1,\,\dot{\bar{1}}|\bar{1}\}\,
\end{equation}
where
\begin{equation}
    \dfrac{f_\alpha^{++}}{f_\alpha}=\dfrac{\bB_{\alpha,(-)}}{\bB_{\alpha,(+)}}\;,
    \quad
    \bB_{\alpha,(\pm)} =
    \prod_{i=1}^{K_\alpha}\left(1-\frac{1}{x(u) x_{\alpha,i}^{\mp}}\right)\;.
\end{equation}
This notation allows us to write the most general ansatz for $Q_{a|k}$ compatible with \eqref{4q} as
\begin{equation}\label{1qsplit}
    Q_\alpha^{{\downarrow}}\propto \mathbb{Q}_{\alpha}\sqrt{f^{+}} h_\alpha^+(u)\;\;,\;\;\alpha\in\{1|1,\;\bar{1}|1,\;\dot{1}|1,\;\dot{\bar{1}}|1\}\;,
\end{equation}
where $h_\alpha$ is a UHP analytic function with no zeros or poles on the main sheet. Due to \eq{4q}, the $h_{\alpha}$ is subject to the constraint
\begin{equation}   h_{1|1}h_{\bar{1}|1}h_{\dot{1}|1}h_{\dot{\bar{1}}|\bar{1}}\propto1\;.
\end{equation}

To find further constraints on $h$'s, we need to consider the individual $\nu$'s, not only their product. From \eq{nudefaba}, we identify the zeros of $\nu$ and write a general ansatz
\begin{equation}\label{nusplit}
    \nu_{1}^{~\bar{2}}\propto\mathbb{Q}^-_{1|1}\mathbb{Q}^{-}_{\bar{1}|\bar{1}}\sqrt{f^{*--}f}\cdot\mathcal{F}\;,
    \quad \nu_{\dot{1}}{}^{\dot{\bar{2}}}\propto\mathbb{Q}^-_{\dot{1}|1}\mathbb{Q}^{-}_{\dot{\bar{1}}|\bar{1}}\sqrt{f^{*--}f}\cdot\mathcal{F}^{-1}\,,
\end{equation}
where ${\cal F}$ has no zeros on the main sheet.
Similarly, for the barred system, we make the following ansatz:
\begin{equation}\label{nusplitbar}
    \nu_{\bar{1}}{}^{2}\propto\mathbb{Q}^-_{1|1}\mathbb{Q}^{-}_{\bar{1}|\bar{1}}\sqrt{f^{*--}f}\cdot\mathcal{\bar{F}}\;,
    \quad
    \nu_{\dot{\bar{1}}}^{~\dot{2}}\propto\mathbb{Q}^-_{\dot{1}|1}\mathbb{Q}^{-}_{\dot{\bar{1}}|\bar{1}}\sqrt{f^{*--}f}\cdot\mathcal{\bar{F}}^{-1}\;.
\end{equation}
From these new expressions, we can also write individual $\tau$'s via \eq{nudefaba} as follows:
\begin{equation}\label{tausplit}
\tau^{1}_{~\bar{2}}\propto\sqrt{\dfrac{f^{*--}}{f}}\mathcal{F}(h_{1|1}h_{\bar{1}|\bar{1}})^{-1}\quad,\quad
\dot\tau^{1}_{~\bar{2}}\propto\sqrt{\dfrac{f^{*--}}{f   }}\mathcal{F}^{-1}(h_{\dot{1}|1}h_{\dot{\bar{1}}|\bar{1}})^{-1}\;.
\end{equation}

Now let us constrain $\mathcal{F}$ and $h_{\alpha}$ using the analytic properties of our QQ-system. Understanding these functions is crucial for identifying all dressing phases in the large volume limit, as we will describe in Section~\ref{subsec:ABAEquations}.

\paragraph{Constraining $h$'s.}
To further constrain $h_\alpha$, we  use \eqref{udaba1}, which relates $Q^{\uparrow}_{\alpha}$ and $Q^{\downarrow}_{\alpha}$. We note that a standard way to construct $Q^{\downarrow}_{\alpha}$ is to take the complex conjugate of $Q^{\downarrow}_{\alpha}$; we will denote this by $*$, just as we did for $f$. We then substitute \eqref{1qsplit} and \eqref{tausplit} into \eqref{udaba1}, along with their barred equivalents, to find
\begin{equation}\label{FH2}
    \mathcal{F}^{+}\propto h_{\bar{1}|\bar{1}}^+h^{*-}_{1|1}\propto \left(h^{+}_{\dot{\bar{1}}|\bar{1}}h_{\dot{1}|1}^{*-}\right)^{-1}\,,
    \quad
    \bar{\mathcal{F}}^+ \propto h^{+}_{1|1}h_{\bar{1}|\bar{1}}^{*-}\propto \left(h^{+}_{\dot{1}|1}h_{\dot{\bar{1}}|\bar{1}}^{*-}\right)^{-1}\,.
\end{equation}
A lot of information can be extracted from equation (\ref{FH2}) by utilising the analytic structure of $h$. We recall that $h_{\alpha}$ are UHP analytic, hence  $h_{\alpha}^*$'s are LHP analytic. For example, $h_{\bar{1}|\bar{1}}^+h^{*-}_{1|1}\propto \left(h^{+}_{\dot{\bar{1}}|\bar{1}}h_{\dot{1}|1}^{*-}\right)^{-1}$
can be written as $h_{\bar{1}|\bar{1}}^+h^{+}_{\dot{\bar{1}}|\bar{1}}\propto \left(h_{\dot{1}|1}^{*-}h^{*-}_{1|1}\right)^{-1}$, which tells us that $h_{1|1}\propto h_{\dot{1}|1}^{-1}$. This is so because
the l.h.s. has no cuts in the complex plane above $-i/2$, and the r.h.s. is analytic below $i/2$; thus, these products are analytic functions without zeros. As the asymptotic behaviour is at most power-like, we conclude that each of the products must be a constant. Following this argument, we get
\begin{equation}
    h_{1|1}\propto h_{\dot{1}|1}^{-1}\,,
    \quad
    h_{\bar{1}|\bar{1}}\propto h_{\dot{\bar{1}}|\bar{1}}^{-1}\;.
\end{equation}
We then simply denote 
\begin{equation}
    h = h_{1|1} \propto \frac{1}{h_{\dot{1}|1}}\,,
    \quad
    \bar{h} = h_{\bar{1}|\bar{1}} \propto \frac{1}{h_{\dot{\bar{1}}|\bar{1}}}\;,
\end{equation}
and so we find
\begin{equation}\label{1q}
\begin{split}
    Q_{1|1}^{{\downarrow}}&\propto \mathbb{Q}_{1|1}\sqrt{f^{\downarrow+}} h^+\;,
    \quad
    Q_{\dot{1}|1}^{{\downarrow}}\propto \mathbb{Q}_{\dot{1}|1}\sqrt{f^{\downarrow+}} (h^+)^{-1}\;, \\
    Q_{\bar{1}|\bar{1}}^{{\downarrow}}&\propto \mathbb{Q}_{\bar{1}|\bar{1}}\sqrt{f^{\downarrow+}} \bar{h}^+\;,
    \quad
    Q_{\dot{\bar{1}}|\bar{1}}^{{\downarrow}}\propto \mathbb{Q}_{\dot{\bar{1}}|\bar{1}}\sqrt{f^{\downarrow+}} (\bar{h}^+)^{-1}\;.
\end{split}
\end{equation}
Let us note that if we are in the symmetric sector, see Section~\ref{sec:SymmetricSector}, then we have $Q_{1|1} \propto Q_{\dot{1}|1}$ and thus trivially  $h,\mathcal{F}$ are just constants. 

\paragraph{Constraining ${\cal F}$'s.}
Note that from \eq{FH2} we have
\beq\label{Fhh}
{\cal F}^+=\bar{h}^+ h^{*-}\,,
\quad
\bar{\mathcal{F}}^+ = h^+\bar{h}^{*-}\,.
\eeq
In this equation, we fixed the arbitrary scale factor in ${\cal F}$ to make it an equality.
We see that for this choice $({\cal F}^+)^{*}=\bar{\mathcal{F}}^{+}$. 

To further constrain $\mathcal{F}$ and $\bar{\mathcal{F}}$, we impose the following two conditions. Firstly, we must ensure that all $\nu$ are mirror (anti-)periodic \eq{nugamma}; secondly, we need to ensure compatibility between the two expressions in \eqref{tausplit}, as we recall that $\dot\tau^{1}_{~\bar{2}}\equiv ({\tau^{1}_{~\bar{2}}})^{[+2]}$.

Imposing $\nu^\gamma =\pm \nu^{[2]}$ gives
\begin{equation}\label{FF1}
\frac{\mathcal{F}^{\gamma}}{\mathcal{F}^{[+2]}}=
\frac{\bar{\mathcal{F}}^{\gamma}}{{\bar{\mathcal{F}}}^{[+2]}}=\pm {\cal S}\equiv\pm
\sqrt{\frac{
{\mathbb Q}_{1|1}^+
{\mathbb Q}_{\bar 1|\bar 1}^+
{\mathbb Q}_{\dot 1|1}^-
{\mathbb Q}_{\bar {\dot 1}|\bar 1}^-
}{
{\mathbb Q}_{1|1}^-
{\mathbb Q}_{\bar 1|\bar 1}^-
{\mathbb Q}_{\dot 1|1}^+
{\mathbb Q}_{\bar {\dot 1}|\bar 1}^+
}}\;,
\end{equation}
while consistency with \eqref{tausplit} implies
\begin{equation}\label{FH1}
\mathcal{F}\mathcal{F}^{[+2]} = C_h (h h^{[2]}) \,(\bar{h} \bar{h}^{[2]}) \;,
\quad
\bar{\mathcal{F}}\bar{\mathcal{F}}^{[+2]} = C_{\bar{h}} (h h^{[2]}) \,(\bar{h} \bar{h}^{[2]})\,,
\end{equation}
where we used the overall normalisation freedom in $\bar{\cal F}$ to ensure the first equality.
The last equation still has an unfixed constant $C_h$. Using (\ref{Fhh}), we find
\begin{equation}\label{hhbarreal}
    C_h = \frac{(hh^{[2]})^*}{h h^{[2]}}\,,
    \quad
    C_{\bar{h}} = \frac{(\bar{h}\bar{h}^{[2]})^*}{\bar{h} \bar{h}^{[2]}}\;.
\end{equation}
Note that we still have the freedom to rescale $h$ and $\bar{h}$ by an arbitrary phase factor, which must be the same for both functions. 

Note that the functions $\mathcal{F}$ and $\bar{\mathcal{F}}$ are both defined by the same system of equations, \eqref{FF1} and \eqref{FH1}. This means that their ratio $R_{\cal F}\equiv{\cal F}/{\bar{\cal F}}$ satisfies $R_{\cal F}^\gamma = R_{\cal F}^{[+2]}=c/R_{\cal F}$, where $c=C_h/C_{\bar{h}}$.
The last equality implies that $R_{\cal F}$ has only quadratic cuts.
Consider the combination $R_{\cal F}/\sqrt{c}+\sqrt{c}/R_{\cal F}$, which is a periodic function and invariant 
under the analytic continuation, which can only be a constant given that there are no poles or zeros in $R_{\cal F}$ and it cannot grow faster than a power.
Thus $R_{\cal F}$ itself is a constant, and it must be $\sqrt{c}$ due to $R_{\cal F}^\gamma=c/R_{\cal F}$, and thus ${\cal F}=\sqrt{c}\bar{\cal F}$. Substituting here \eqref{Fhh}, we can again compare the analyticity properties of both sides of the equation and conclude that
\begin{equation}
    \bar{h}=c^{1/4}\; h\;.
\end{equation}
Finally, recall that we can still rescale $h$ by an arbitrary phase factor. It is convenient to set \begin{equation} C_h=1~,~~\text{then}~,~~C_{\bar{h}}=c^{-1}\equiv e^{-2i\theta}~~,~\theta\in \mathbb{R}\;.
\end{equation}

\paragraph{The resulting constraints.}
Let us here summarise the conclusions we can draw from our study of $h$ and $\mathcal{F}$. First, we recall that we managed to write all $h_{\alpha}$ in terms of $h\equiv h_{1|1}$. It is convenient to package this object as
\beq\la{Hdef}
H\equiv h^{++}h=(h^{++}h)^*\;.
\eeq
The reason being that \eq{hhbarreal} implies that $H$ is a real function, and since $h$ is UHPA, it has only one cut on the real axis. 

Using this new object, the constraining equations for $\mathcal{F}$ become
\begin{equation}\label{FHsys}
    {\cal F}^{\gamma}=\pm{\cal F}^{[+2]}{\cal S}\;\;,\;\;
{\cal F}{\cal F}^{[+2]} =e^{i\theta} H^2
\;\;,\;\;{\cal F}^{+}=e^{i\theta/2} h^+ (h^*)^-\;.
\end{equation}
In Appendix~\ref{app:HAppendix} we solve the system \eq{FHsys}. With this, we have managed to constrain the free parameters entering into our ansatz for $Q_{\alpha}$. 

\subsection{Finding $\bP_A$ and $\bQ_I$ }
The relevant Q-functions among $\bP_A$ and $\bQ_{I}$ in the ABA limit are $\bP_1$ and $\bQ_1$. They are distinguished by the fact that $\bP_1$ becomes exponentially small, while $\bQ_1$ becomes exponentially large; see \eqref{pqlarge}. As motivated by the results of \cite{Cavaglia:2021eqr,Ekhammar:2021pys,Ekhammar:2024kzp}, we introduce the following notations
\begin{equation}
    {\bBullet}_{,(\pm)} = \bB_{1|1,(\pm)}  \bB_{\dot{1}|1,(\pm)}\;,
    \quad
   {\bBulletBar}_{,(\pm)} = \bB_{\bar{1}|\bar{1},(\pm)}  \bB_{\dot{\bar{1}}|\bar{1},(\pm)}
\end{equation}
as well as the following new functions $S$, parametrising $\bP$ and $\bQ$, without reducing generality
\begin{align}\label{pqaba1}
\bP_1 &\propto  x^{-L}R_{\tilde{1}}B_{\tilde{\bar{1}}}{\bBullet}_{,(-)}S\;,
\quad
\bQ_1\propto  x^{L}R_1B_{1}\dfrac{f}{{\bBulletBar}_{,(+)}}\dfrac{1}{S'}\;,\\
\bP_{\bar{1}} &\propto x^{-L}R_{\bar{1}}B_{1}{\bBulletBar}_{,(-)}\bar{S}\;,
\quad
\bQ_{\bar{1}} \propto  x^{L}R_{\tilde{\bar{1}}}B_{\tilde{1}}\dfrac{f}{{\bBullet}_{,(+)}}\dfrac{1}{\bar{S}'},
\end{align}
where the functions $R$ encode the roots of $\bP$ and $\bQ$ on the first sheet. 
We then define $S,S',\bar{S},\bar{S}'$ as functions without zeros and poles on the first sheet. The zeros of $R$ and $B$ are called auxiliary roots. We have explicitly
\begin{equation}\label{Rdef}
    R_{a} = \prod_{i = 1}^{\mathcal{K}_a}(x- y_{a,i})\;,
    \quad
    B_{a} = \prod_{i = 1}^{\mathcal{K}_a}\(\frac{1}{x} - y_{a,i}\)\,,
    \quad
    a\in\{1,~\tilde{1},~\bar{1},~\tilde{\bar{1}}\}\,.
\end{equation}
The functions $S,$ and $\bar{S}$ are constrained by the $QQ$-relations. In particular, let us consider (\ref{eq:bQbP1QQ}) which in the large volume limit becomes
\begin{equation}
    \dfrac{S}{S'}R_{\tilde{1}}B_{\tilde{\bar{1}}}R_1B_{\bar{1}}f\dfrac{{\bBullet}_{,(-)}}{{\bBulletBar}_{,(+)}}=\mathbb{Q}_{1|1}^+\mathbb{Q}_{\dot{1}|1}^+f^{++}-\mathbb{Q}_{1|1}^-\mathbb{Q}_{\dot{1}|1}^-f\,.
\end{equation}
With the use of \eqref{Bpm} and \eqref{fud} it simplifies to
\begin{equation}\label{aba1}
\dfrac{S}{S'}R_{\tilde{1}}B_{\tilde{\bar{1}}}R_1B_{\bar{1}}=\bR_{1|1(+)}\bR_{\dot{1}|1(+)}\bB_{\bar{1}|\bar{1}(-)}\bB_{\dot{\bar{1}}|\bar{1}(-)}-\bR_{1|1(-)}\bR_{\dot{1}|1(-)}\bB_{\bar{1}|\bar{1}(+)}\bB_{\dot{\bar{1}}|\bar{1}(+)}.  
\end{equation}
From this equation, it is clear that $S/S'$ is a rational function of $x$. Now, writing down the barred version of \eqref{aba1} we furthermore find that 
\begin{equation}\label{eq:continueS}
    \left(S/S'\right)\left(\frac{1}{x}\right)\propto (\bar{S}/\bar{S}')\left(x\right)\,.
\end{equation}
This equation allows us to constrain the ratio $S/S'$. Indeed, by definition $S/S'$ cannot have poles or zeros for finite $|x|\geq 1$, and by \eqref{eq:continueS} the same is true for $|x|\leq 1, x\neq 0$. Thus,
\begin{equation}
    S' = x^{-K_{S}}\,S\,,
    \quad
    \bar{S}' = x^{K_{S}}\bar{S}\,,
\end{equation}
where we have imposed an exact equality (as opposed to $\propto$) since $S$ is only defined up to a constant.

To fix $S$ and $\bar{S}$, recall that we require \eqref{pmu1aba}, relating ${\bf P}_1^\gamma$ to ${\bf Q}_{\bar 1}$. Substituting the ansatz for ${\bf P}_1$ and ${\bf Q}_{\bar{1}}$, as well as the relations \eqref{wmuaba} and \eqref{1q}, results in two relations for $S$ and $\bar{S}$. These are
\begin{equation}
    S^{\gamma}\bar{S}\propto \frac{1}{x^{K_{S}}}\dfrac{\bR_{1|1(+)}\bR_{\dot{1}|1(+)}}{\bR_{1|1(-)}\bR_{\dot{1}|1(-)}}\fu^{--}\fd^{++}\;,
    \quad
    \bar{S}^{\gamma}S\propto x^{K_{S}}\dfrac{\bR_{\bar{1}|\bar{1}(+)}\bR_{\dot{\bar{1}}|\bar{1}(+)}}{\bR_{\bar{1}|\bar{1}(-)}\bR_{\dot{\bar{1}}|\bar{1}(-)}}\fu^{--}\fd^{++}\;.
\end{equation}
The system is solved using the following substitution
\begin{equation}
S=\sqrt{\dfrac{{\bBullet}_{,(+)}}{{\bBullet}_{,(-)}}}\sigma_{\bullet}\,\rho\ \rho_{K}\;,
    \quad
    \bar{S}=\sqrt{\dfrac{{\bBulletBar}_{,(+)}}{{\bBulletBar}_{,(-)}}}\sigma_{\bullet}\,\bar{\rho}\,\bar{\rho}_{K}\,,
\end{equation}
where the functions $\sigma_{\bullet}$ and $\rho,\bar{\rho}$ are defined exactly in the same way as in the AdS$_3\times $S$^3\times$T$^4$ \cite{Ekhammar:2024kzp}. They satisfy the functional relations
\begin{align}\label{rhophase}
\sigma^{\gamma}_{\bullet}\sigma_{\bullet} &\propto \fd^{++}\fu^{--}\,,
\quad
\rho^{\gamma} \bar{\rho} \propto \sqrt{\frac{{\rBullet}_{,(+)}}{{\rBullet}_{,(-)}}\frac{{\bBulletBar}_{,(-)}}{{\bBulletBar}_{,(+)}}}\,,
\quad
\bar{\rho}^{\gamma} \rho \propto \sqrt{\frac{\rBulletBarP}{\rBulletBarM}\frac{\bBulletM}{\bBulletP}}\;,\\
\rho^{\gamma}_K\bar{\rho}_K &= x^{-K_S}\,,
\quad
\bar{\rho}^{\gamma}_K\rho_K = x^{K_S} \label{eq:rhoK}\,.
\end{align}
One can show that \eqref{eq:rhoK}, supplemented with the condition that $\rho_{S} \simeq 1$, implies that $K_S=0$ and $\rho_{K}\propto \bar{\rho}_{K} \propto 1$. This precise calculation is presented in Appendix B.2 in \cite{Ekhammar:2024kzp}, and we will therefore not repeat it here.

Now that $S$ and $\bar{S}$ are known, the final expressions for $\bP$ and $\bQ$ become
\begin{align}
\bP_1 &\propto x^{-L}R_{\tilde{1}}B_{\tilde{\bar{1}}}\sqrt{\bB_{\bullet,(+)}\bB_{\bullet,(-)}}\sigma_{\bullet}\rho\;,
\quad
\bQ_1\propto x^{L}R_1B_{\bar{1}}\sqrt{\dfrac{\bB_{\bullet,(-)}}{\bB_{\bullet,(+)}}}\dfrac{f}{B_{\bar{\bullet},(+)}}\dfrac{1}{\sigma_{\bullet}\rho}\,,\label{q1aba} \\
\bP_{\bar{1}} &\propto x^{-L}R_{\bar{1}}B_{1}\sqrt{\bB_{\bar{\bullet},(+)}\bB_{\bar{\bullet},(-)}}\sigma_{\bullet}\bar{\rho}\;,
\quad
\bQ_{\bar{1}}\propto x^{L}R_{\tilde{\bar{1}}}B_{\tilde{1}}\sqrt{\dfrac{\bfB_{\bar{\bullet},(-)}}{\bB_{\bar{\bullet},(+)}}}\dfrac{f}{\bB_{\bullet,(+)}}\dfrac{1}{\sigma_{\bullet}\bar{\rho}}\;.\label{q1abab}
\end{align}
As we have now introduced all building blocks needed for the ABA equations, we can define, in the next section, a particular combination that will be identified with the dressing phases in the S-matrix bootstrap approach.

\paragraph{Quantum numbers and Bethe roots.}

With the explicit Q-functions at hand, we can now relate the $\algosp(4|2)^{\oplus 2}$ group charges (\ref{Eqn:quasiclass}) to the number of roots of our $Q$-functions. We do so by comparing the asymptotic behaviour of our large volume solution with \eqref{asymp}.

From the explicit expressions \eqref{1q}, \eqref{q1aba}, and \eqref{q1abab}, we find that
\begin{align*}
    \bP_{1} \sim u^{-L+\mathcal{K}_{\tilde{1}}}\,,
    \quad
    Q_{1|1} \sim u^{K_{1|1}+\gamma/2}
    \quad
    Q_{\dot{1}|1} \sim u^{K_{\dot{1}|1}+\gamma/2}\,,
    \\
    \bP_{\bar{1}} \sim u^{-L+\mathcal{K}_{\bar{1}}}\,,
    \quad
    Q_{\bar{1}|\bar{1}} \sim u^{K_{\bar{1}|\bar{1}}+\gamma/2}
    \quad
    Q_{\dot{\bar{1}}|\bar{1}} \sim u^{K_{\dot{\bar{1}}|\bar{1}}+\gamma/2}\,.
\end{align*}
Then, upon using $ \mathcal{K}_{\tilde{1}} = K_{1|1}+K_{\dot{1}|1}-\mathcal{K}_{1}-1$, 
we obtain
\begin{align}
M_{1} &= L+\mathcal{K}_1-K_{1|1}-K_{\dot{1}|1}+1\,,
&
M_{2} &= K_{\dot{1}|1}-K_{1|1}\,, \nonumber \\
M_{\bar{1}} &= L-\mathcal{K}_{\bar{1}}\,, &
    M_{\bar{2}} &= K_{\dot{\bar{1}}|\bar{1}}-K_{\bar{1}|\bar{1}}\,, \\
\hat{M}_1 &= 1+L+\mathcal{K}_1+\gamma\,,
&
\hat{M}_{\bar{1}} &= K_{\bar{1}|\bar{1}} + K_{\dot{\bar{1}}|\bar{1}}+L-\mathcal{K}_{\bar{1}}+\gamma\,. \nonumber
\end{align}
The above six equations, along with $\bP$ and $\bQ$ asymptotics \eqref{MaMab} and \eqref{MaMab2}, are enough to find the group charges
\begin{align}
    \Delta &= L + \gamma +\frac{K_{\bar{1}|\bar{1}}+K_{\dot{\bar{1}}|\bar{1}}}{2} +\frac{\mathcal{K}_1-\mathcal{K}_{\bar{1}}}{2}\,, \nonumber \\
    S &= \frac{K_{\bar{1}|\bar{1}}+K_{\dot{\bar{1}}|\bar{1}}}{2} -\frac{\mathcal{K}_1+\mathcal{K}_{\bar{1}}}{2}\,, \nonumber \\
    J_1 &= L-\frac{K_{1|1}+K_{\dot{1}|1}}{2}+\frac{\mathcal{K}_1-\mathcal{K}_{\bar{1}}}{2}\,, \\
    K_1 &= \frac{K_{1|1}+K_{\dot{1}|1}}{2}-\frac{\mathcal{K}_1+\mathcal{K}_{\bar{1}}}{2}\,, \nonumber  \\
    J_2 &= \frac{K_{\dot{1}|1}-K_{1|1}}{2}+\frac{K_{\dot{\bar{1}}|\bar{1}}-K_{\bar{1}|\bar{1}}}{2} \,, \nonumber  \\
    K_2 &=  \frac{K_{1|1}-K_{\dot{1}|1}}{2}+\frac{K_{\dot{\bar{1}}|\bar{1}}-K_{\bar{1}|\bar{1}}}{2}\,.\nonumber 
\end{align}
This nicely shows that it is indeed only $\Delta$ that develops an anomalous part.

\subsection{The Dressing Phases \label{sec:DressingPhasesMain}}
Let us provide more details regarding the functions $h, \sigma_{\bullet},\rho,\bar{\rho}$ that appear in our final expressions \eqref{1q}, \eqref{q1aba}, and \eqref{q1abab}. We will attempt to identify these objects as the basic building blocks of the dressing phases that appear in the S-matrix approach. 

\paragraph{$\sigma_{
\bullet}$ and the BES phase.}
The function $\sigma_{\bullet}$ has appeared many times before in the QSC context; it is a basic building block of the so-called BES phase. Since the right-hand side of \eqref{rhophase} is a product over all the massive excitations, it is natural to also write $\sigma_{\bullet}$ in this fashion. For reasons that will soon become clear, we will not consider $\sigma_{\bullet}$, but rather the ratio $\sigma^+_{\bullet}/\sigma^-_{\bullet}$. We find
\begin{equation}\label{eq:SigmaBullet}
    \frac{\sigma_{\bullet}(x^+)}{\sigma_{\bullet}(x^-)} = \prod_{\alpha\in \mathcal{I}} \prod_{k=1}^{K_{\alpha}} \sigmaBES(x,x_{\alpha,k})\;,
    \quad
    \mathcal{I} = \{1|1,\dot{1}|1,\bar{1}|\bar{1},\dot{\bar{1}}|\bar{1}\}\,.
\end{equation}
where $\sigma_{\text{BES}}$ is the famous Beisert-Eden-Staudacher-phase \cite{Beisert:2006ez}. 

\paragraph{$\rho$ and $\bar{\rho}$.}
The functions $\rho$ and $\bar{\rho}$ first appeared in the proposed AdS$_3\times$S$^3 \times$T$^4$ QSC \cite{Cavaglia:2021eqr,Ekhammar:2021pys}. Just as $\sigma_{\bullet}$, they are written as a product over the momentum-carrying excitations. We write
\begin{equation}\label{eq:RhoBarRho}
\begin{split}
    \rho &= \prod_{\alpha \in \mathcal{I}_2}\prod_{k=1}^{K_{\alpha}} \varrho(x,x_{\alpha,k}) \prod_{\alpha \in \mathcal{I}_{\bar{2}}}\prod_{k=1}^{K_{\alpha}} \bar{\varrho}(x,x_{\alpha,k})\,,
    \quad
    \bar{\rho} = \prod_{\alpha \in \mathcal{I}_2}\prod_{k=1}^{K_{\alpha}} \bar{\varrho}(x,x_{\alpha,k}) \prod_{\alpha \in \mathcal{I}_{\bar{2}}}\prod_{k=1}^{K_{\alpha}} \varrho(x,x_{\alpha,k})\,,
\end{split}
\end{equation}
where  $\mathcal{I}_2 = \{1|1,\dot{1}|1\}, ~\mathcal{I}_{\bar{2}} = \{\bar{1}|\bar{1},\dot{\bar{1}}|\bar{1}\}$. Here, the newly introduced building blocks satisfy the discontinuity equations \cite{Ekhammar:2024kzp}
\begin{equation}
    \varrho(x^{\gamma},v)\bar{\varrho}(x,v) \propto \sqrt{\frac{x-y^{-}}{x-y^{+}}}\,,
    \quad
    \bar{\varrho}(x^{\gamma},v)\bar{\varrho}(x,v) \propto \sqrt{\frac{1/x-y^{+}}{1/x-y^{-}}}\,,
    \quad
    v = g\;\(y+\frac{1}{y}\)\,.
\end{equation}
It will be convenient to combine $\sigma_{\bullet},\rho$ and $\bar{\rho}$. We follow \cite{Ekhammar:2024kzp} and define
\begin{equation}\label{eq:BBPhases}
    \sigma^{\bullet\bullet}(x,y) = \sigmaBES(x,y)\frac{\varrho(x^+,y)}{\varrho(x^-,y)} \,,
    \quad
    \tilde{\sigma}^{\bullet\bullet}(x,y) = \sigmaBES(x,y) \frac{\bar{\varrho}(x^+,y)}{\bar{\varrho}(x^-,y)}\,.
\end{equation}

\paragraph{The H-phase.}
Just as for the other phases above, the defining equation for $H$ (defined in \eq{FHsys} and \eq{Hdef}) implies that it can be naturally written as a product over momentum-carrying roots, explicitly 
\begin{equation}
    H = \prod_{\alpha \in \{1|1,\bar{1}|\bar{1}\}}\prod_{k=1}^{K_{\alpha}}e^{\mathcal{H}(x,x_{\alpha,k})}\prod_{\alpha \in \{\dot{1}|1,\dot{\bar{1}}|\bar{1}\} }\prod_{k=1}^{K_{\alpha}}e^{-\mathcal{H}(x,x_{\alpha,k})}\;.
\end{equation}
The discontinuity equation for $\mathcal{H}(x,y)$ is significantly more complicated compared to $\varrho,\bar{\varrho}$; we work out the details in Appendix~\ref{app:HAppendix}. The final result is
\begin{equation}\label{eq:HDisc}
{{\cal H}^{\gamma}}-{\cal H}=
\frac{1}{2}\log\(\frac{u-v+\ii/2}{u-v-\ii/2}\)
-\sum_{n=0}^\infty
2(-1)^n
\[{\cal H}^{[-2n-2]}-{\cal H}^{[+2n+2]}\]\,,
\end{equation}
with $\mathcal{H}= \mathcal{H}(x,y), \mathcal{H}^{\gamma} = \mathcal{H}(x^{\gamma},y)$, and we finally introduce the associated dressing phase
\begin{equation}\label{eq:HDef}
   \hat{\sigma}(x,y) = \exp\left(\mathcal{H}(x^+,y)-\mathcal{H}(x^-,y)\right)\,.
\end{equation}
We also discuss massless dressing phases in Appendix~\ref{app:HAppendix}.

\subsubsection{Crossing Equations}

\begin{figure}[h]
    \centering
    \begin{tikzpicture}
        \draw[black,thick] (-2,-2) rectangle (2,2);
        \draw[black,thick] (-1,0.5)--(1,0.5);
        \filldraw (-1,0.5) circle (2pt);
        \filldraw (1,0.5) circle (2pt);
        \draw[black,thick] (-1,-0.5)--(1,-0.5);
        \filldraw (-1,-0.5) circle (2pt);
        \filldraw (1,-0.5) circle (2pt);
        \draw[<-,red] (-0.6,1) .. controls (0,1) and (0,-1) .. (-0.6,-1);
        \draw[red] (-1.5,0) .. controls (-1.2,1) .. (-0.6,1);
        \draw[<-,red] (-0.6,-1) .. controls (-1.2,-1) .. (-1.5,0);
        \filldraw[red] (-1.5,0) circle (2pt);
    \end{tikzpicture}
    \caption{The path $\bar{\gamma}_c$ depicted in the $u$-plane. Note that it first crosses the branch-cut of $x^{+}$ and thereafter the cut of $x^{-}$.}
    \label{fig:full-crossing}
\end{figure}
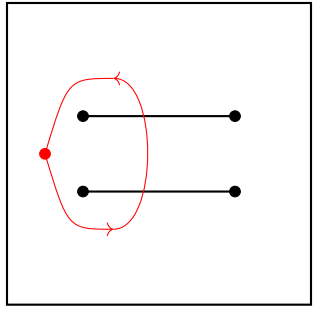

We now proceed to compute the so-called crossing equations for our phases. We note that the dressing phases are functions with cuts at $(-2g,2g)\pm \frac{\ii}{2}$. The crossing equations are obtained by analytically continuing the phases across these two cuts, following the path $\bar{\gamma}_c$ defined in Figure~\ref{fig:full-crossing}.

The crossing for $\sigma^{\bullet\bullet}$ and $\tilde{\sigma}^{\bullet\bullet}$ has been studied previously, see for example \cite{Ekhammar:2024kzp}, the result is
\begin{equation}
\begin{split}
\label{sigmacrossing}
[\sigma^{\bullet\bullet}(x^{\bar\gamma_c},y)
\tilde{\sigma}^{\bullet\bullet}(x,y)]^2&=
\frac{(y^-)^4 \left(x^+-y^+\right) \left(x^--y^+\right) \left(y^+ x^+-1\right)^2}
{(y^+)^4 \left(x^--y^-\right) \left(x^+-y^-\right) \left(y^- x^+-1\right)^2}\,,\\
[\tilde{\sigma}^{\bullet\bullet}(x^{\bar\gamma_c},y)
\sigma^{\bullet\bullet}(x,y)]^2&=\frac{(y^-)^4 \left(x^--y^+\right)^2 \left(y^+ x^--1\right) \left(y^+ x^+-1\right)}
{(y^+)^4 \left(x^--y^-\right)^2 \left(y^- x^--1\right) \left(y^- x^+-1\right)}\;.
\end{split}
\end{equation}
For the new piece $\hat{\sigma}$, we find in Appendix~\ref{app:HAppendix} that the discontinuity relation \eqref{eq:HDisc} results in
\begin{equation}\label{hatcrossing}
    \hat{\sigma}(x^{\bar{\gamma}_c},y)\hat{\sigma}(x,y)=\sqrt{\dfrac{x^--y^+}{x^+-y^-}\dfrac{1-\frac{1}{x^-y^+}}{1-\frac{1}{x^+y^-}}}\,.
\end{equation}

\subsection{The ABA Equations}\label{subsec:ABAEquations}
Having found the large volume limit for $\bP_1,\bQ_1,Q_{\dot{1}|1}$ and $Q_{1|1}$, we can now find the corresponding Asymptotic Bethe Ansatz by plugging these expressions into the exact Bethe Ansatz Equations. We start with the grading in (\ref{eq:BetheEquationsGrading1}):
\begin{equation}\label{ABAGrading1}
    \dfrac{Q_{1|1}^{[2]}\bQ_1^{-}}{Q_{1|1}^{[-2]}\bQ_1^+}\bigg|_{u:\;Q_{1|1}=0}=-1\;,
    \quad
    \dfrac{Q_{\dot{1}|1}^+Q_{1|1}^+}{Q_{\dot{1}|1}^-Q_{1|1}^-}\bigg|_{u:\;\bQ_{1}=0}=1\;,
    \quad
    \dfrac{Q_{\dot{1}|1}^{[2]}\bQ_1^-}{Q_{\dot{1}|1}^{[-2]}\bQ_1^+}\bigg|_{u:\;Q_{\dot{1}|1}=0}=-1\;,
\end{equation}
and plug \eqref{1q} and \eqref{q1aba} to find the following ABA equations
\begin{align}\label{aba_1}
        -\(\dfrac{x^+_{1|1,k}}{x^-_{1|1,k}}\)^{L}&=\dfrac{\mathbb{Q}_{1|1}^{[+2]}}{\mathbb{Q}_{1|1}^{[-2]}}\,\dfrac{R_1^-B_{\bar{1}}^-}{R_1^+B_{\bar{1}}^+}\,\sqrt{\dfrac{\bBulletBarP^{+}\bBulletBarM^{+}}{\bBulletBarP^{-}\bBulletBarM^{-}}}\,\frac{\sigma^+_{\bullet}\rho^+H^+}{\sigma^-_{\bullet}\rho^- H^-}\Bigg|_{x=x_{1|1,k}}\,,\\
        \label{aba_2}
        1&=\dfrac{\rBulletP \bBulletBarM}{\rBulletM \bBulletBarP}\Biggl|_{x=y_{1,k}}\,,\\\label{aba_3}
        -\(\dfrac{x^+_{\dot{1}|1,k}}{x^-_{\dot{1}|1,k}}\)^{L}&=\dfrac{\mathbb{Q}_{\dot{1}|1}^{[+2]}}{\mathbb{Q}_{\dot{1}|1}^{[-2]}}\cdot\dfrac{R_1^-B_{\bar{1}}^-}{R_1^+B_{\bar{1}}^+}\,\sqrt{\dfrac{\bBulletBarP^{+}\bBulletBarM^{+}}{\bBulletBarP^{-}\bBulletBarM^{-}}}\,\frac{\sigma^+_{\bullet}\rho^+H^-}{\sigma^-_{\bullet}\rho^- H^+}\Bigg|_{x=x_{\dot{1}|1,k}}\,.
\end{align}
We will consider the second copy in a different grading, the reason for this is simply to ensure that only the auxiliary roots $y_{1,k},\bar{y}_{1,k}$ appear while $y_{\tilde{1},k},y_{\tilde{\bar{1}},k}$ are absent. From the definition of $\bP_{\bar{1}}$ (\ref{eq:bQDef}) it follows that 
\begin{equation}\label{BetheEqnGrading2}
    1=\dfrac{\bP_{\bar{1}}^+Q_{\dot{\bar{1}}|\bar{1}}^{[-2]}}{\bP_{\bar{1}}^-Q_{\dot{\bar{1}}|\bar{1}}^{[+2]}}\Biggl|_{u:\;Q_{\bar{1}|\bar{1}}=0}\;,
    \quad
1=\dfrac{Q_{\dot{\bar{1}}|\bar{1}}^{+}Q_{\bar{1}|\bar{1}}^{+}}{Q_{\dot{\bar{1}}|\bar{1}}^{-}Q_{\bar{1}|\bar{1}}^{-}}\Biggl|_{u:\;\bP_{\bar{1}}=0}\;,
    \quad
    1=\dfrac{\bP_{\bar{1}}^+Q_{\bar{1}|\bar{1}}^{[-2]}}{\bP_{\bar{1}}^-Q_{\bar{1}|\bar{1}}^{[+2]}}\Biggl|_{u:\;Q_{\dot{\bar{1}}|\bar{1}}=0}\;.
\end{equation}
Plugging in \eqref{1q} and \eqref{q1abab} gives
\begin{align}\label{abab_1}
    -\(\dfrac{x_{\bar{1}|\bar{1},k}^+}{x_{\bar{1}|\bar{1},k}^-}\)^{L}&=\dfrac{\bR^-_{\dot{\bar{1}}|\bar{1},(-)}}{\bR^+_{\dot{\bar{1}}|\bar{1},(+)}}\dfrac{\bB^+_{\bar{1}|\bar{1},(+)}}{\bB^-_{\bar{1}|\bar{1},(-)}}\,\dfrac{B_1^+R_{\bar{1}}^+}{B_1^-R_{\bar{1}}^-}\,\sqrt{\dfrac{\bBulletP^+\bBulletP^-}{\bBulletM^+ \bBulletM^-}}\frac{\sigma^+_{\bullet}\bar{\rho}^+H^+}{\sigma^-_{\bullet}\bar{\rho}^-H^-}\Bigg|_{x=x_{\bar{1}|\bar{1},k}}\\\label{abab_2}
    1&=\dfrac{\rBulletBarP\bBulletM}{\rBulletBarM\bBulletP}\Biggl|_{x=y_{\bar{1},k}}\,,\\
    \label{abab_3}
    -\(\dfrac{x_{\dot{\bar{1}}|\bar{1},k}^+}{x_{\dot{\bar{1}}|\bar{1},k}^-}\)^{L} &= \dfrac{\bR^-_{\bar{1}|\bar{1},(-)}}{\bR^+_{\bar{1}|\bar{1},(+)}}\dfrac{\bB^+_{\dot{\bar{1}}|\bar{1},(+)}}{\bB^-_{\dot{\bar{1}}|\bar{1},(-)}}\dfrac{B_1^+R_{\bar{1}}^+}{B_1^-R_{\bar{1}}^-}\,\sqrt{\dfrac{\bBulletP^+\bBulletP^-}{\bBulletM^+ \bBulletM^-}}\,\frac{\sigma^+_{\bullet}\bar{\rho}^+H^{-}}{\sigma^-_{\bullet}\bar{\rho}^+ H^+}\Bigg|_{x=x_{\dot{\bar{1}}|\bar{1},k}}\,.
\end{align}

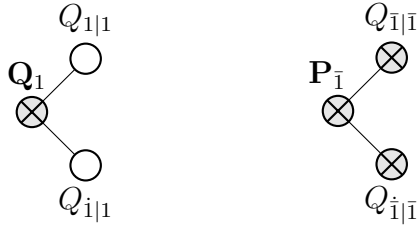
\begin{figure}
\begin{center}
\begin{tikzpicture}
\node[] (n1) at (-2,0) {\begin{tikzpicture}[cross/.style={path picture={ 
\draw[black]
(path picture bounding box.south east) -- (path picture bounding box.north west) (path picture bounding box.south west) -- (path picture bounding box.north east);
}}]
  % your two other lines
  \draw[] (0,0)--({1/sqrt(2)},{1/sqrt(2)});
  \draw[] (0,0)--({1/sqrt(2)},{-1/sqrt(2)});

  % circle at (1/√2,1/√2)
  \draw[fill=white,thick]
       ({1/sqrt(2)},{1/sqrt(2)}) circle (0.2) 
       node[anchor=south,yshift=4] {$Q_{1|1}$};

  % crossed circle at (0,0), now with gray background
  \draw[fill=gray!20,cross,thick]
       (0,0) circle (0.2) 
       node[anchor=south,yshift=4,xshift=-2] {$\bQ_{1}$};

  % circle at (1/√2,–1/√2)
  \draw[fill=white,thick]
       ({1/sqrt(2)},-{1/sqrt(2)}) circle (0.2) 
       node[anchor=north,yshift=-4]{$Q_{\dot{1}|1}$};
\end{tikzpicture}};
\node[] (n2) at (2,0) {
\begin{tikzpicture}[cross/.style={path picture={ 
\draw[black]
(path picture bounding box.south east) -- (path picture bounding box.north west) (path picture bounding box.south west) -- (path picture bounding box.north east);
}}]
  % your two other lines
  \draw[] (0,0)--({1/sqrt(2)},{1/sqrt(2)});
  \draw[] (0,0)--({1/sqrt(2)},{-1/sqrt(2)});

  % circle at (1/√2,1/√2)
  \draw[fill=gray!20,cross,thick]
       ({1/sqrt(2)},{1/sqrt(2)}) circle (0.2) 
       node[anchor=south,yshift=4] {$Q_{\bar{1}|\bar{1}}$};

  % crossed circle at (0,0), now with gray background
  \draw[fill=gray!20,cross,thick]
       (0,0) circle (0.2) 
       node[anchor=south,yshift=4,xshift=-4] {$\bP_{\bar{1}}$};

  % circle at (1/√2,–1/√2)
  \draw[fill=gray!20,cross,thick]
       ({1/sqrt(2)},-{1/sqrt(2)}) circle (0.2) 
       node[anchor=north,yshift=-4]{$Q_{\dot{\bar{1}}|\bar{1}}$};
    \end{tikzpicture}};
\end{tikzpicture}
\captionof{figure}{We depict the grading of the ABA-equations as well as the Q-functions that explicitly appear. The left grading corresponds to \eqref{aba_1},\eqref{aba_2},\eqref{aba_3} while the right encodes \eqref{abab_1},\eqref{abab_2},\eqref{abab_3}. The choice of grading is purely a matter of convenience. We have chosen Q-functions so that only the auxiliary roots $y_{k}$ appear, while $\tilde{y}_k$ are absent.}
\label{GradingPicture}
\end{center}
\end{figure}
Now, this system can be compared with the ABA equations derived in \cite{Borsato:2012ss}, using a very different S-matrix bootstrap approach; we recall all the relevant equations in Appendix~\ref{app:ABAEquations}. To begin with, we find an exact match for \eqref{aba_2} and \eqref{abab_2} with  \cite{Borsato:2012ss}. The remaining equations in the ABA also match our proposal structurally but feature unknown, so-called dressing phases: $S^{\alpha\beta}(x,y)\;, \alpha,\beta = \{1,3,\bar{1},\bar{3}\}$ \cite{Borsato:2012ss}. Upon identifying factors evaluated at the same roots, we find

\begin{align}\label{eq:SDressingPhases}
    S^{11}(x,y) &= \frac{1-\frac{1}{x^+y^-}}{1-\frac{1}{x^-y^+}}\sigma^{\bullet\bullet}(x,y)\hat{\sigma}(x,y)\;,
    &
    S^{13}(x,y)  &= \sigma^{\bullet\bullet}(x,y) \hat{\sigma}^{-1}(x,y)
    \\
    S^{1\bar{1}}(x,y) &= \sqrt{\frac{1-\frac{1}{x^+ y^-}}{1-\frac{1}{x^- y^+}}}\tilde{\sigma}^{\bullet\bullet}(x,y)\hat{\sigma}(x,y)\;,
    &
     S^{1\bar{3}}(x,y) &= \sqrt{\frac{1-\frac{1}{x^+ y^-}}{1-\frac{1}{x^- y^+}}}\tilde{\sigma}^{\bullet\bullet}(x,y)\hat{\sigma}^{-1}(x,y)\;,
\end{align}
and the remaining phases are given as $S^{33}=S^{\bar{1}\bar{1}}=S^{\bar{3}\bar{3}}=S^{11}\,, S^{31}=S^{13}=S^{\bar{1}\bar{3}}=S^{\bar{3}\bar{1}}$, $S^{1\bar{1}} = S^{\bar{1}1} = S^{3\bar{3}} = S^{\bar{3}3}$, and $S^{1\bar{3}} = S^{\bar{3}1} = S^{\bar{1}3} = S^{3\bar{1}}$. Let us point out that the process of deducing  \eqref{eq:SDressingPhases} from the Bethe equations is, to some extent, ambiguous. For example, factors that cancel out 
in the product over all roots (e.g., due to the momentum conservation condition \eq{mom}) can be distributed freely. We have fixed the exact proportionality to facilitate easier comparison with the literature.

There are no proposals for the exact form of $S^{\alpha\beta}$  in the literature; however, their crossing equations have been proposed in \cite{Borsato:2012ss}. Here, the crossing equation simply means an equation constraining $S^{\alpha\beta}$ upon analytic continuation along $\bar{\gamma}_{c}$. recall that this contour was defined in Figure~\ref{fig:full-crossing}.  Simply substituting  \eqref{sigmacrossing} and \eqref{hatcrossing} into \eqref{eq:SDressingPhases}, found from QSC, we get 
\begin{align}
    \scalebox{0.9}{$S^{11}(x^{\bar{\gamma}_c},y)S^{\bar{1}1}(x,y) = \frac{1-\frac{1}{x^- y^-}}{1-\frac{1}{x^+ y^-}}\sqrt{\frac{1-\frac{1}{x^+y^+}}{1-\frac{1}{x^- y^-}}}$}\;,
    \quad
    \scalebox{0.9}{$S^{\bar{1}1}(x^{\bar{\gamma}_c},y)S^{11}(x,y) = \sqrt{\frac{y^-}{y^+}}\frac{x^- - y^+}{x^+ - y^+}\sqrt{\frac{x^+ - y^+}{x^- - y^+}}$}\;,\\
    \scalebox{0.9}{$S^{13}(x^{\bar{\gamma}_c},y)S^{\bar{1}3}(x,y) = \frac{y^-}{y^+}\frac{1-\frac{1}{x^-y^-}}{1-\frac{1}{x^-y^+}}\sqrt{\frac{1-\frac{1}{x^+y^+}}{1-\frac{1}{x^-y^-}}}$}\;, 
    \quad
    \scalebox{0.9}{$S^{\bar{1}3}(x^{\bar{\gamma}_c},y)S^{13}(x,y) = \sqrt{\frac{y^-}{y^+}}\frac{x^+-y^-}{x^+-y^+}\sqrt{\frac{x^+-y^+}{x^--y^-}}$}\;,
\end{align}
which matches perfectly with the expressions from \cite{Borsato:2012ss} assuming one interprets the crossing transformation in that paper as being in the opposite direction of $\bar{\gamma}_c$. Such a change of orientation was also necessary for AdS$_3\times$S$^3\times$T$^{4}$, see Appendix B in \cite{Borsato:2013hoa} for a detailed discussion. We discuss this change of orientation further in Appendix~\ref{app:ABAEquations}. This highly non-trivial check grants some credence to the hope that our proposal can give the spectrum of free strings on AdS$_3\times$S$^3\times$S$^3 \times$S$^1$.

\subsection{Braiding Unitarity of the Dressing Phases} \label{sec:UnitarityIssues} 

Normally, in the S-matrix bootstrap approach, one constrains the dressing phases not only by crossing equations but also by braiding unitarity; that is, for example, $S^{11}(x,y)S^{11}(y,x)=1$. It is not a priori obvious that the phases arising in the ABA limit of QSC will satisfy this requirement. In \cite{Ekhammar:2024kzp}, it was shown that this will hold for $\sigma^{\bullet\bullet}(u,v)$ and $\tilde{\sigma}^{\bullet\bullet}(u,v)$. 
The situation with $\hat{\sigma}(u,v)$ turns out to be more complicated.
Firstly, no solution has been presented in the S-matrix bootstrap literature that satisfies both. Furthermore, we also found that the equations derived from QSC (which are normally more constraining than the crossing equations alone) are in tension with unitarity while 
reproducing the crossing equation (or vice versa, depending on the interpretation of the factors in the ABA limit), as discussed in Appendix~\ref{app:HAppendix}. 

In the QSC, we can only compute $H^+/H^-$, which is a product of individual factors, and not the individual dressing phases. Previously, in this section, we decomposed
\begin{equation}
    \frac{H^+}{H^-} = \frac{\prod_{\alpha \in \{1|1,\bar{1}|\bar{1}\}}\prod_{k=1}^{K_{\alpha}} \hat{\sigma}(u,x_{\alpha,k})}{\prod_{\alpha \in \{\dot{1}|1,\dot{\bar{1}}|\bar{1}\}}\prod_{k=1}^{K_{\alpha}} \hat{\sigma}(u,x_{\alpha,k})}\,.
\end{equation}
It is therefore natural to ask whether there exists another splitting that would yield phases satisfying braiding unitarity. In Appendix~\ref{app:HAppendix}, we show that this is the case, but it requires the momentum-carrying roots to satisfy, in general, a non-trivial constraint; see \eqref{alrestrict} and \eqref{eq:alphaDef}. These constraints are automatically satisfied in the symmetric sector, essentially because in this sector $H^+\big/H^- = 1$. The price to pay is that the resulting unitary phase, even in the symmetric sector, will no longer satisfy the crossing equation \eqref{hatcrossing}. We have not been able to find, nor rule out, a more complicated splitting that provides both the correct crossing equation and braiding unitarity.
This issue definitely deserves future investigation.

At the same time, in the symmetric sector, we expect that one should really only look at the product $S^{11}(u,v)S^{13}(u,v)$. This product satisfies both crossing and braiding unitarity; thus, our equations appear perfectly consistent with the literature.

\section{Conclusions}
\label{Sec:conclusions}
In this paper, we propose a quantum spectral curve (QSC) for AdS$_3\times$S$^3\times$S$^3\times$S$^1$ when the radii of the two S$^3$ are equal. Our proposal is based on symmetry considerations and amounts to connecting two separate $\algosp(4|2)$ QQ-systems through analytic continuation, a procedure which closely follows that of the AdS$_3\times$S$^3\times$T$^4$ QSC case \cite{Ekhammar:2021pys,Cavaglia:2021eqr}. 

To test our proposal, we investigate the large-volume limit of the QSC, where the spectrum is expected to be given by the Asymptotic Bethe Ansatz \cite{Borsato:2012ss}. When comparing our QSC equations with the ABA, we found agreement with the general structure and successfully reproduced all the crossing equations for the product of the dressing phases appearing in the Bethe ansatz. However, we were not able to split the product of dressing phases arising in the ABA equations in the non-symmetric case into individual dressing phases that satisfy all the assumptions made in the S-matrix bootstrap approach---namely, the crossing equations and braiding unitarity. To the best of our knowledge, no such factors are currently known to exist in the S-matrix literature.

In the symmetric sector, however, we do reproduce all previously known results. For this sector, we furthermore extended the ABA in Appendix~\ref{MasslessABA} to also include massless modes, following \cite{Ekhammar:2024kzp}. 

The issue regarding the dressing phases warrants further investigation. It may be that additional ingredients are needed to define the QSC beyond the symmetric sector. Alternatively, the problem could be an artefact of a possible Stokes phenomenon in the ABA limit (i.e. non-commutativity of the large volume limit and analytic continuations), in which case extra care is needed when taking the large-volume limit.

The most pressing task at present is to demonstrate that our QSC has non-trivial solutions for finite quantum numbers, at least perturbatively and numerically, at weak coupling. We believe the techniques developed in \cite{Cavaglia:2022xld} should be able to accomplish this with minor modifications, and we hope that recent developments in AdS$_5\times$S$^5$ will help extend numerics to strong coupling as well \cite{Ekhammar:2024rfj}. Unfortunately, since the dual CFT is unknown, we currently lack data for comparison at weak coupling, and it appears unlikely that this will change in the near future. At strong coupling, there is more hope for the immediate future. The AdS Virasoro-Shapiro amplitude programme \cite{Alday:2023mvu} has recently yielded new predictions for AdS$_3\times$S$^3\times$T$^4$ \cite{Chester:2024wnb,Jiang:2025oar}, and the extension to AdS$_3\times$S$^3\times$S$^3\times$S$^1$ appears to be within reach.

In this paper, we have focused on understanding AdS$_3\times$S$^3\times$S$^3\times$S$^1$ supported by pure RR flux; however, the pure NSNS case is also highly interesting. In particular, much more is known about the dual CFT, especially in the case of minimal flux \cite{Eberhardt:2019niq}, but also beyond \cite{sriprachyakul2024spacetimedilatonrmads3times,sriprachyakul2024superstringsnearconformalboundary,Gaberdiel:2024dva}. In AdS$_3\times$S$^3\times$T$^4$, recent advances on both the integrability side \cite{Frolov:2024pkz,Frolov:2025uwz,Frolov:2025tda} and the dual CFT side \cite{Gaberdiel:2023lco,Gaberdiel:2024dfw,Gaberdiel:2024nge} give hope that we will soon be able to compare explicit results. It would be very exciting to extend these results and methods to AdS$_3\times$S$^3\times$S$^3\times$S$^1$.

It would be interesting to attempt to generalise our construction away from equal radii. This would presumably require gluing together two $D(2,1|\alpha)$ QQ-systems. The first challenge would be to construct these QQ-systems; to the best of our knowledge, they are not known even for a rational spin chain. We expect that these systems will feature QQ-relations with shifts of $\alpha\,\ii$ and $(1-\alpha)\,\ii$. Having two different shift values may be in tension with analytic continuation; resolving this would be highly interesting.

\paragraph{Note added.} As this work was nearing completion, we became aware of overlapping work by Andrea Cavaglià,  Rouven Frassek, Nicolò  Primi, and Roberto Tateo, some of which were presented by Nicolò Primi at IGST 2025 \cite{Primi2025AdS3S3S3S1}. We thank them for interesting discussions, especially during IGST 2025, and for agreeing to coordinate submissions.

\section*{Acknowledgements}
We are grateful for helpful discussions with   
A.~Cavaglià,  N.~Primi,
B.~Stefański, R.~Tateo and C.~Thull. F.C and S.E are grateful to the participants of the Workshop \textit{Workshop on Higher-d Integrability in Favignana
(Italy)} in June 2025 for stimulating discussions.

The work of N.G. was
supported by the European Research Council (ERC) under the European Union’s Horizon 2020 research and innovation programme – 60 – (grant agreement No. 865075) EXACTC. N.G.’s research is supported in part by the Science Technology \& Facilities Council (STFC) under the grants ST/P000258/1 and ST/X000753/1. The work of S.E. was conducted with funding awarded
by the Swedish Research Council, Vetenskapsrådet, grant VR 2024-00598. The work of F.C. is funded by the STFC Research Council grant ST/Y509279/1. The work of B.S. is supported by the STFC Research Council grant ST/T000759/1.

\newpage
\appendix

\section{Conventions}\label{app:conventions}

In this appendix, we summarise our conventions for the various explicit matrix representations used throughout the paper.
\paragraph{Raising and lowering indices.}
We use indices $i,a,\dot{a}=\dot{1},\dot{2}$ and introduce the fully antisymmetric tensors
\begin{equation}
    \ep^{\dot{a}\dot{b}}=\ep_{\dot{a}\dot{b}}=\ep_{ab}=\ep^{ab}=\ep^{ij}=\ep_{ij}=\begin{pmatrix}
    0&1\\-1&0
\end{pmatrix},
\end{equation}
using which we raise and lower indices according to $\psi^{a} = \epsilon^{ab}\psi_b\,, \psi_{a} = -\epsilon_{ab}\psi^{b}\,.$
The equivalent matrices for the vector representation indexed by $A=1,2,3,4$ and $I=1,2,3$ are
\begin{equation}
    \eta^{AB}=\eta_{AB}=\begin{pmatrix}
0&0&0&1\\0&0&1&0\\0&1&0&0\\1&0&0&0
\end{pmatrix}\;,
\quad
\rho^{IJ}=\rho_{IJ}=\begin{pmatrix}
    0&0&1\\0&1&0\\1&0&0
\end{pmatrix}\;.
\end{equation}

\paragraph{$\algso_4$ conventions.}
We map between spinors and vectors of $\algso_4$ using $\sigma^{\dot{a}b}_A$, explicitly written as
\begin{align}
\si_1^{\dot{a}b}&=\begin{pmatrix}
-1&0\\0&0
\end{pmatrix}~,~\si_2^{\dot{a}b}=\begin{pmatrix}
0&0\\1&0
\end{pmatrix}~,~\si_3^{\dot{a}b}=\begin{pmatrix}
0&1\\0&0
\end{pmatrix}~,~\si_4^{\dot{a}b}=\begin{pmatrix}
0&0\\0&1
\end{pmatrix}, \\
\si_1^{a\dot{b}}&=\begin{pmatrix}
-1&0\\0&0
\end{pmatrix}~,~\si_2^{a\dot{b}}=\begin{pmatrix}
0&1\\0&0
\end{pmatrix}~,~\si_3^{a\dot{b}}=\begin{pmatrix}
0&0\\1&0
\end{pmatrix}~,~\si_4^{a\dot{b}}=\begin{pmatrix}
0&0\\0&1
\end{pmatrix}\;.
\end{align}
We also use $(\sigma_A)_{a\dot{b}} = \epsilon_{ac}\epsilon_{\dot{b}\dot{d}}\left(\sigma_{A}\right)^{c\dot{d}}$.

\paragraph{$\mathfrak{sp}_2$ conventions.}
Analogous map for $\algsp_2$ is performed using matrices $\Sigma_I, I=1,2,3$. They take the following form
\begin{equation}
    (\Sigma_1)^{~j}_i=\begin{pmatrix}
0&0\\1&0
\end{pmatrix}~,~(\Sigma_2)^{~j}_i=\dfrac{1}{\sqrt{2}}\begin{pmatrix}
1&0\\0&-1
\end{pmatrix}~,~(\Sigma_3)^{~j}_i=\begin{pmatrix}
0&1\\0&0
\end{pmatrix}.
\end{equation}
Note then that $(\Sigma_I)_{ij} \equiv -\epsilon_{jk}(\Sigma_I)_{j}{}^{k}$ and $\left(\Sigma_I\right)^{ij} = \epsilon^{ik} \left(\Sigma_I\right)_{k}{}^{j}$ are symmetric, explicitly
\begin{align}\label{Sigmasym}
    (\Sigma_1)_{ij} &=\begin{pmatrix}
0&0\\0&1
\end{pmatrix}~,~(\Sigma_2)_{ij}=\dfrac{1}{\sqrt{2}}\begin{pmatrix}
0&1\\1&0\end{pmatrix}~,~(\Sigma_3)_{ij}=\begin{pmatrix}
-1&0\\0&0
\end{pmatrix}\,, \\
(\Sigma_1)^{ij} &= \begin{pmatrix}
1&0\\0&0
\end{pmatrix}\,,
\quad
(\Sigma_2)^{ij}=-\dfrac{1}{\sqrt{2}}\begin{pmatrix}
0&1\\1&0\end{pmatrix}~,~(\Sigma_3)^{ij}=\begin{pmatrix}
0&0\\0&-1
\end{pmatrix}\,.\nonumber
\end{align}
\paragraph{Useful relations.}
We widely use the so-called Fierz identities
\begin{equation}
\begin{split}\label{eq:Fierz}
(\sigma_A)^{a\dot{b}}(\sigma^A)_{\dot{c}d}=-\delta^a_d\delta^{\dot{b}}_{\dot{c}}\,,~~~&~~~(\Sigma_I)^{ij}(\Sigma^I)_{kl}=-\delta^i_l\delta^j_k-\dfrac{1}{2}\ep^{ij}\ep_{kl}\,, \\
(\si_A)_{a\dot{b}}(\si^B)^{\dot{b}a}=-\delta_A^B,~~~&~~~(\Sigma_I)_{kl}(\Sigma^J)^{lk}=-\delta_I^J\,.
\end{split}
\end{equation}

\section{Additional Consequence of the QQ-system}\label{App:QQ}
In this appendix, we collect further algebraic relations that follow from those of our $\algosp(4|2)$ QQ-system presented in Section~\ref{sec:OSPQSC}.

\subsection{Fermionic QQ-relation.}
In this section, we derive the fermionic QQ-relation
\begin{equation}\label{PQ_1}
\hat{\bQ}_{kl}\bP_{\dot{a}b}=-Q^+_{\dot{a}|l}Q^+_{b|k}+Q^-_{b|l}Q^-_{\dot{a}|k}\;.
\end{equation}
We start from the definitions of $\hat{\bQ}$ and $\bP$ which give in product
\begin{equation}
    \hat{\bQ}_{kl}\bP_{\dot{e}f}=(Q^{\dot{c}|j})^+Q_{b|k}^+\ep_{jl}\ep^{ab}\bP_{\dot{c}a}\bP_{\dot{e}f}\;.
\end{equation}
To simplify it, we note that for any $2\times2$ matrix with unit determinant
\begin{equation}\label{detP}
    \bP_{\dot{a}b}\bP_{\dot{c}d}-\bP_{\dot{c}b}\bP_{\dot{a}d}=\ep_{\dot{a}\dot{c}}\ep_{bd}\,.
\end{equation}
Substituting this results in
\begin{equation}
   \hat{\bQ}_{kl}\bP_{\dot{e}f}=-Q_{\dot{e}|l}^+Q_{f|k}^+-Q_{f|l}^-Q_{b|k}^+\ep^{ab}\bP_{\dot{e}a} \,.
\end{equation}
The second term here is simplified using once more (\ref{SpinorPQ1}) and (\ref{detP}). As a result
\begin{equation}\label{qq3spin}
    \hat{\bQ}_{kl}\bP_{\dot{e}f}=-Q_{\dot{e}|l}^+Q_{f|k}^++Q_{\dot{e}|k}^-Q_{f|l}^-\;.
\end{equation}

\subsection{Vector Representations of Q-function}\label{app:Vector}

In this section, we present some useful equations in vector representation. We begin by introducing
\begin{equation}
Q_{A|I} = -\sigma_{A}^{a\dot{b}} \Sigma^{ij}_{I} Q_{a|i} Q_{\dot{b}|j}\;,
\quad
Q_{A|\circ} = -\sigma_{A}^{a\dot{b}} \ep^{ij} Q_{a|i} Q_{\dot{b}|j}\;,
\end{equation}
using which we can rewrite \eqref{qq3spin} as
\begin{equation}
    \bQ_I\bP_A=Q_{A|I}^+-Q_{A|I}^-~~,~~\bQ_\circ\bP_A=Q_{A|\circ}^++Q_{A|\circ}^-\,,
\end{equation}
Furthermore, in vector notation, it is easy to rotate $\bP$ into $\bQ$, we find
\begin{equation}
    \bQ_I=\mp Q_{A|I}^{\pm}\bP^{A}\;,
    \quad 
    \bQ_\circ=- Q_{A|\circ}^{\pm}\bP^{A}\,,
\end{equation}
which can also be inverted to give
\begin{equation}\label{PQ Rel}
    \bP_{A} = \bQ_{I}\rho^{IJ}Q^{-}_{A|J} + \frac{1}{2}\bQ_{\circ}Q^{-}_{A|\circ}\,.
\end{equation}
Let us recall the constraint $\bP_{A}\bP^{A}=-2$, by rotating $\bP$ into $\bQ$ here we find that 
\begin{equation}
    \bQ_{I} \rho^{IJ}\bQ_{J} -\frac{1}{2} \bQ^2_{\circ} = -2\;.
\end{equation}

\subsection{Constraints on $\mathcal{A}\mathcal{A}$ and $\mathcal{B}\mathcal{B}$.}\label{app:asymp}

In this section, we constrain $\mathcal{A}_{A}\mathcal{A}^{A}$ and $\mathcal{B}_I\mathcal{B}^{I}$. We start from
\begin{equation}\label{eq:AsymptoticsApp}
    \bP_A \simeq \mathcal{A}_A\,u^{-M_A}\,,
    \quad
    \bQ_I \simeq \mathcal{B}_I\,u^{\hat{M}_I-1}\,.
\end{equation}
Recall that $\bP_A\bP^{A}=-2$, which implies
\begin{equation}\label{eq:AAConstraint1}
    \sum_B \mathcal{A}_B \mathcal{A}^B = -2.
\end{equation}
We then recall that
\begin{equation}
    Q^{\pm}_{A|I}\bP^{A} = \mp \bQ_{I}\,,
\end{equation}
which upon requiring \eqref{eq:AsymptoticsApp} implies
\begin{equation}\label{eq:AAConstraint2}
    \sum_B \frac{\mathcal{A}_B \mathcal{A}^B}{M_B + \hat{M}_K} = 0\;.
\end{equation}
Combining \eqref{eq:AAConstraint1} and \eqref{eq:AAConstraint2} we solve the above to find
\begin{align}
    &\mathcal{A}_1\mathcal{A}^1 = \mcA_4 \mcA^4 = \frac{-M_1^2 + \hat{M}_1^2}{M_1^2 - M_2^2}\;,
    \quad
    \mathcal{A}_2\mathcal{A}^2=\mathcal{A}_3\mathcal{A}^3 = \frac{M_2^2 - \hat{M}_1^2}{M_1^2 - M_2^2}\;.
\end{align}

We can follow a similar path to find $\mathcal{B}_K \mathcal{B}^K$. We make use of the following QQ-relations
\begin{align}
    \bP_A = Q^{-}_{A|I}\bQ^I + \frac{\bQ_\circ Q^{-}_{A|\circ}}{2}\;,
    \quad
    -2 = \bQ_I \bQ^I - \frac{1}{2} \bQ_{\circ}^2\;,
\end{align}
which, upon taking asymptotics into account, give the following constraints
\begin{equation}
    \sum_{I=1}^{3} \frac{\mathcal{B}_I \mathcal{B}^I}{M_A + \hat{M}_I}=\frac{M_A}{2}\;,
\end{equation}
These can be solved to find
\begin{align}
    \mathcal{B}_1\mathcal{B}^1 = \mathcal{B}_3\mathcal{B}^3 = \frac{(M^2_1 - \hat{M}^2_1) (M^2_2 - \hat{M}^2_1)}{4 \hat{M}_1^2}\,,
    \quad
    \mathcal{B}_2\mathcal{B}^2 = -\frac{M_1^2 M_2^2}{2 \hat{M}_1^2}\,.
\end{align}

\section{The $\bP\mu$ and $\bQ\omega$ systems}

\subsection{Derivation of $\bQ\tau$}
Let us explain in more detail here the computation leading to the $\bQ\tau$ system in spinor notation. Taking the difference between (\ref{qq3spin}) and its analytic continuation along $\bar{\gamma}$ gives
\begin{equation}
(\hat{\bQ}_{kl})^{\bar{\gamma}}(\bP_{\dot{a}b})^{\bar{\gamma}}-\hat{\bQ}_{kl}\bP_{\dot{a}b}=-(Q_{\dot{a}|l}^{-})^{\bar{\gamma}}(Q_{b|k}^{-})^{\bar{\gamma}}+Q_{\dot{a}|l}^{-}Q_{b|k}^{-}.
\end{equation}
Using \eqref{ud} and contracting with $F_{\bar{l}}^{~n}F_{\bar{k}}^{~m}$ results in the discontinuity relation
\begin{equation}(\tau_{\bar{k}}^{~m++})^{\bar{\gamma}}(\tau_{\bar{l}}^{~n})^{\bar{\gamma}}-\tau_{\bar{k}}^{~m++}\tau_{\bar{l}}^{~n}=\hat{\bQ}_{\bar{k}\bar{l}}^{\downarrow}(\hat{\bQ}^{mn\downarrow})^{\bar{\gamma}}-(\hat{\bQ}_{\bar{k}\bar{l}}^{\downarrow})^{\gamma}\hat{\bQ}^{mn\downarrow}.\end{equation}

\subsection{Mirror-Periodicity}\label{QwPm Appendix}
In this appendix, we show the mirror-periodicity condition:
\begin{equation}
    \left((\nu_{\dot{c}}^{~\dot{\bar{e}}}\nu_a^{~\bar{f}})^{[2]}\right)^{\bar{\gamma}}=\nu_{\dot{c}}^{~\dot{\bar{e}}}\nu_a^{~\bar{f}}\,.
\end{equation}
We start from $\nu$ definition through $\tau$, see \eqref{nudef}, this implies
\begin{equation}
\begin{cases}\label{numirrsys}(\nu_{\dot{c}}^{~\dot{\bar{e}}}\nu_a^{~\bar{f}})^{++\bar{\gamma}}=Q_{\dot{c}|l}^{+}Q_{a|k}^{+}Q^{\dot{\bar{e}}|\bar{l}+}Q^{\bar{f}|\bar{k}+}((\tau^{k}_{~\bar{k}})^{[+2]}\tau^l_{~\bar{l}})^{\bar{\gamma}}~~\\~~\nu_{\dot{c}}^{~\dot{\bar{e}}}\nu_a^{~\bar{f}}=Q_{\dot{c}|l'}^{-}Q_{a|k'}^{-}Q^{\dot{\bar{e}}|\bar{l}'-}Q^{\bar{f}|\bar{k}'-}\tau^{k'}_{~\bar{k}'}(\tau^{l'}_{~\bar{l}'})^{[+2]}
\end{cases}
\end{equation}
The first line can be simplified using the $\tau$ discontinuity relation (\ref{tautau}):
$$
    (\nu_{\dot{c}}^{~\bar{\dot{e}}}\nu_a^{~\bar{f}})^{++\bar{\gamma}}=Q_{\dot{c}|l}^{+}Q_{a|k}^{+}Q^{\bar{\dot{e}}|\bar{l}+}Q^{\bar{f}|\bar{k}+}\((\tau^{k}_{~\bar{k}})^{[+2]}\tau^l_{~\bar{l}}+\hat{\bQ}^{kl}(\hat{\bQ}_{\bar{k}\bar{l}})^{\bar{\gamma}}-(\hat{\bQ}^{kl})^{\gamma}\hat{\bQ}_{\bar{k}\bar{l}}\)
$$
and then further expanded using (\ref{qgammabar}) to remove $\bar{\gamma}$:
$$
    (\nu_{\dot{c}}^{~\bar{\dot{e}}}\nu_a^{~\bar{f}})^{++\bar{\gamma}}=Q_{\dot{c}|l}^{+}Q_{a|k}^{+}Q^{\bar{\dot{e}}|\bar{l}+}Q^{\bar{f}|\bar{k}+}\Bigl((\tau^{k}_{~\bar{k}})^{[+2]}\tau^l_{~\bar{l}}-\hat{\bQ}^{kl}(\tau_{~\bar{k}}^{l'})^{[+2]}\tau_{~\bar{l}}^{k'}\hat{\bQ}_{l'k'}-$$
    \begin{equation}\label{numirr1}
    -(\hat{\bQ}^{kl})^{\gamma}\hat{\bQ}_{\bar{k}\bar{l}}+(\hat{\bQ}^{l'k'})^{\gamma}\hat{\bQ}_{l'k'}\hat{\bQ}_{\bar{k}\bar{l}}\hat{\bQ}^{kl}\Bigr)\,.
\end{equation}
The second line should be brought exactly to this form. As it is now, it contains several $Q^-$ functions, all of which should somehow be replaced by $Q^+$. This can be achieved by considering one of the $QQ$-relation (\ref{qq3spin}) rewritten as follows
$$Q_{b|l}^-Q_{\dot{d}|k}^{-}=Q_{\dot{d}|l'}^+Q_{b|k'}^{+}(\delta^{k'}_{k}\delta^{l'}_{l}-\hat{\bQ}_{kl}
\hat{\bQ}^{k'l'})\,.$$
Plugging it into the second line of  (\ref{numirrsys}) will reproduce (\ref{numirr1}) without any further simplifications.

\subsection{$\bQ\omega$ System in Vector Notation}

In this section, we rewrite the $\bQ\omega$-system using vector notation.  To do so we define 3 new objects: $\omega$, $\psi$ and $\chi$ by
\begin{align}
    \omega_I^{~\bar{J}} &= -(\Sigma_I)^{ji}(\tau_i^{~\bar{i}})^{[+2]}\tau_j^{~\bar{j}}(\Sigma^{\bar{J}})_{\bar{i}\bar{j}}\,,
    \\
    \psi_I &= -(\Sigma_I)^{ji}(\tau_i^{~\bar{i}})^{[+2]}\tau_j^{~\bar{j}}\ep_{\bar{i}\bar{j}}\,,
    \\
    \chi &= \dfrac{1}{2}\ep_{\bar{i}\bar{j}}(\tau_{i}^{~\bar{i}})^{[+2]}\tau_{j}^{~\bar{j}}\ep^{ij}.
\end{align}
The $2\ii$ periodicity of $\tau$ manifests itself here as
\begin{equation}\label{Qommirr}(\omega_I^{~\bar{J}})^{[+2]}=\omega_I^{~\bar{J}}~~,~~\psi^{[+2]}_I=-\psi_I~~,~~\chi^{[+2]}=\chi.
\end{equation}
Then contracting the indices in equation (\ref{qtau}) with $(\Sigma_I)^{ij}$ gives
\begin{equation}\label{qw1}(\bQ_I^{\downarrow})^{\gamma}=\omega_I^{~\bar{J}}\bQ_{\bar{J}}^{\downarrow}+\dfrac{1}{2}\psi_I\bQ_{\circ}^{\downarrow}~~,~~(\bQ_{\bar{I}}^{\downarrow})^{\gamma}=\omega_{\bar{I}}^{~J}\bQ_{J}^{\downarrow}+\dfrac{1}{2}\psi_{\bar{I}}\bQ_{\bar{\circ}}^{\downarrow}\,.\end{equation}
and
\begin{equation}\label{qw2}
(\bQ_{\circ}^{\downarrow})^{\gamma}=\chi \bQ^{\downarrow}_{\circ}-\psi^{\bar{I}}\bQ_{\bar{I}}^{\downarrow}~~,~~(\bQ_{\bar{\circ}}^{\downarrow})^{\gamma}=\bar{\chi} \bQ^{\downarrow}_{\bar{\circ}}-\psi^{I}\bQ_{I}^{\downarrow}.
\end{equation}
The discontinuity equation \eqref{tautau} splits into
$$
(\omega_{\bar{I}}^{~J})^{\bar{\gamma}}-\omega_{\bar{I}}^{~J}=-\bQ_{\bar{I}}^{\downarrow}(\bQ^{J\downarrow})^{\bar{\gamma}}+(\bQ_{\bar{I}}^{\downarrow})^{\gamma}\bQ^J,
$$
\begin{equation}\label{qw}
(\psi_{\bar{I}})^{\bar{\gamma}}-\psi_{\bar{I}}=-\bQ_{\bar{I}}^{\downarrow}(\bQ_{\circ}^{\downarrow})^{\bar{\gamma}}+(\bQ_{\bar{I}}^{\downarrow})^{\gamma}\bQ_{\circ}^{\downarrow},\end{equation}
$$
\chi^{\bar{\gamma}}-\chi=\dfrac{1}{2}(\bQ_{\bar{\circ}}^{\downarrow}(\bQ_{\circ}^{\downarrow})^{\bar{\gamma}}-(\bQ_{\bar{\circ}}^{\downarrow})^{\gamma}\bQ_{\circ}^{\downarrow}).
$$
Similar relations for the second $\algosp({4|2})$ copy are obtained from (\ref{qw}) by switching barred and unbarred indices.
\subsection{$\bP \mu$ System}
In a similar fashion to $\omega$, we introduce $\mu$ as the analogue of $\nu$ in the vector representation. We define matrix $\mu$ as
\begin{equation}
    \mu_A^{~\bar{B}}=-(\sigma_A)^{a\dot{c}}\nu_a^{~\bar{f}}\nu_{\dot{c}}^{~\bar{\dot{e}}}(\sigma^{\bar{B}})_{\bar{f}\bar{\dot{e}}}\,,
\end{equation}
and note that it satisfies the mirror periodicity condition
\begin{equation}\label{muppA}
(\mu_A^{~\bar{B}})^{[+2]}=(\mu_A^{~\bar{B}})^{\gamma},
\end{equation}
The vector form of relation (\ref{pmu1}) is then
\begin{equation}
\bP_A^{\gamma}=(\mu_{A}^{~~\bar{B}})^{\gamma}\bP_{\bar{B}}.
\end{equation}
Alternatively, it can be inverted to give $\bP_{\bar{B}}^{\bar{\gamma}}=\mu^{A}_{~~\bar{B}}\bP_A$. The discontinuity equation for $\nu$ becomes
\begin{equation}\label{pmu}
(\mu_A^{~~\bar{B}})^{\gamma}-\mu_{A}^{~~\bar{B}}=\bP_A(\bP^{\bar{B}})^{\bar{\gamma}}-(\bP_A)^{\gamma}\bP^{\bar{B}}.
\end{equation}

\section{The ABA equations}\label{app:ABAEquations}
As in previous iterations of the QSC, we will test our proposal against the ABA derived independently in \cite{Borsato:2012ss}. The ABA contains a total of $4$ different types of momentum carrying massive roots $\{x_k\}_{k=1}^{K_1},\{z_k\}_{k=1}^{K_3},\{\bar{x}_k\}_{k=1}^{\bar{K}_1},\{\bar{z}_k\}_{k=1}^{\bar{K}_3}$ as well as $2$ auxiliary roots $\{y_{k}\}_{k=1}^{K_2},\{\bar{y}_{k}\}_{k=1}^{K_{\bar{2}}}$

The ABA equations associated with the left system read
\begin{align}\label{theABA_1}
    \Bigl(\dfrac{x^+}{x^-}\Bigr)^{L} &= \dfrac{\bfR_{1|1(+)}^{+}}{\bfR_{1|1(-)}^{-}}\cdot\dfrac{R_1^-B_{\bar{1}}^-}{R_1^+B_{\bar{1}}^+}\cdot\sqrt{\dfrac{\bfB_{\bar{\bullet}(-)}^+}{\bfB_{\bar{\bullet}(+)}^-}}\times\\
    &\times\prod_{k=1}^{K_{1|1}}S^{11}(x,x_{1|1,k})\prod_{k=1}^{K_{\dot{1}|1}}S^{13}(x,x_{\dot{1}|1,k})\prod_{k=1}^{K_{\bar{1}|\bar{1}}}S^{1\bar{1}}(x,x_{1|1,k})\prod_{k=1}^{K_{\dot{\bar{1}}|\bar{1}}}S^{1\bar{3}}(x,x_{\dot{\bar{1}}|\bar{1},k})\Bigg|_{x=x_{1|1,k}}\,,\\
    \label{theABA_2}
    1&=\dfrac{\bfR_{\bullet(+)}\bfB_{\bar{\bullet}(-)}}{\bfR_{\bullet(-)}\bfB_{\bar{\bullet}(+)}}\Bigg|_{x=y_{1,k}}\,,\\\label{theABA_3}
    \Bigl(\dfrac{x^+}{x^-}\Bigr)^{L}&=\dfrac{\bfR_{\dot{1}|1(+)}^{+}}{\bfR_{\dot{1}|1(-)}^{-}}\cdot\dfrac{R_1^-B_{\bar{1}}^-}{R_1^+B_{\bar{1}}^+}\cdot\sqrt{\dfrac{\bfB_{\bar{\bullet}(-)}^+}{\bfB_{\bar{2}(+)}^-}}\times\\
    &\times\prod_{k=1}^{K_{\dot{1}|1}}S^{33}(x,x_{\dot{1}|1,k})\prod_{k=1}^{K_{1|1}}S^{31}(x,x_{1|1,k})\prod_{k=1}^{K_{\bar{1}|\bar{1}}}S^{3 \bar{1}}(x,x_{\bar{1}|\bar{1},k})\prod_{k=1}^{K_{\dot{\bar{1}}|\bar{1}}}S^{3\bar{3}}(x,x_{\dot{\bar{1}}|\bar{1},k})\Bigg|_{x=x_{\dot{1}|1,k}}\,.
\end{align}
While the right system reads
\begin{align}\label{theABA_4}
    \Bigl(\dfrac{x^+}{x^-}\Bigr)^{L}&=\dfrac{\bfR_{\dot{\bar{1}}|\bar{1}(-)}^{-}}{\bfR_{\dot{\bar{1}}|\bar{1}(+)}^{+}}\cdot\dfrac{R_{\bar{1}}^+B_{1}^+}{R_{\bar{1}}^-B_{1}^-}\,\sqrt{\dfrac{\bfB_{\bullet,(+)}^-}{\bfB_{\bullet,(-)}^+}}\times\\&
    \times\prod_{k=1}^{K_{\bar{1}|\bar{1}}}S^{\bar{1}\bar{1}}(x,x_{\bar{1}|\bar{1},k})\prod_{k=1}^{K_{\dot{\bar{1}}|\bar{1}}}S^{\bar{1}\bar{3}}(x,x_{\dot{\bar{1}}|\bar{1},k})\prod_{k=1}^{K_{1|1}}S^{\bar{1} 1}(x,x_{1|1,k})\prod_{k=1}^{K_{1|\dot{1}}}S^{\bar{1}3}(x,x_{\dot{1}|1,k})\Bigg|_{R_{\bar{1}|\bar{1}}=0}\,,\\\label{theABA_5}
    &1=\dfrac{\bfB_{\bullet(+)}\bfR_{\bar{\bullet}(-)}}{\bfB_{\bullet(-)}\bfR_{\bar{\bullet}(+)}}\Biggl|_{x=y_{\bar{1},k}}\\\label{theABA_6}
        \Bigl(\dfrac{x^+}{x^-}\Bigr)^{L}&=\dfrac{\bfR_{\bar{1}|\bar{1},(-)}^{-}}{\bfR_{\dot{\bar{1}}|\bar{1},(+)}^{+}}\cdot\dfrac{R_{\bar{1}}^+B_{1}^+}{R_{\bar{1}}^-B_{1}^-}\cdot\sqrt{\dfrac{\bfB_{\bullet,(+)}^-}{\bfB_{\bullet,(-)}^+}}\times\\&
        \times\prod_{k=1}^{K_{1|1}}S^{\bar{3}1}(x,x_{1|1,k})\prod_{k=1}^{K_{\dot{1}|1}}S^{\bar{3}3}(x,x_{\dot{1}|1,k})\prod_{k=1}^{K_{\bar{1}|\bar{1}}}S^{\bar{3}\bar{1}}(x,x_{\bar{1}|\bar{1},k})\prod_{k=1}^{K_{\dot{\bar{1}}|\bar{1}}}S^{\bar{3}\bar{3}}(x,x_{\dot{\bar{1}}|\bar{1},k})\Bigg|_{x=x_{\bar{1}|\bar{1},k}}\,
\end{align}

\paragraph{Crossing equations}
In \cite{Borsato:2012ss}, the following crossing equations were written down
\begin{align}
    S^{11}_{xy}S^{1\bar{1}}_{x\bar{y}} &= \frac{x^- - y^+}{x^- - y^-}\sqrt{\frac{x^+}{x^-}}\sqrt{\frac{x^- -y^-}{x^+ - y^+}}
    \;,
    &
    S^{11}_{x\bar{y}}S^{1\bar{1}}_{x y} &= \frac{1-\frac{1}{x^+ y^+}}{1-\frac{1}{x^+ y^-}}\sqrt{\frac{1-\frac{1}{x^- y^-}}{1-\frac{1}{x^+ y^+}}} \\
    S^{31}_{xy} S^{3\bar{1}}_{x\bar{y}} &= \frac{x^+ - y^-}{x^- - y^-}\sqrt{\frac{x^+}{x^-}}\sqrt{\frac{x^- - y^-}{x^+ - y^+}}
    &
    S^{31}_{x\bar{y}} S^{3\bar{1}}_{xy} &= \frac{x^+}{x^-}\frac{1-\frac{1}{x^+ y^+}}{1-\frac{1}{x^- y^+}}\sqrt{\frac{1-\frac{1}{x^- y^-}}{1-\frac{1}{x^+ y^+}}}
\end{align}

To map these to our crossing equations, we need to slightly massage these expressions. First, we use unitarity to rewrite the equations as 
\begin{align}
    S^{\bar{1}1}_{\bar{x}y}S^{11}_{xy} = \frac{x^- - y^-}{x^+-y^-}\sqrt{\frac{y^-}{y^+}}\sqrt{\frac{x^+ - y^+}{x^- - y^-}}
    \quad
    S^{11}_{\bar{x}y}S^{\bar{1}1}_{xy} = \frac{1-\frac{1}{x^- y^+}}{1-\frac{1}{x^+ y^+}}\sqrt{\frac{1-\frac{1}{x^+ y^+}}{1-\frac{1}{x^- y^-}}} \\
    S^{\bar{1}3}_{\bar{x}y}S^{13}_{xy} = \frac{x^- - y^-}{x^--y^+}\sqrt{\frac{y^-}{y^+}}\sqrt{\frac{x^+ - y^+}{x^- - y^-}}
    \quad
    S^{13}_{\bar{x}y}S^{\bar{1}3}_{xy} = \frac{y^-}{y^+}\frac{1-\frac{1}{x^+ y^-}}{1-\frac{1}{x^+ y^+}}\sqrt{\frac{1-\frac{1}{x^+ y^+}}{1-\frac{1}{x^- y^-}}}
\end{align}
Next, we need to understand what the analytic continuation $\bar{x}$ means in our language. It was pointed out in \cite{Borsato:2013hoa} that this analytic continuation should be interpreted as $\gamma_{c} = \bar{\gamma}_c^{-1}$. Thus we apply $\bar{\gamma}_c$ and find
\begin{align}
    S^{11}(x^{\bar{\gamma}_c},y)S^{\bar{1}1}(x,y) &= \frac{1-\frac{1}{x^- y^-}}{1-\frac{1}{x^+ y^-}}\sqrt{\frac{1-\frac{1}{x^+y^+}}{1-\frac{1}{x^- y^-}}}\;,
    \\
    S^{\bar{1}1}(x^{\bar{\gamma}_c},y)S^{11}(x,y) &= \sqrt{\frac{y^-}{y^+}}\frac{x^- - y^+}{x^+ - y^+}\sqrt{\frac{x^+ - y^+}{x^- - y^+}}\;,\\
    S^{13}(x^{\bar{\gamma}_c},y)S^{\bar{1}3}(x,y) &= \frac{y^-}{y^+}\frac{1-\frac{1}{x^-y^-}}{1-\frac{1}{x^-y^+}}\sqrt{\frac{1-\frac{1}{x^+y^+}}{1-\frac{1}{x^-y^-}}}\;, \\
    S^{\bar{1}3}(x^{\bar{\gamma}_c},y)S^{13}(x,y) &= \sqrt{\frac{y^-}{y^+}}\frac{x^+-y^-}{x^+-y^+}\sqrt{\frac{x^+-y^+}{x^--y^-}}\;,
\end{align}

\section{The Massless ABA}\label{MasslessABA}

In this appendix, we generalise the large volume considerations to incorporate massless degrees of freedom. We propose, in analogy with AdS$_3\times$S$^3\times$T$^4$, that they manifest themselves as zeros of $\mu$ sitting on the cut $\left(-2g,2g\right)$. To distinguish the cut from above and below, it is convenient to use Zhukovsky parameters. We write $\{z_i\}_{i=1}^{N}$ and encode them in
\begin{equation}
    \kappa=\prod_{i=1}^N(x-z_i)\,,
    \quad
    \kappa^{*}=\prod_{i=1}^N(x-\dfrac{1}{z_i})\;.
\end{equation}
We now proceed to modify the ABA construction of Section~\ref{Sect:ABA}, largely following the previous analysis for AdS$_{3}\times $S$^3\times $T$^4$ QSC in \cite{Ekhammar:2024kzp}.

\subsection{The Function $F$}
We begin with generalising the function $F$, defined in the case of purely massive excitations in \eqref{abaf}. It now takes the form
\begin{equation}\label{masslessF^2}
    F^2=\dfrac{\mu}{(\mu)^{[+2]}}\dfrac{\mathbb{Q^+}}{\mathbb{Q}^-}\dfrac{\kappa^{*}}{\kappa}\,,
    \quad 
    F(\infty)=\pm1\,.
\end{equation}
 As in the massive case, using the definition of $\mu$, see \eqref{muQQQQomega}, we find
\begin{equation}\label{mlessQQQQ}
F^2 \propto\dfrac{Q_{1|1}^{-}Q_{\bar{1}|\bar{1}}^{-}Q_{\dot{1}|1}^{-}Q_{\dot{\bar{1}}|\bar{1}}^{-}}{Q_{1|1}^{+}Q_{\bar{1}|\bar{1}}^{+}Q_{\dot{1}|1}^{+}Q_{\dot{\bar{1}}|\bar{1}}^{+}}\dfrac{\mathbb{Q}^+}{\mathbb{Q}^-}\dfrac{\kappa^{*}}{\kappa}\,,
\quad
F^2 \propto \dfrac{Q_{1|1}^{*-}Q_{\bar{1}|\bar{1}}^{*-}Q_{\dot{1}|1}^{*-}Q_{\dot{\bar{1}}|\bar{1}}^{*-}}{Q_{1|1}^{*+}Q_{\bar{1}|\bar{1}}^{*+}Q_{\dot{1}|1}^{*+}Q_{\dot{\bar{1}}|\bar{1}}^{*+}}\dfrac{\mathbb{Q}^+}{\mathbb{Q}^-}\dfrac{\kappa^{*}}{\kappa}\,.
\end{equation}
As a consequence, $F$ is analytic both in the upper and lower halves of the complex plane. Its analytic continuation under the cut on the real line is
\begin{equation}
    FF^{\gamma}=FF^{\bar{\gamma}}=\pm\dfrac{\mathbb{Q}^+}{\mathbb{Q}^-}\dfrac{1}{\prod_{i=1}^{N}z_i}.
\end{equation}
These equations define a scalar Riemann-Hilbert problem which, along with $F(\infty)=1$, fixes the function $F$ uniquely and imposes a cyclicity constraint on the roots. The solution to this problem is given as
\begin{equation}\label{masslessF}
    F=\dfrac{\mathbf{B}_{(+)}}{\mathbf{B}_{(-)}}\,,
    \quad
    \prod_{i=1}^{M}\dfrac{x^+_i}{x^-_i}\prod_{k=1}^{N} z_k=1\,.
\end{equation}

\subsection{Finding $\mu$ and $\omega$}
Now with an exact expression for $F$ \eq{masslessF}, equation (\ref{masslessF^2}) can be solved to find $\mu$
\begin{equation}\label{mlessmu}
\mu\propto \mathbb{Q}^-\mathcal{M}f f^{*[-2]} \prod_{n=0}^{\infty} \kappa^{[2n]} \kappa^{*[-2n-2]}\;.
\end{equation}
The solution is defined up to an arbitrary function $\mathcal{M}$ which is periodic for $N$ even and anti-periodic for $N$ odd, as follows from carefully treating the infinite product in \eqref{mlessmu}. It is constrained by the mirror-periodicity condition $\mu^{\gamma}=\mu^{[+2]}$ which translates into
\begin{equation}\label{mlessmirror}
    \mathcal{M^{\gamma}}= (-x)^{N}\mathcal{M}\,.
\end{equation}
When solving \eqref{mlessmirror}
we will introduce cuts unless $N$ is an even number; we will hence assume that this is the case. The solution to \eqref{mlessmirror} is given by
\begin{equation}\label{eq:SolvingM}
    \mathcal{M} \propto p\prod_{n=-\infty}^{+\infty}\Bigl(\dfrac{1}{x^{[2n]}}\Bigr)^{N/2}\,,
    \quad
    p^{\gamma}=p\,.
\end{equation}
We will keep the (anti)-periodic function $p(x)$ here unidentified for now. 

Just as in the massive case, using the explicit form of $F$ we find from \eqref{mlessQQQQ}
\begin{equation}\label{mlessQQQQ_1}
    Q_{1|1}^{-}Q_{\dot{1}|1}^{-}Q_{\bar{1}|\bar{1}}^{-}Q_{\dot{\bar{1}}|\bar{1}}\propto\mathbb{Q}^{-}f^{2}\prod_{n=0}^{+\infty}\left(\dfrac{\kappa}{\kappa^*}\right)^{[2n]}\;,
\end{equation}
which by \eqref{muQQQQomega} implies 
\begin{equation}\label{mlessom}
    \omega\propto\mathcal{M}\dfrac{(f^*)^{[-2]}}{f}\prod_{-\infty}^{+\infty}(\kappa^*)^{[2n]}\;.
\end{equation}

\subsection{Finding $Q_{a|i}$}
We now proceed to modify the general ansatz for $Q_{a|i}$ \eq{1qsplit} to also include massless roots. To do so, we introduce some notation that will soon become convenient. We write
\begin{equation}\label{fcirc}
f_\circ^2=\prod_{n=0}^{+\infty}\dfrac{\kappa^{[2n]}}{(\kappa^*)^{[2n]}}\;,
\quad
\mathcal{I}^2 = \mathcal{M} \prod_{n=-\infty}^{\infty}(\kappa^{\star})^{[2n]} \frac{f_{\circ}}{f_{\circ}^{*,--}}\,,
\end{equation}
and then we find
\begin{equation}\label{mless1qsplit}
    Q_\alpha^{{\downarrow}}\propto \mathbb{Q}_{\alpha}\sqrt{f^{+}f^{+}_\circ} h_\alpha^+(u)\,,
    \quad 
    \alpha\in\{1|1,\;\bar{1}|1,\;\dot{1}|1,\;\bar{\dot {1}}|1\}\,,
\end{equation}
where $h_{\alpha}$ are constrained by (\ref{mlessQQQQ_1}) to satisfy
\begin{equation}
h_{1|1}h_{\bar{1}|1}h_{\dot{1}|1}h_{\bar{\dot{1}}|\bar{1}} \propto 1\,.
\end{equation}
Following the massive analysis we split
\begin{equation}\label{mlessnusplit}
    \nu_{1}^{~\bar{2}}\propto\mathbb{Q}^-_{1|1}\mathbb{Q}^{-}_{\bar{1}|\bar{1}}\sqrt{f^{*--}ff_\circ^{*--}f_\circ}\cdot\mathcal{I}\mathcal{F}\,,
    \quad
    \nu_{\dot{1}}{}^{\dot{\bar{2}}}\propto\mathbb{Q}^-_{\dot{1}|1}\mathbb{Q}^{-}_{\dot{\bar{1}}|\bar{1}}\sqrt{f^{*--}ff_\circ^{*--}f_\circ}\cdot\mathcal{I}\mathcal{F}^{-1}\,,
\end{equation}
and find $\tau$ via \eq{nudefaba} as
\begin{equation}\label{mlesstauI}
\tau^{1}_{~\bar{2}}\propto\sqrt{\dfrac{f^{*--}f^{*--}_\circ}{ff_\circ}}\cdot\mathcal{I}\mathcal{F} h_{1|1}^{-1}h_{\bar{1}|\bar{1}}^{-1}~~,~~
\dot\tau^{1}_{~\bar{2}}\propto\sqrt{\dfrac{f^{*--}f^{*--}_\circ}{ff_\circ}}\cdot\mathcal{I}\mathcal{F}^{-1} h^{-1}_{\dot{1}|1}h^{-1}_{\dot{\bar{1}}|\bar{1}}\;.\end{equation}
The same relations also hold for the barred counterparts. Finding the ABA expressions for $Q_{a|i}$ now amounts to finding the functions $h_{\alpha}$ and $\mathcal{F}$, just as for the massive case. Using \eqref{udaba1} results in a series of relations:
\begin{equation}\label{mlessFH2}
    \mathcal{F}^{+}\mathcal{I}^+\propto h_{\bar{1}|\bar{1}}^+h^{*-}_{1|1}\propto(h_{\dot{\bar{1}}|\bar{1}}^{+}h^{*-}_{\dot{1}|1})^{-1}~~,~~\bar{\mathcal{F}}^{+}\mathcal{I}^+\propto h_{\bar{1}|\bar{1}}^{*-}h^{+}_{1|1}\propto (h_{\dot{\bar{1}}|\bar{1}}^{*-}h^{+}_{\dot{1}|1})^{-1}\;.
    \end{equation}
Looking at initial splitting (\ref{mless1qsplit}) it is apparent that $h_{\alpha}$ are UHPA, since $Q_{a|i}^{\downarrow}$ are. Following the same reasoning as in the massive case, we arrive at
\begin{equation}
    h_{1|1}\propto h_{\dot{1}|1}^{-1}~~,~~h_{\bar{1}|\bar{1}}\propto h_{\dot{\bar{1}}|\bar{1}}^{-1}\;.
\end{equation}
Denoting $h\equiv h_{1|1}~,~\bar{h}\equiv h_{\bar{1}|\bar{1}}$, the expression for the $Q$-functions  (\ref{mless1qsplit}) simplifies
\begin{equation}\label{mless1q}
\begin{split}
    Q_{1|1}&\propto \mathbb{Q}_{1|1}\sqrt{f^{+}f^{+}_\circ} h^+\;,
    \quad
    Q_{\dot{1}|1}\propto \mathbb{Q}_{\dot{1}|1}\sqrt{f^{+}f^{+}_\circ} (h^+)^{-1}\;, \\
    Q_{\bar{1}|\bar{1}}&\propto \mathbb{Q}_{\bar{1}|\bar{1}}\sqrt{f^{+}f^{+}_\circ} \bar{h}^+\;,
    \quad
    Q_{\dot{\bar{1}}|\bar{1}}\propto \mathbb{Q}_{\dot{\bar{1}}|\bar{1}}\sqrt{f^{+}f^{+}_\circ} (\bar{h}^+)^{-1}\;.
\end{split}
\end{equation}
To proceed further, we now search for an appropriate choice of the function $p$.
\paragraph{Fixing $p$}
To fix $p$, we start by slightly reshuffling the terms in $\omega$ to write
\begin{equation}
    \omega \propto \mathcal{M}\dfrac{(f^*)^{[-2]}}{f}\prod_{-\infty}^{+\infty}(\kappa^*)^{[2n]} = \dfrac{(f^*f_{\circ}^*)^{[-2]}}{f f_{\circ}} \times \mathcal{M}\prod_{-\infty}^{+\infty}(\kappa^*)^{[2n]} \frac{f_{\circ}}{f^{*[-2]}_\circ} = \dfrac{(f^*f_{\circ}^*)^{[-2]}}{f f_{\circ}} \mathcal{I}^2\,.
\end{equation}
And the same notation results in
\begin{equation}
    \mu \propto \betheQ^- f f_{\circ} \left(f^{*}f^{*}_{\circ}\right)^{[-2]} \mathcal{I}^2\;.
\end{equation}
We notice that
\begin{equation}
\mathcal{I}^2=\mathcal{M}\prod_{n=-\infty}^{\infty}{(\kappa^{*})^{[2n]}}\frac{f_\circ}{f^{*--}_\circ} \propto p^2 \prod_{k=1}^{N}\sinh{\pi(u-\theta_k)}\;.
\end{equation}
where $\theta_k = g(z_k+\frac{1}{z_k})$. The above identity follows from writing out all $\kappa$'s in $\mathcal{I}$ explicitly in terms of massless roots $z_k$, see \cite{Ekhammar:2024kzp} for more details.

We immediately observe that in general $\mathcal{I}$ has exponential asymptotics. Then so do $\mu$ and $\omega$, but from experience with other instances of the QSC, we expect that these Q-functions must have powerlike asymptotics. To fix this issue, we follow \cite{Ekhammar:2024kzp} and choose the so far arbitrary function $p$ such that $\mathcal{I}=1$. Notice that this forces the massless roots to be $2$ times degenerate in order not to introduce cuts in $p^2$. We will soon see that the roots are actually even further degenerate in the symmetric sector.

\paragraph{Fixing $\mathcal{F}$ and $h$.} As we will now show, the defining relations for $h$ and $\mathcal{F}$ are very similar to the massive-case. We start by normalizing $\mathcal{F}$ and $
\bar{\mathcal{F}}$ so that
\beq\label{FhhApp}
{\cal F}^+=\bar{h}^+ h^{*-}\,,
\quad
\bar{\mathcal{F}}^+ = h^+\bar{h}^{*-}\,.
\eeq
Then the first constraint on $\mathcal{F}$, derived by applying mirror-periodicity to \eq{mlessnusplit}, is
\begin{equation}\label{mless2}
\frac{\mathcal{F}^{\gamma}}{\mathcal{F}^{[+2]}}=
\frac{\bar{\mathcal{F}}^{\gamma}}{{\bar{\mathcal{F}}}^{[+2]}}={\cal S}\equiv\pm
\sqrt{\frac{
{\mathbb Q}_{1|1}^+
{\mathbb Q}_{\bar 1|\bar 1}^+
{\mathbb Q}_{\dot 1|1}^-
{\mathbb Q}_{\bar {\dot 1}|\bar 1}^-
}{
{\mathbb Q}_{1|1}^-
{\mathbb Q}_{\bar 1|\bar 1}^-
{\mathbb Q}_{\dot 1|1}^+
{\mathbb Q}_{\bar {\dot 1}|\bar 1}^+
}}\times C_{\circ}^{-1}\;.
\end{equation}
Here $C_{\circ}=(\prod_{k=1}^{N}z^2_k)^{1/4}$ is the only contribution of the massless factors $f_{\circ}$. The second constraint on $\mathcal{F}$, which we derive by applying $\tau^{[+2]}=\dot{\tau}$ to \eq{mlesstauI} is
\begin{equation}\label{FH1appendix}
\mathcal{F}\mathcal{F}^{[+2]} = C_h (h h^{[2]}) \,(\bar{h} \bar{h}^{[2]}) \;,
\quad
\bar{\mathcal{F}}\bar{\mathcal{F}}^{[+2]} = C_{\bar{h}} (h h^{[2]}) \,(\bar{h} \bar{h}^{[2]})\;.
\end{equation}
Since the single change as compared to the massive case (\ref{FHsys}) is in the constant $C_\circ$, it is straightforward to check that $\bar{h}=e^{i \theta/2} h$ and the crossing is only slightly modified 
 \begin{equation}\label{mlesscrossing}
    \left[\dfrac{(H^{+})^{\bar{\gamma}_{+}}}{H^{-}}\right]^{\bar{\gamma}_-}= \frac{C_\circ}{ {\cal B}^+{\cal B}^{\gamma+} {\cal B}^-{\cal B}^{\gamma-}}
  \frac{H^-}{H^+}~~~,~~H\equiv h^{[+2]}h\;.
\end{equation}
Denoting our newly discovered phase in the massive limit \eq{eq:HDef} as $\hat{\sigma}_{\bullet}$, we see that the total phase $\hat{\sigma}$ receives a massless contribution defined by a very simple crossing relation
\begin{equation}
    \hat{\sigma}=\hat{\sigma}_{\bullet}\hat{\sigma}_{\circ}\,,
    \quad (\hat{\sigma}_\circ)^{\bar{\gamma}_c}\hat{\sigma}_{\circ}=\left(\prod_{k=1}^{N}z^2_k\right)^{1/4}\,.
\end{equation}
This result might look suspicious at first, but it has some justification. Based on experience with other QSC's functions $f_{\circ}$ can be viewed as a massless limit of $f$, i.e limit $x^{\pm}_k\to z^{\pm1}_k$. Notably, this transformation sends ratio of polynomials $\mathbb{Q}^{+}/\mathbb{Q}^{-}$ to constant. So the r.h.s of \eq{mless2} should become constant in the massless limit, which is exactly what we find.

\subsection{Restricting to the Symmetric Sector}
As we explained in the main text, it appears that our proposal faces various additional difficulties away from the symmetric sector. To avoid dealing with these problems, as well as other technical difficulties, we will henceforth assume that we are in the symmetric sector, see Section \ref{sec:SymmetricSector} for details. An important simplification is that the functions $h$ and $\mathcal{F}$ drop out, as can be seen from setting dotted and non-dotted Q-functions equal in \eqref{mless1q}.

Let us summarise the most important expressions in the symmetric sector, the $Q_{a|i}$ functions are
\begin{equation}\label{mlessqsplitsymsec}
 Q_{1|1}\propto \mathbb{Q}_{1|1}\sqrt{f^{+}f^{+}_\circ}~~,~~Q_{\bar{1}|\bar{1}}\propto \mathbb{Q}_{\bar{1}|\bar{1}}\sqrt{f^{+}f^{+}_\circ}\;,
\end{equation}
while $\nu$ and $\tau$ are given as
\begin{equation}
    \nu_{1}^{~\bar{2}}\propto\mathbb{Q}^-_{1|1}\mathbb{Q}^{-}_{\bar{1}|\bar{1}}\sqrt{f^{*--}ff_\circ^{*--}f_\circ}~~,~~\tau^{1}_{~\bar{2}}\propto\sqrt{\dfrac{f^{*--}f^{*--}_\circ}{ff_\circ}}\;.
\end{equation}
We notice that $\sqrt{f_{\circ}}$ in general has a fourth-order branch-cut. However, no such cut should be present in $\nu$, and we are led to deduce that all massless roots are four times degenerate. A similar multiplicity of massless roots occurred in \cite{Ekhammar:2024kzp}. Let us introduce $K_{\circ}=N/4$, and write
\begin{equation}  \kappa=\varkappa^4~,~\varkappa=\prod_{k=1}^{K_{\circ}}(x-z_k)\;.
\end{equation}

\subsection{Finding $\bP$ and $\bQ$}
To find $\bP$ and $\bQ$, we start from the following general ansatz, motivated by our massive analysis
\begin{align}\label{mlesspqaba1}
    \bP_1\propto x^{-L}R_{\tilde{1}}B_{\tilde{\bar{1}}}S~~,~~\bQ_1\propto x^{L+K_{\circ}}R_1B_{\bar{1}}\mathcal{T}\dfrac{\mathbf{B}_{\bullet(-)}}{\mathbf{B}_{\bar{\bullet}(+)}}\dfrac{ff_\circ}{S \varkappa^2}\,,\\
\label{mlesspqaba2}
    \bP_{\bar{1}}\propto x^{-L}R_{\bar{1}}B_{1}\bar{S}~~,~~\bQ^{\downarrow}_{\bar{1}}\propto x^{L+K_{\circ}}R_{\tilde{\bar{1}}}B_{\tilde{1}}\bar{\mathcal{T}}\dfrac{\mathbf{B}_{\bar{\bullet}(-)}}{\mathbf{B}_{\bullet(+)}}\dfrac{ff_\circ}{\bar{S}\varkappa^2},
\end{align}
Then the $QQ$-relation \eq{eq:bQbP1QQ} for barred and unbarred systems becomes

\begin{equation}\label{mlessqq3}
    x^{K_{\circ}}B_{\tilde{1}}R_{\tilde{\bar{1}}}B_1R_{\bar{1}}\bar{\mathcal{T}}=\mathbf{R}_{\bar{\bullet}(+)}\mathbf{B}_{\bullet(-)}\varkappa^{2*}-\mathbf{R}_{\bar{\bullet}(-)}\mathbf{B}_{\bullet(+)}\varkappa^2,
\end{equation}
and
\begin{equation}
    x^{K_{\circ}}R_{\tilde{1}}B_{\tilde{\bar{1}}}R_1B_{\bar{1}}\mathcal{T}=\mathbf{R}_{\bullet(+)}\mathbf{B}_{\bar{\bullet}(-)}\varkappa^{2*}-\mathbf{R}_{\bullet(-)}\mathbf{B}_{\bar{\bullet}(+)}\varkappa^2.
\end{equation}
As in the massive case, these equations are related by analytic continuation, implying that 
\begin{equation}\label{mlessTbT}
    \mathcal{T}(x) = \bar{\mathcal{T}}\left(\frac{1}{x} \right)
\end{equation}
The functions $\mathcal{T}$ and $\bar{\mathcal{T}}$ are rational functions with no zeros or poles for $|x|>1$, and hence also for $|x|<1$ using \eq{mlessTbT}, and $|x|=1$ by \eq{mlessqq3}. We then conclude that $\mathcal{T}$ is some power of $x$
\begin{equation}
    \mathcal{T}=x^{K_{\mathcal{T}}}~~~,~~~\bar{\mathcal{T}}=x^{K_{\bar{\mathcal{T}}}}=\frac{1}{x^{K_{\mathcal{T}}}}
\end{equation}
With similar arguments as in the AdS$_3\times $S$^3\times $T$^4$ case, it can be demonstrated that $K_{\mathcal{T}}=0$. To fix $S$ and $\bar{S}$ we consider \eqref{pmu1aba} and substitute all the large volume expressions for all the $Q$-functions involved, namely \eq{mlessom},\eq{mless1q}, \eq{mlesspqaba1} and \eq{mlesspqaba2}. As a result
\begin{equation}\label{SSmassless}
    S^\gamma\bar{S}\propto\dfrac{\mathbf{R}_{\bullet(+)}\mathbf{B}_{\bar{\bullet}(-)}}{x^{-K_{\circ}}\varkappa^2}f^{[+2]}f^{*,[-2]}f_\circ^{[+2]}f_\circ^{*,[-2]}\;.
\end{equation}
The same equations also hold by "applying" bar. As before, we split $S$ and $\bar{S}$ into a few constituents that will later become the various dressing phases. To simplify the coming expressions we parameterize $S$ and $\bar{S}$ in analogy with the expression from \cite{Ekhammar:2024kzp}, this gives
\begin{equation}
    S = \sqrt{\frac{\mathbf{B}_{\bullet(+)}\mathbf{B}_{\bullet(-)}}{x^{-4K_{\circ}}(\varkappa\varkappa^{*})^2}} \sigma_{\bullet} \sigma_{\circ} \rho_\bullet\rho_{\circ}\,,
    \quad
    \bar{S} = \sqrt{\frac{\mathbf{B}_{\bar{\bullet},(+)}\mathbf{B}_{\bar{\bullet},(-)}}{x^{-4K_{\circ}}(\varkappa\varkappa^{*})^2}} \sigma_{\bullet} \sigma_{\circ} \bar{\rho}_\bullet \rho_{\circ}\,. 
\end{equation}
Here $\sigma_{\bullet},~\rho_\bullet\equiv\rho$ are the same dressing phases defined from our analysis of the massive case, see \eq{rhophase}. In order to satisfy \eqref{SSmassless} the "massless" pieces $\sigma_\circ$ and $\rho_{\circ}$ are defined as
\begin{equation}
    \sigma^{\gamma}_{\circ}\sigma_{\circ} \propto f^{++}_\circ f^{*--}_\circ\,,
    \quad \rho_{\circ}^{\gamma}\rho_{\circ}\propto\dfrac{(\varkappa^{*})^2}{\prod_{k=1}^{K_{\circ}}x/z_k}\;.
\end{equation}
These factors have been studied before \cite{Ekhammar:2024kzp} and the resulting phases are also connected to Sine-Gordon dressing phases \cite{Torrielli:2025lgu}.

Upon plugging in the explicit expressions for $S,\bar{S}$, the final asymptotic expressions for $\bP,\bQ$ takes the form
\begin{align}\label{masslessPQ1}
    \bP_1\propto x^{-L}R_{\tilde{1}}B_{\tilde{\bar{1}}}\sqrt{\frac{\mathbf{B}_{\bullet(+)}\mathbf{B}_{\bullet(-)}}{x^{-4K_{\circ}}(\varkappa\varkappa^{*})^2}} \sigma_{\bullet} \sigma_{\circ} \rho_\bullet\rho_\circ\,,
    \quad
    \bQ_1\propto x^{L-K_{\circ}}R_1B_{\bar{1}}\sqrt{\dfrac{\mathbf{B}_{\bullet(-)}}{\mathbf{B}_{\bullet(+)}}}\dfrac{f f_{\circ}}{\mathbf{B}_{\bar{\bullet}(+)}}\dfrac{1}{\sigma_{\bullet} \sigma_{\circ} \rho_\bullet\rho_\circ}\dfrac{\varkappa^{*}}{\varkappa}\;, \\
\label{masslessPQ2}
    \bP_{\bar{1}}\propto x^{-L}R_{\bar{1}}B_{1}\sqrt{\frac{\mathbf{B}_{\bar{\bullet},(+)}\mathbf{B}_{\bar{\bullet},(-)}}{x^{-4K_{\circ}}(\varkappa\varkappa^{*})^2}} \sigma_{\bullet} \sigma_{\circ} \bar{\rho}_\bullet \rho_\circ\,,
    \quad
    \bQ_{\bar{1}}\propto x^{L-K_{\circ}}R_{\tilde{\bar{1}}}B_{\tilde{1}}\sqrt{\dfrac{\mathbf{B}_{\bar{\bullet}(-)}}{\mathbf{B}_{\bar{\bullet}(+)}}}\dfrac{ff_{\circ}}{\mathbf{B}_{\bullet(+)}}\dfrac{1}{\sigma_{\bullet} \sigma_{\circ} \bar{\rho}_\bullet\rho_\circ}\dfrac{\varkappa^{*}}{\varkappa}\;.
\end{align}

\subsection{Symmetric Sector Bethe Equations with Massless Modes}
We can derive the massive ABA equations following the analysis presented in the main part of the paper. To be precise, we use \eqref{ABAGrading1} and \eqref{BetheEqnGrading2} using the  $Q$-functions \eqref{masslessPQ1} and  \eqref{masslessPQ2} as well as \eqref{mlessqsplitsymsec}. 

Using \eqref{ABAGrading1}] we find for the first copy of $\algosp_{4|2}$ $Q$-system 
\begin{align}\label{mlessaba_1}
        -\left(\dfrac{x^+_{1|1,k}}{x^-_{1|1,k}}\right)^{L-K_{\circ}}&=\dfrac{\mathbb{Q}_{1|1}^{[+2]}}{\mathbb{Q}_{1|1}^{[-2]}}\,\dfrac{R_1^-B_{\bar{1}}^-}{R_1^+B_{\bar{1}}^+}\,\sqrt{\dfrac{\bBulletBarP^{+}\bBulletBarM^{+}}{\bBulletBarP^{-}\bBulletBarM^{-}}}\,\frac{\sigma^+_{\bullet}\rho^+_{\bullet}\sigma^+_{\circ}\rho^+_{\circ}}{\sigma^-_{\bullet}\rho^-_{\bullet}\sigma^-_{\circ}\rho^-_{\circ} }\frac{H^+}{H^-}\Bigg|_{x=x_{1|1,k}}\,,\\
        \label{mlessaba_2}
        1&=\dfrac{\rBulletP \bBulletBarM}{\rBulletM \bBulletBarP}\(\dfrac{\varkappa^{*}}{\varkappa}\)^2\Biggl|_{x=y_{1,k}}\;.
\end{align}
For the second copy, we use \eqref{BetheEqnGrading2} to obtain
\begin{align}\label{mlessabab_1}
    -\left(\dfrac{x_{\bar{1}|\bar{1},k}^+}{x_{\bar{1}|\bar{1},k}^-}\right)^{L-2K_{\circ}}&=\dfrac{R^-_{\dot{\bar{1}}|\bar{1},(-)}}{R^+_{\dot{\bar{1}}|\bar{1},(+)}}\dfrac{B^+_{\bar{1}|\bar{1},(+)}}{B^-_{\bar{1}|\bar{1},(-)}}\,\dfrac{B_1^+R_{\bar{1}}^+}{B_1^-R_{\bar{1}}^-}\,\sqrt{\dfrac{\bBulletP^+\bBulletP^-}{\bBulletM^+ \bBulletM^-}}\frac{\sigma^+_{\bullet}\bar{\rho}_{\bullet}^+\sigma^+_{\circ}\rho^+_{\circ}}{\sigma^-_{\bullet}\bar{\rho}_{\bullet}^-\sigma^-_{\circ}\rho^-_{\circ}}\(\dfrac{\varkappa^{-}}{\varkappa^{*+}}\)^2\Bigg|_{x=x_{\bar{1}|\bar{1},k}}\\\label{mlessabab_2}
    1&=\dfrac{\rBulletBarP\bBulletM}{\rBulletBarM\bBulletP}\(\dfrac{\varkappa^{*}}{\varkappa}\)^2\Biggl|_{x=y_{\bar{1},k}}\;.
\end{align}
These results are very similar to those in the main section. The most novel part is to find the massless middle node equation. We start by considering \eq{pmu1} for both the barred and unbarred system:
\begin{equation}\label{mlessaba}
 \bP_{1}=\nu_1^{~\bar{f}}\nu_{\dot{1}}^{~\dot{\bar{e}}}\bP_{\dot{\bar{e}}\bar{f}}^{\bar{\gamma}}~~,~~\bP_{\bar{1}}=\nu_{\bar{1}}^{~f}\nu_{\dot{\bar{1}}}^{~\dot{e}}\bP_{\dot{e}f}^{\bar{\gamma}}\;.
\end{equation}
Both sums here have four terms even in the large volume limit. However, in the symmetric sector $\nu_{1}^{~\bar{2}}$ and $\nu_{\dot{1}}^{~\dot{\bar{2}}}$ are identified, and will share the same sets of roots. The relations above then simplify significantly when evaluated at points $z_k$
\begin{equation}\label{mlessABA_1}
    \bP_{1}=(\nu_1^{~\bar{1}})^2\bP_{\bar{1}}^{\bar{\gamma}}~~,~~\bP_{\bar{1}}=(\nu_{\bar{1}}^{~1})^2\bP_{1}^{\bar{\gamma}}~~~,~~~x=z_k\;,
\end{equation}
Where all the dotted quantities have been eliminated using (\ref{nusymsec}). From the definition of $\nu$, see \eq{nudef}, it follows that in the ABA-limit we find
\begin{equation}
    \nu_1^{~\bar{1}}\propto Q_{1|1}^{-}Q_{\bar{2}|\bar{1}}^{-}\tau^{1}_{~\bar{2}}~~,~~\nu_{\bar{2}}^{~2}\propto Q_{\bar{2}|\bar{1}}^{-}Q_{1|1}^{-}\tau^{\bar{1}}_{~2}\;,
\end{equation}
and using the definition of $\tau$ \eq{taudef} we obtain
\begin{equation}
    \dfrac{\nu_1^{~\bar{1}}}{\nu_{\bar{2}}^{~2}}=\dfrac{\tau^{1}_{~\bar{2}}}{\tau^{\bar{1}}_{~2}}=\dfrac{F^1_{~\bar{2}}}{F^{\bar{1}}_{~2}}\dfrac{\Omega^{\bar{2}}_{~\bar{2}}}{\Omega^{2}_{~2}}=\dfrac{\bar{\alpha}}{\alpha}\equiv\zeta
\end{equation}
Here $\alpha$ and $\bar{\alpha}$ are off-diagonal components of constant matrices $F^k_{~\bar{m}}$ and $F^{\bar{p}}_{~q}$, and the ratio $\Omega^{\bar{2}}{}_{\bar{2}}/\Omega^2{}_{2}$ is equal to one in the ABA limit. Since the determinant of $\nu$ is equal to unity
\begin{equation}
\pm1=\nu_{\bar{1}}^{~1}\nu_{\bar{2}}^{~2}\Bigr|_{x=z_k}
\end{equation}
Substituting this into the ratio of $\nu$ functions above results in 
\begin{equation}
\pm\zeta=\nu_{\bar{1}}^{~1}\nu_{1}^{~\bar{1}}\Bigr|_{x=z_k}
\end{equation}
Taking the product of the two relations in \eq{mlessABA_1}, we can now get rid of all the $\nu$ factors
\begin{equation}\label{mlessaba1}
    \pm\zeta^{2}=\dfrac{\bP_{1}\bP_{\bar{1}}}{\bP_{1}^{\bar{\gamma}}\bP_{\bar{1}}^{\bar{\gamma}}}\Bigr|_{x=z_k}
\end{equation}
Finally, substituting \eq{masslessPQ2} here we find
\begin{equation}
    \pm\zeta^2z_k^{4\left(L-K_{\circ}\right)}=\left(\dfrac{\sigma_{\bullet}\sigma_{\circ}\rho_{\circ}}{\sigma_{\bullet}^{\bar{\gamma}}\sigma_{\circ}^{\bar{\gamma}}\rho_{\circ}^{\bar{\gamma}}}\right)^{2}\dfrac{\rho_{\bullet}\bar{\rho}_{\bullet}}{\rho_{\bullet}^{\bar{\gamma}}\bar{\rho}_{\bullet}^{\bar{\gamma}}}\sqrt{\left(\prod_{k=1}^{M}x^+_{k}x^-_{k}\right)\dfrac{\bfB_{(+)}\bfB_{(-)}}{\bfR_{(+)}\bfR_{(-)}}}\times\dfrac{R_{\tilde{1}}B_{\tilde{\bar{1}}}R_{\bar{1}}B_{1}}{B_{\tilde{1}}R_{\tilde{\bar{1}}}B_{\bar{1}}R_{1}}
\end{equation}
This is our final result of this appendix and constitutes our proposal for a massless Bethe equation. 

\section{Dressing Phases}\label{app:HAppendix}
In this appendix, we provide additional details regarding the new novel dressing phase $\hat{\sigma}(u,v)$ defined in \eqref{eq:HDef}. To find this phase, we must address the functional equations \eqref{FHsys} that control $\mathcal{F}$ and $h$. We recall those equations here, and slightly generalise them by writing
\begin{equation}\label{eq:FHsysApp}
    \mathcal{F}^{\gamma} = e^{\ii \mathcal{N}} \mathcal{F}^{[2]}\mcS\,,
    \quad
    \mcF \mcF^{[2]} = e^{i \theta}H^2\,,
    \quad
    \mathcal{F}^+ = e^{i \theta/2}h^+ (h^*)^-\,.
\end{equation}
where
\begin{equation}
   \mcS = \sqrt{\frac{
{\mathbb Q}_{1|1}^+
{\mathbb Q}_{\bar 1|\bar 1}^+
{\mathbb Q}_{\dot 1|1}^-
{\mathbb Q}_{\bar {\dot 1}|\bar 1}^-
}{
{\mathbb Q}_{1|1}^-
{\mathbb Q}_{\bar 1|\bar 1}^-
{\mathbb Q}_{\dot 1|1}^+
{\mathbb Q}_{\bar {\dot 1}|\bar 1}^+
}}\,,
    \quad
    H\equiv h h^{[2]}=(h h^{[2]})^*\,,
\end{equation}
and we perfectly recover \eqref{FHsys} upon setting $e^{\ii \mathcal{N}}=\pm 1$. As we will show in this appendix, and already discussed in Section~\ref{sec:UnitarityIssues}, the newly introduced $\mathcal{N}$ plays an important part in understanding the unitarity of the dressing phase $\hat{\sigma}$. We recall that this phase is related to the ratio $h^{[3]}/h^-=H^+/H^-$ that appears in the ABA equations, see Section~\ref{subsec:ABAEquations}.

\paragraph{Solving the system.} 
Combining the last two equations of \eqref{eq:FHsysApp} reveals that 
\begin{equation}
    H = hh^{[2]}=h^{*}h^{*[-2]} = H^{\star}\;.
\end{equation}
It then follows that $H$ is real and has only a single cut on the real line, since $h$ is UHPA and so $h^{*}$ must be LHPA. In terms of $H$, the first two equations of \eqref{eq:FHsysApp} becomes 
\begin{equation}\label{FHH}
\mathcal{F}^{\gamma}\mathcal{F}=e^{\ii \mathcal{N}}e^{i \theta}H^2\mathcal{B}\mathcal{B}^{\gamma}\,,
\quad 
\mathcal{F}\mathcal{F}^{[+2]}=e^{i \theta}H^2\,.
\end{equation}
It is convenient to express $h,h^{*}$ and $\mathcal{F}$  in terms of $H$. From $H=hh^{[+2]}$ we can express $h$ as a formal product
\begin{equation}
    h=\prod_{n=0}^\infty \left(H^{[+2n]}\right)^{(-1)^n}\;\;,\;\;
h^*=\prod_{n=0}^\infty \left(H^{[-2n]}\right)^{(-1)^n}\;.
\end{equation}
or equivalently
\begin{equation}
    h = \prod_{n=0}^{\infty} \frac{H^{[4n]}}{H^{[4n+2]}}\,,
    \quad
    h^{\star} = \prod_{n=0}^{\infty} \frac{H^{[-4n]}}{H^{[-4n-2]}}\,.
\end{equation}
Then, using the last relation in \eqref{eq:FHsysApp} gives
\begin{equation}\label{eq:FInfProd}
    \mathcal{F}=e^{i \theta/2}H\prod_{n=0}^\infty \left(\frac{H^{[-2n-2]}}{H^{[+2n+2]}}\right)^{(-1)^n}\;.
\end{equation}
Since $H$ only has a single cut on a real line, the above expression implies
\begin{equation}\label{eq:FNoCut}
    \dfrac{\mathcal{F}}{H}=\dfrac{\mathcal{F}^{\gamma}}{H^{\gamma}}\;.
\end{equation}
Eliminating  $\mathcal{F}^{\gamma}$  using (\ref{FHH}) results in
\begin{equation}\label{Hgamma}
    \dfrac{H^{\gamma}}{H}= e^{\ii \mathcal{N}}e^{i \theta}\dfrac{H^2\mcS}{\mathcal{F}^2}~~~,~~~\dfrac{H^{\bar{\gamma}}}{H}=e^{-\ii \mathcal{N}}e^{-i \theta}\dfrac{\mathcal{F}^2}{H^2\mcS}\;,
\end{equation}
where the second equation is acquired by analytically continuing the first one along $\bar{\gamma}$. Notably, the terms on the r.h.s of these expressions have no cuts on the real line. We can clearly see that $H^{\gamma}\neq H^{\bar{\gamma}}$, so our proposed splitting for $Q_{a|i}$ and $\nu$ will have non-quadratic cuts even in the ABA limit. This is a novelty compared to known examples of the QSC.

\paragraph{Deriving the crossing equation.}
We would like to derive a crossing equation for $h^{[+3]}/h^{-}$, which in terms of $H$ means
\begin{equation}
    \dfrac{h^{[+3]}}{h^{-}}=\dfrac{H^{+}}{H^{-}}\;,
\end{equation}
Our crossing contour is taken to be as depicted in Figure~\ref{fig:full-crossing}. We derive crossing in two steps in accordance with two parts of the contour $\bar{\gamma}_c$. First, taking an analytic continuation of the above along $\bar{\gamma}_+$ results in
\begin{equation}
    \dfrac{(H^{+})^{\bar{\gamma}_{+}}}{H^{-}}=e^{-\ii \mathcal{N}}e^{-i \theta}
  \frac{({\cal F}^{+})^2}{H^+H^- \mcS^{+}}\;,
\end{equation}
where the rightmost equation of (\ref{Hgamma}) has been used. To apply $\bar{\gamma}_-$ we slightly rearrange the terms in the above relation and use the fact that $\mathcal{F}^{[2]}/H = e^{i \theta} H/\mathcal{F}$ and hence does not have a cut due to \eqref{eq:FNoCut}. Using this fact, we have
\begin{equation}
\left[\dfrac{(H^{+})^{\bar{\gamma}_{+}}}{H^{-}}\right]^{\bar{\gamma}_-}=e^{-i \theta}\left[\frac{H^-({\cal F}^{+}/ H^-)^2}{H^+ \mcS^+}\right]^{\bar\gamma_-}=e^{-i \theta}e^{-\ii\mathcal{N}}\left[H^{-}\right]^{\bar\gamma_-}\frac{{(\cal F}^{+})^2}{(H^-)^2 H^+ \mcS^+}\;,
\end{equation}
whereupon we can once again use \eqref{Hgamma}. After some simplifications using the rightmost equation in (\ref{FHH}), we arrive to
\begin{equation}
    \left[\dfrac{(H^{+})^{\bar{\gamma}_{+}}}{H^{-}}\right]^{\bar{\gamma}_-}= e^{-2\ii \mathcal{N}}\frac{1}{\mcS^+\mcS^-}
  \frac{H^-}{H^+}\;.
\end{equation}
As a result, the additional phase factor $\hat{\sigma}\equiv H^{+}/H^{-}$, which will appear in BAE, satisfies the following crossing
\begin{equation}
    (\hat{\sigma})^{\bar{\gamma}_c}\hat{\sigma}=e^{-2\ii\mathcal{N}}\sqrt{\dfrac{\mathbb{Q}_{1|1}^{[-2]}\mathbb{Q}_{\bar{1}|\bar{1}}^{[-2]}}{\mathbb{Q}_{1|1}^{[+2]}\mathbb{Q}_{\bar{1}|\bar{1}}^{[+2]}}\cdot\dfrac{\mathbb{Q}_{\dot{1}|1}^{[+2]}\mathbb{Q}_{\dot{\bar{1}}|\bar{1}}^{[+2]}}{\mathbb{Q}_{\dot{1}|1}^{[-2]}\mathbb{Q}_{\dot{\bar{1}}|\bar{1}}^{[-2]}}}\;,
\end{equation}
where notation in terms of functions $\mathbb{Q}$ has been restored recalling (\ref{FF1}). We see that for $e^{\ii \mathcal{N}}=\pm 1$ we find \eqref{hatcrossing}.

\subsection{Finding $H$ Explicitly.}
In this subsection, we obtain $H$ as a perturbative series in $g$. Our starting point is \eqref{Hgamma}, which we rewrite using \eqref{eq:FInfProd} as
\begin{equation}
    H^{\gamma} = e^{\ii \mathcal{N}} H \mathcal{S}\big/ \left(\prod_{n=0}^{\infty} \left(H^{[-2n-2]}/H^{[2n+2]} \right)^{(-1)^n}\right)^{2}\,.
\end{equation}
In order to solve for $H$ we take the logarithm of both sides
\beq\label{halfcrosH}
\log{H^{\gamma}}-\log H=
\log \mathcal{S}+\ii \mathcal{N}
-\sum_{n=0}^\infty 2(-1)^n \log\left(\frac{H^{[-2n-2]}}{H^{[+2n+2]}}\right)
\eeq
In this form, it is clear that the equation is, in fact, a linear equation for $\log H$. Let us define an elementary building block ${\cal H}(x,y)$ such that
\begin{equation}\label{Hsplit}
    \log H(x)=\sum_{{\scriptscriptstyle\alpha\in\{1|1\,,\bar{1}|\bar{1}\}}}\sum_{j=1}^{K_{\alpha}} {\cal H}(x,x_{\alpha,j})-\sum_{{\scriptscriptstyle\beta\in\{\dot{1}|1\,,\dot{\bar{1}}|\bar{1}\}}}\sum_{k=1}^{K_{\beta}} {\cal H}(x,x_{\beta,k})\;,
\end{equation}
which reduces  (\ref{halfcrosH}) to
\beq\label{halfcrosHxy}
{{\cal H}^{\gamma}}-{\cal H}=
\log\(\sqrt{\frac{x-y^-}{x-y^+}\frac{1-\frac{1}{xy^-}}{1-\frac{1}{xy^+}}}\)
-\sum_{n=0}^\infty
2(-1)^n
\[{\cal H}^{[-2n-2]}-{\cal H}^{[+2n+2]}\]+f(u)\alpha(y^\pm)\;.~~~~~~~
\eeq
Where we have introduced an arbitrary function $\alpha(y)$, constrained by \eqref{halfcrosH} to satisfy
\begin{equation}\label{alrestrict}
    \sum_{{\scriptscriptstyle\alpha\in\{1|1\,,\bar{1}|\bar{1}\}}}\sum_{j=1}^{K_{\alpha}} \alpha(x_{\alpha,j})-\sum_{{\scriptscriptstyle\beta\in\{\dot{1}|1\,,\dot{\bar{1}}|\bar{1}\}}}\sum_{k=1}^{K_{\beta}} \alpha(x_{\beta,k})=\ii \mcN\;.
\end{equation}
We now describe how to perturbatively solve for ${\cal H}$. The main idea is to write an ansatz
\begin{equation}
    \mathcal{H} = \mathcal{H}^{(0)} + \mathcal{H}^{(1)} + \mathcal{H}^{(2)} + \dots 
\end{equation}
where 
\begin{equation}
    \left(\mathcal{H}^{(p)}\right)^{\gamma} - \mathcal{H}^{(p)} = \mathcal{O}(g^{p})\,,
    \quad
    \mathcal{H}^{(p)} = \mathcal{O}(g^{p+1})\,. 
\end{equation}
which will allow us to solve \eqref{halfcrosHxy} order by order in $g$.

We start from $\mathcal{O}(g^0)$, then we can neglect the summation term in the r.h.s. of \eqref{halfcrosHxy}, that is
\beq
\({{\cal H}^{(0)}}\)^\gamma-{\cal H}^{(0)}=
\log\(\sqrt{\frac{x-y^-}{x-y^+}}\sqrt{\frac{1-\frac{1}{xy^-}}{1-\frac{1}{xy^+}}}\)\;.
\eeq
Since $\mathcal{H}^{(0)}$ has only a single cut on the real line and is bounded at infinity, the above Riemann-Hilbert problem has the following unique solution
\begin{equation}
    \mathcal{H}^{(0)}=\int_{-2g}^{2g}\dfrac{dw}{4\pi i}\dfrac{1}{w-u}\log\(\dfrac{w-v+i/2}{w-v-i/2}\)\;,
\end{equation}
where $u$ and $v$ are defined via Zhukovsky map (\ref{P_laurent}) as $x=x(u)$ and $y=x(v)$. The solution expands at weak coupling as
\begin{equation}
\begin{split}
{\cal H}^{(0)}=&\dfrac{ig}{\pi u}\log\(\dfrac{2v-i}{2v+i}\)\\
&+\quad\dfrac{g^3}{3\pi u^2}\(\dfrac{4i}{u}\log\(\dfrac{2v-i}{2v+i}\)+\dfrac{16(1+4v(u+v))}{(1+4v^2)^2}\)+\dots
\end{split}
\end{equation}
We then proceed to the next order in \eqref{halfcrosHxy}, which becomes
\begin{equation}
    [{{\cal H}^{(1)}]^{\gamma}}-{\cal H}^{(1)}=
-\sum_{n=0}^\infty
2(-1)^n
\[{\cal H}^{(0)[-2n-2]}-{\cal H}^{(0)[+2n+2]}\]\;,
\end{equation}
which we can solve with the same method as before
\begin{equation}
    {\cal H}^{(1)}(u)=-\sum_{n=0}^\infty
2(-1)^n\int_{-2g}^{2g}\dfrac{dw}{2\pi i}\dfrac{1}{w-u}\[{\cal H}^{(0)[-2n-2]}(w)-{\cal H}^{(0)[+2n+2]}(w)\]\,.
\end{equation}
We repeat this procedure several times to calculate higher order corrections $\mathcal{H}^{(k)}$ to reach the desired accuracy in the coupling constant $g$, resulting in 
\begin{equation}
\begin{split}
\mathcal{H} &= g\frac{i\log \left(\frac{2 v-i}{2 v+i}\right)}{\pi 
   u}+g^2\frac{8 i
   \log (2) \log \left(\frac{2 v-i}{2 v+i}\right)}{\pi ^2
   u}\\
&+g^3 \left(\frac{16 \left(4 u v+4
   v^2+1\right)}{3 \pi  \left(4 u v^2+u\right)^2}+\frac{4 i
   \left(48 u^2 \log ^2(2)+\pi ^2\right) \log \left(\frac{2
   v-i}{2 v+i}\right)}{3 \pi ^3 u^3}\right)\\
&+g^4 \left(\frac{512 v \log (2)}{3 u \left(4 \pi  v^2+\pi
   \right)^2}+\frac{16 i \log \left(\frac{2 v-i}{2
   v+i}\right) \left(u^2 \left(96 \log ^3(2)-3 \pi ^2 \zeta
   (3)\right)+\pi ^2 \log (4)\right)}{3 \pi ^4
   u^3}\right)+\dots\,.
\end{split}
\end{equation}
Finally, let us attempt to fix the arbitrary function $\alpha$ by demanding that $\hat{\sigma}$ is unitary. Including $\alpha$ we find
\begin{equation}\label{Hweak}
    {\cal H}(u,v) =\sum_{k}{\cal{H}}^{(k)}(u,v)+\alpha(v)\beta(u)\;.
\end{equation}
where $\beta$ must satisfy
\begin{equation}
    {\beta^{\gamma}}-\beta=
1
-\sum_{n=0}^\infty
2(-1)^n
\[\beta^{[-2n-2]}-\beta^{[+2n+2]}\]\,.
\end{equation}
This equation can be solved in the same manner as before. The weak coupling result is
\begin{equation}
\begin{split}
    \beta &= \frac{2 i}{\pi
    u}g+\frac{16 i \log (2)}{\pi ^2 u}g^2+\frac{8
   i g^3 \left(48 u^2 \log ^2(2)+\pi ^2\right)}{3 \pi ^3
   u^3}\\
    &\quad \quad \quad +
    \frac{32 i g^4 \left(u^2 \left(96 \log ^3(2)-3 \pi ^2 \zeta
   (3)\right)+\pi ^2 \log (4)\right)}{3 \pi ^4 u^3}\,.
\end{split}
\end{equation}
Requiring that the dressing phase $\hat{\sigma}=H^{+}/H^{-}$ is unitary, i.e
\begin{equation}
    \hat{\sigma}(u,v)\hat{\sigma}(v,u)=1\;,
\end{equation}
we can determine $\alpha$ by evaluating at $u=v$ to find
 \begin{equation}
     \alpha(v)=-\dfrac{{\cal H}(v+\frac{i}{2},v)-{\cal H}(v-\frac{i}{2},v)}{\beta(v+\frac{i}{2},v)-\beta(v-\frac{i}{2},v)}\;.
\end{equation}
While we cannot prove at the moment that this indeed guarantees unitarity to all orders, we do observe that it indeed holds to a high loop order. The weak coupling expansion for $\alpha$ is then
\begin{equation}\label{eq:alphaDef}
    \alpha(v)=-\frac{1}{2} \log \left(\frac{2 v-i}{2
   v+i}\right)+\frac{32 i v}{\left(4
   v^2+1\right)^2}g^2-\frac{512 i v \log (2)}{3 \pi  \left(4
   v^2+1\right)^2}g^3+\mathcal{O}(g^4)\,.
\end{equation}
We find the weak coupling expansion for a unitary solution $\mathcal{H}^{u}$ is
\begin{equation}
\begin{split}
    \mathcal{H}^{u} &=\frac{16 g^3 \left(-8 u v+4 v^2+1\right)}{3 \pi 
   \left(4 u v^2+u\right)^2} \\
   &-\frac{64 g^5 \left(192 u^3 v \left(4
   v^2-1\right)-4 u^2 \left(48 v^4+8 v^2-1\right)+8 u v \left(4
   v^2+1\right)^2-3 \left(4 v^2+1\right)^3\right)}{15 \pi  \left(4 u
   v^2+u\right)^4}\\
   &-\frac{256 g^6 \zeta (3) \left(8 u v-4 v^2-1\right)}{3 \pi ^2 u^2
   \left(4 v^2+1\right)^2}\,.
\end{split}
\end{equation}

\bibliographystyle{JHEP}
\bibliography{refs}

\makeatletter \@ifundefined{Sphere}{\newcommand{\Sphere}{{S}{}}}{} \@ifundefined{AdS}{\newcommand{\AdS}{{AdS}{}}}{} \@ifundefined{CFT}{\newcommand{\CFT}{{CFT}{}}}{} \@ifundefined{CP}{\newcommand{\CP}{CP}}{} \@ifundefined{Torus}{\newcommand{\Torus}{{T}{}}}{} \@ifundefined{superN}{\newcommand{\superN}{\mathcal{N}}}{} \@ifundefined{grpOSp}{\newcommand{\grpOSp}{\text{OSp}}}{} \@ifundefined{grpPSU}{\newcommand{\grpPSU}{\text{PSU}}}{} \@ifundefined{grpSU}{\newcommand{\grpSU}{\text{SU}}}{} \@ifundefined{grpU}{\newcommand{\grpU}{\text{U}}}{} \@ifundefined{grpD}{\newcommand{\grpD}{\text{D}}}{} \@ifundefined{grpSL}{\newcommand{\grpSL}{\text{SL}}}{} \@ifundefined{grpSp}{\newcommand{\grpSp}{\text{Sp}}}{} \@ifundefined{grpUSp}{\newcommand{\grpUSp}{\text{USp}}}{} \@ifundefined{grpSO}{\newcommand{\grpSO}{\text{SO}}}{} \@ifundefined{grpO}{\newcommand{\grpO}{\text{O}}}{} \@ifundefined{algOSp}{\newcommand{\algOSp}{\text{osp}}}{} \@ifundefined{algPSU}{\newcommand{\algPSU}{\text{psu}}}{}
  \@ifundefined{algSU}{\newcommand{\algSU}{\text{su}}}{} \@ifundefined{algSp}{\newcommand{\algSp}{\text{sp}}}{} \@ifundefined{algSL}{\newcommand{\algSL}{\text{sl}}}{} \@ifundefined{algGL}{\newcommand{\algGL}{\text{gl}}}{} \@ifundefined{algU}{\newcommand{\algU}{\text{u}}}{} \@ifundefined{algSO}{\newcommand{\algSO}{\text{so}}}{} \@ifundefined{algO}{\newcommand{\algO}{\text{o}}}{} \@ifundefined{Integers}{\newcommand{\Integers}{\text{Z}}}{} \@ifundefined{Reals}{\newcommand{\Reals}{\text{R}}}{} \@ifundefined{Complex}{\newcommand{\Complex}{\text{C}}}{} \makeatother

\providecommand{\href}[2]{#2}\begingroup\raggedright\begin{thebibliography}{10}

\bibitem{Maldacena:1997re}
J.~M. Maldacena, \emph{The large {N} limit of superconformal field theories and supergravity}, {\emph{Adv. Theor. Math. Phys.} {\bfseries 2} (1998) 231} [\href{https://arxiv.org/abs/hep-th/9711200}{{\ttfamily hep-th/9711200}}].

\bibitem{Witten:1998qj}
E.~Witten, \emph{Anti-de {S}itter space and holography}, {\emph{Adv. Theor. Math. Phys.} {\bfseries 2} (1998) 253} [\href{https://arxiv.org/abs/hep-th/9802150}{{\ttfamily hep-th/9802150}}].

\bibitem{Gubser:1998bc}
S.~S. Gubser, I.~R. Klebanov and A.~M. Polyakov, \emph{Gauge theory correlators from non-critical string theory}, \href{https://doi.org/10.1016/S0370-2693(98)00377-3}{\emph{Phys. Lett.} {\bfseries B428} (1998) 105} [\href{https://arxiv.org/abs/hep-th/9802109}{{\ttfamily hep-th/9802109}}].

\bibitem{Metsaev:1998it}
R.~R. Metsaev and A.~A. Tseytlin, \emph{Type {IIB} superstring action in {$\AdS_5 \times \Sphere^5$} background}, \href{https://doi.org/10.1016/S0550-3213(98)00570-7}{\emph{Nucl. Phys.} {\bfseries B533} (1998) 109} [\href{https://arxiv.org/abs/hep-th/9805028}{{\ttfamily hep-th/9805028}}].

\bibitem{Minahan:2002ve}
J.~A. Minahan and K.~Zarembo, \emph{The {B}ethe-ansatz for {$\superN = 4$} super {Y}ang-{M}ills}, {\emph{JHEP} {\bfseries 0303} (2003) 013} [\href{https://arxiv.org/abs/hep-th/0212208}{{\ttfamily hep-th/0212208}}].

\bibitem{Beisert:2003jj}
N.~Beisert, \emph{The complete one-loop dilatation operator of {$\superN = 4$} super {Y}ang-{M}ills theory}, \href{https://doi.org/10.1016/j.nuclphysb.2003.10.019}{\emph{Nucl. Phys.} {\bfseries B676} (2004) 3} [\href{https://arxiv.org/abs/hep-th/0307015}{{\ttfamily hep-th/0307015}}].

\bibitem{Beisert:2005fw}
N.~Beisert and M.~Staudacher, \emph{Long-range {$\grpPSU(2,2|4)$} {B}ethe ansaetze for gauge theory and strings}, \href{https://doi.org/10.1016/j.nuclphysb.2005.06.038}{\emph{Nucl. Phys.} {\bfseries B727} (2005) 1} [\href{https://arxiv.org/abs/hep-th/0504190}{{\ttfamily hep-th/0504190}}].

\bibitem{Gromov:2009tv}
N.~Gromov, V.~Kazakov and P.~Vieira, \emph{Integrability for the full spectrum of planar {AdS/CFT}}, \href{https://doi.org/10.1103/PhysRevLett.103.131601}{\emph{Phys. Rev. Lett.} {\bfseries 103} (2009) 131601} [\href{https://arxiv.org/abs/0901.3753}{{\ttfamily 0901.3753}}].

\bibitem{Bombardelli:2009ns}
D.~Bombardelli, D.~Fioravanti and R.~Tateo, \emph{Thermodynamic {B}ethe ansatz for planar {AdS/CFT}: a proposal}, \href{https://doi.org/10.1088/1751-8113/42/37/375401}{\emph{J. Phys.} {\bfseries A42} (2009) 375401} [\href{https://arxiv.org/abs/0902.3930}{{\ttfamily 0902.3930}}].

\bibitem{Arutyunov:2009ur}
G.~Arutyunov and S.~Frolov, \emph{{Thermodynamic Bethe Ansatz for the AdS(5) x S(5) Mirror Model}}, \href{https://doi.org/10.1088/1126-6708/2009/05/068}{\emph{JHEP} {\bfseries 05} (2009) 068} [\href{https://arxiv.org/abs/0903.0141}{{\ttfamily 0903.0141}}].

\bibitem{Beisert:2010jr}
N.~Beisert et~al., \emph{Review of {AdS/CFT} integrability: An overview}, \href{https://doi.org/10.1007/s11005-011-0529-2}{\emph{Lett. Math. Phys.} {\bfseries 99} (2012) 3} [\href{https://arxiv.org/abs/1012.3982}{{\ttfamily 1012.3982}}].

\bibitem{Gromov:2017blm}
N.~Gromov, \emph{{Introduction to the Spectrum of $N=4$ SYM and the Quantum Spectral Curve}},  \href{https://arxiv.org/abs/1708.03648}{{\ttfamily 1708.03648}}.

\bibitem{Levkovich-Maslyuk:2019awk}
F.~Levkovich-Maslyuk, \emph{{A review of the AdS/CFT Quantum Spectral Curve}}, \href{https://doi.org/10.1088/1751-8121/ab7137}{\emph{J. Phys. A} {\bfseries 53} (2020) 283004} [\href{https://arxiv.org/abs/1911.13065}{{\ttfamily 1911.13065}}].

\bibitem{Kazakov:2018ugh}
V.~Kazakov, \emph{{Quantum Spectral Curve of $\gamma$-twisted ${\cal N}=4$ SYM theory and fishnet CFT}},  \href{https://arxiv.org/abs/1802.02160}{{\ttfamily 1802.02160}}.

\bibitem{Gukov:2004ym}
S.~Gukov, E.~Martinec, G.~W. Moore and A.~Strominger, \emph{The search for a holographic dual to {$\AdS_3 \times \Sphere^3 \times \Sphere^3 \times \Sphere^1$}}, {\emph{Adv. Theor. Math. Phys.} {\bfseries 9} (2005) 435} [\href{https://arxiv.org/abs/hep-th/0403090}{{\ttfamily hep-th/0403090}}].

\bibitem{Elitzur:1998mm}
S.~Elitzur, O.~Feinerman, A.~Giveon and D.~Tsabar, \emph{String theory on {$\AdS_3 \times \Sphere^3 \times \Sphere^3 \times \Sphere^1$}}, \href{https://doi.org/10.1016/S0370-2693(99)00101-X}{\emph{Phys. Lett.} {\bfseries B449} (1999) 180} [\href{https://arxiv.org/abs/hep-th/9811245}{{\ttfamily hep-th/9811245}}].

\bibitem{Dolan2002}
L.~Dolan, \emph{Type IIB String Theory on AdS3{\texttimes}S3{\texttimes}T4}, pp.~219--226.
\newblock Springer US, Boston, MA, 2002.
\newblock 10.1007/0-306-47094-2\_20.

\bibitem{deBoer:1998ip}
J.~de~Boer, \emph{Six-dimensional supergravity on {$\Sphere^3 \times \AdS_3$} and 2d conformal field theory}, \href{https://doi.org/10.1016/S0550-3213(99)00160-1}{\emph{Nucl. Phys.} {\bfseries B548} (1999) 139} [\href{https://arxiv.org/abs/hep-th/9806104}{{\ttfamily hep-th/9806104}}].

\bibitem{Giveon:1998ns}
A.~Giveon, D.~Kutasov and N.~Seiberg, \emph{Comments on string theory on {$\AdS_3$}}, {\emph{Adv. Theor. Math. Phys.} {\bfseries 2} (1998) 733} [\href{https://arxiv.org/abs/hep-th/9806194}{{\ttfamily hep-th/9806194}}].

\bibitem{Giveon:2001up}
A.~Giveon and D.~Kutasov, \emph{Notes on {$\AdS_3$}}, \href{https://doi.org/10.1016/S0550-3213(01)00573-9}{\emph{Nucl. Phys.} {\bfseries B621} (2002) 303} [\href{https://arxiv.org/abs/hep-th/0106004}{{\ttfamily hep-th/0106004}}].

\bibitem{Giveon:2003ku}
A.~Giveon and A.~Pakman, \emph{More on superstrings in {$\AdS_3 \times N$}}, {\emph{JHEP} {\bfseries 0303} (2003) 056} [\href{https://arxiv.org/abs/hep-th/0302217}{{\ttfamily hep-th/0302217}}].

\bibitem{Babichenko:2009dk}
A.~Babichenko, B.~Stefa{\'n}ski, jr. and K.~Zarembo, \emph{Integrability and the {AdS${}_{3}$/CFT${}_{2}$} correspondence}, \href{https://doi.org/10.1007/JHEP03(2010)058}{\emph{JHEP} {\bfseries 1003} (2010) 058} [\href{https://arxiv.org/abs/0912.1723}{{\ttfamily 0912.1723}}].

\bibitem{OhlssonSax:2011ms}
O.~Ohlsson~Sax and B.~Stefa{\'n}ski, jr., \emph{Integrability, spin-chains and the {AdS${}_{3}$/CFT${}_{2}$} correspondence}, \href{https://doi.org/10.1007/JHEP08(2011)029}{\emph{JHEP} {\bfseries 1108} (2011) 029} [\href{https://arxiv.org/abs/1106.2558}{{\ttfamily 1106.2558}}].

\bibitem{Zarembo:2010sg}
K.~Zarembo, \emph{Strings on semisymmetric superspaces}, \href{https://doi.org/10.1007/JHEP05(2010)002}{\emph{JHEP} {\bfseries 1005} (2010) 002} [\href{https://arxiv.org/abs/1003.0465}{{\ttfamily 1003.0465}}].

\bibitem{Zarembo:2010yz}
K.~Zarembo, \emph{Algebraic curves for integrable string backgrounds},  \href{https://arxiv.org/abs/1005.1342}{{\ttfamily 1005.1342}}.

\bibitem{Cagnazzo:2012se}
A.~Cagnazzo and K.~Zarembo, \emph{{B}-field in {AdS${}_{3}$/CFT${}_{2}$} correspondence and integrability}, \href{https://doi.org/10.1007/JHEP11(2012)133, 10.1007/JHEP04(2013)003}{\emph{JHEP} {\bfseries 1211} (2012) 133} [\href{https://arxiv.org/abs/1209.4049}{{\ttfamily 1209.4049}}].

\bibitem{Borsato:2013qpa}
R.~Borsato, O.~Ohlsson~Sax, A.~Sfondrini, B.~Stefa{\'n}ski, jr. and A.~Torrielli, \emph{The all-loop integrable spin-chain for strings on {$\AdS_3 \times \Sphere^3 \times \Torus^4$}: the massive sector}, \href{https://doi.org/10.1007/JHEP08(2013)043}{\emph{JHEP} {\bfseries 1308} (2013) 043} [\href{https://arxiv.org/abs/1303.5995}{{\ttfamily 1303.5995}}].

\bibitem{Borsato:2014hja}
R.~Borsato, O.~Ohlsson~Sax, A.~Sfondrini and B.~Stefa{\'n}ski, jr, \emph{The complete {$\AdS_3 \times \Sphere^3 \times \Torus^4$} worldsheet {S}-matrix}, \href{https://doi.org/10.1007/JHEP10(2014)066}{\emph{JHEP} {\bfseries 1410} (2014) 66} [\href{https://arxiv.org/abs/1406.0453}{{\ttfamily 1406.0453}}].

\bibitem{Borsato:2016kbm}
R.~Borsato, O.~Ohlsson~Sax, A.~Sfondrini and B.~Stefa{\'n}ski, jr., \emph{On the spectrum of {$\AdS_3 \times \Sphere^3 \times \Torus^4$} strings with {R}amond-{R}amond flux}, \href{https://doi.org/10.1088/1751-8113/49/41/41LT03}{\emph{J. Phys.} {\bfseries A49} (2016) 41LT03} [\href{https://arxiv.org/abs/1605.00518}{{\ttfamily 1605.00518}}].

\bibitem{Ekhammar:2021pys}
S.~Ekhammar and D.~Volin, \emph{Monodromy bootstrap for {$\grpSU(2|2)$} quantum spectral curves: from hubbard model to {$\AdS_3/\CFT_2$}}, \href{https://doi.org/10.1007/JHEP03(2022)192}{\emph{JHEP} {\bfseries 03} (2022) 192} [\href{https://arxiv.org/abs/2109.06164}{{\ttfamily 2109.06164}}].

\bibitem{Cavaglia:2021eqr}
A.~Cavagli{\`a}, N.~Gromov, B.~Stefa{\'n}ski, jr. and A.~Torrielli, \emph{Quantum spectral curve for {AdS${}_3$/CFT${}_2$}: a proposal}, \href{https://doi.org/10.1007/JHEP12(2021)048}{\emph{JHEP} {\bfseries 12} (2021) 048} [\href{https://arxiv.org/abs/2109.05500}{{\ttfamily 2109.05500}}].

\bibitem{Frolov:2021bwp}
S.~Frolov and A.~Sfondrini, \emph{Mirror thermodynamic {B}ethe ansatz for {$\AdS_3/\CFT_2$}}, \href{https://doi.org/10.1007/JHEP03(2022)138}{\emph{JHEP} {\bfseries 03} (2022) 138} [\href{https://arxiv.org/abs/2112.08898}{{\ttfamily 2112.08898}}].

\bibitem{Cavaglia:2022xld}
A.~Cavagli\`a, S.~Ekhammar, N.~Gromov and P.~Ryan, \emph{Exploring the quantum spectral curve for {AdS${}_3$/CFT${}_2$}},  \href{https://arxiv.org/abs/2211.07810}{{\ttfamily 2211.07810}}.

\bibitem{Borsato:2012ud}
R.~Borsato, O.~Ohlsson~Sax and A.~Sfondrini, \emph{A dynamic {$\algSU(1|1)^2$} {S}-matrix for {AdS${}_{3}$/CFT${}_{2}$}}, \href{https://doi.org/10.1007/JHEP04(2013)113}{\emph{JHEP} {\bfseries 1304} (2013) 113} [\href{https://arxiv.org/abs/1211.5119}{{\ttfamily 1211.5119}}].

\bibitem{Borsato:2012ss}
R.~Borsato, O.~Ohlsson~Sax and A.~Sfondrini, \emph{All-loop {B}ethe ansatz equations for {AdS${}_{3}$/CFT${}_{2}$}}, \href{https://doi.org/10.1007/JHEP04(2013)116}{\emph{JHEP} {\bfseries 1304} (2013) 116} [\href{https://arxiv.org/abs/1212.0505}{{\ttfamily 1212.0505}}].

\bibitem{Borsato:2015mma}
R.~Borsato, O.~Ohlsson~Sax, A.~Sfondrini and B.~Stefa{\'n}ski, jr., \emph{The {$\AdS_3 \times \Sphere^3 \times \Sphere^3 \times \Sphere^1$} worldsheet {S} matrix}, \href{https://doi.org/10.1088/1751-8113/48/41/415401}{\emph{J. Phys.} {\bfseries A48} (2015) 415401} [\href{https://arxiv.org/abs/1506.00218}{{\ttfamily 1506.00218}}].

\bibitem{Gromov:2013pga}
N.~Gromov, V.~Kazakov, S.~Leurent and D.~Volin, \emph{Quantum spectral curve for {AdS${}_{5}$/CFT${}_{4}$}}, \href{https://doi.org/10.1103/PhysRevLett.112.011602}{\emph{Phys. Rev. Lett.} {\bfseries 112} (2014) 011602} [\href{https://arxiv.org/abs/1305.1939}{{\ttfamily 1305.1939}}].

\bibitem{Gromov:2014caa}
N.~Gromov, V.~Kazakov, S.~Leurent and D.~Volin, \emph{Quantum spectral curve for arbitrary state/operator in {AdS${}_{5}$/CFT${}_{4}$}}, \href{https://doi.org/10.1007/JHEP09(2015)187}{\emph{JHEP} {\bfseries 09} (2015) 187} [\href{https://arxiv.org/abs/1405.4857}{{\ttfamily 1405.4857}}].

\bibitem{Cavaglia:2014exa}
A.~Cavagli{\`a}, D.~Fioravanti, N.~Gromov and R.~Tateo, \emph{Quantum spectral curve of the {$\superN=6$} supersymmetric {C}hern-{S}imons theory}, \href{https://doi.org/10.1103/PhysRevLett.113.021601}{\emph{Phys. Rev. Lett.} {\bfseries 113} (2014) 021601} [\href{https://arxiv.org/abs/1403.1859}{{\ttfamily 1403.1859}}].

\bibitem{Bombardelli:2017vhk}
D.~Bombardelli, A.~Cavagli{\`a}, D.~Fioravanti, N.~Gromov and R.~Tateo, \emph{The full quantum spectral curve for {AdS${}_{4}$/CFT${}_{3}$}},  \href{https://arxiv.org/abs/1701.00473}{{\ttfamily 1701.00473}}.

\bibitem{Gromov:2015wca}
N.~Gromov, F.~Levkovich-Maslyuk and G.~Sizov, \emph{Quantum spectral curve and the numerical solution of the spectral problem in {AdS${}_{5}$/CFT${}_{4}$}}, \href{https://doi.org/10.1007/JHEP06(2016)036}{\emph{JHEP} {\bfseries 06} (2016) 036} [\href{https://arxiv.org/abs/1504.06640}{{\ttfamily 1504.06640}}].

\bibitem{LevkovichMaslyuk:2011ty}
F.~Levkovich-Maslyuk, \emph{Numerical results for the exact spectrum of planar {AdS${}_{4}$/CFT${}_{3}$}}, \href{https://doi.org/10.1007/JHEP05(2012)142}{\emph{JHEP} {\bfseries 1205} (2011) 142} [\href{https://arxiv.org/abs/1110.5869}{{\ttfamily 1110.5869}}].

\bibitem{Giombi_2018}
S.~Giombi and S.~Komatsu, \emph{Exact correlators on the wilson loop in $ \mathcal{N}=4$ sym: localization, defect cft, and integrability}, \href{https://doi.org/10.1007/jhep05(2018)109}{\emph{Journal of High Energy Physics} {\bfseries 2018} (2018) }.

\bibitem{Bianchi:2014laa}
M.~S. Bianchi, L.~Griguolo, M.~Leoni, S.~Penati and D.~Seminara, \emph{{BPS} {W}ilson loops and {B}remsstrahlung function in {ABJ(M)}: a two loop analysis}, \href{https://doi.org/10.1007/JHEP06(2014)123}{\emph{JHEP} {\bfseries 1406} (2014) 123} [\href{https://arxiv.org/abs/1402.4128}{{\ttfamily 1402.4128}}].

\bibitem{Gromov:2014eha}
N.~Gromov and G.~Sizov, \emph{Exact slope and interpolating functions in {ABJM} theory}, \href{https://doi.org/10.1103/PhysRevLett.113.121601}{\emph{Phys. Rev. Lett.} {\bfseries 113} (2014) 121601} [\href{https://arxiv.org/abs/1403.1894}{{\ttfamily 1403.1894}}].

\bibitem{Gromov:2014bva}
N.~Gromov, F.~Levkovich-Maslyuk, G.~Sizov and S.~Valatka, \emph{Quantum spectral curve at work: From small spin to strong coupling in {$\superN = 4$} {SYM}}, \href{https://doi.org/10.1007/JHEP07(2014)156}{\emph{JHEP} {\bfseries 1407} (2014) 156} [\href{https://arxiv.org/abs/1402.0871}{{\ttfamily 1402.0871}}].

\bibitem{Cavaglia:2022yvv}
A.~Cavagli\`a, N.~Gromov, J.~Julius and M.~Preti, \emph{{Integrated correlators from integrability: Maldacena-Wilson line in $ \mathcal{N} $ = 4 SYM}}, \href{https://doi.org/10.1007/JHEP04(2023)026}{\emph{JHEP} {\bfseries 04} (2023) 026} [\href{https://arxiv.org/abs/2211.03203}{{\ttfamily 2211.03203}}].

\bibitem{Cavaglia:2022qpg}
A.~Cavagli\`a, N.~Gromov, J.~Julius and M.~Preti, \emph{{Bootstrability in defect CFT: integrated correlators and sharper bounds}}, \href{https://doi.org/10.1007/JHEP05(2022)164}{\emph{JHEP} {\bfseries 05} (2022) 164} [\href{https://arxiv.org/abs/2203.09556}{{\ttfamily 2203.09556}}].

\bibitem{Caron-Huot:2022sdy}
S.~Caron-Huot, F.~Coronado, A.-K. Trinh and Z.~Zahraee, \emph{{Bootstrapping $ \mathcal{N} $ = 4 sYM correlators using integrability}}, \href{https://doi.org/10.1007/JHEP02(2023)083}{\emph{JHEP} {\bfseries 02} (2023) 083} [\href{https://arxiv.org/abs/2207.01615}{{\ttfamily 2207.01615}}].

\bibitem{Caron-Huot:2024tzr}
S.~Caron-Huot, F.~Coronado and Z.~Zahraee, \emph{{Bootstrapping $\mathcal{N} = 4$ sYM correlators using Integrability and Localization}},  \href{https://arxiv.org/abs/2412.00249}{{\ttfamily 2412.00249}}.

\bibitem{Bercini:2022jxo}
C.~Bercini, A.~Homrich and P.~Vieira, \emph{{Structure constants in N=4 supersymmetric Yang-Mills theory and separation of variables}}, \href{https://doi.org/10.1103/PhysRevD.110.L121901}{\emph{Phys. Rev. D} {\bfseries 110} (2024) L121901} [\href{https://arxiv.org/abs/2210.04923}{{\ttfamily 2210.04923}}].

\bibitem{Basso:2022nny}
B.~Basso, A.~Georgoudis and A.~K. Sueiro, \emph{{Structure Constants of Short Operators in Planar N=4 Supersymmetric Yang-Mills Theory}}, \href{https://doi.org/10.1103/PhysRevLett.130.131603}{\emph{Phys. Rev. Lett.} {\bfseries 130} (2023) 131603} [\href{https://arxiv.org/abs/2207.01315}{{\ttfamily 2207.01315}}].

\bibitem{Basso:2025mca}
B.~Basso and A.~Georgoudis, \emph{{Exploring Structure Constants in Planar $\mathcal{N} = 4$ SYM: From Small Spin to Strong Coupling}},  \href{https://arxiv.org/abs/2504.07922}{{\ttfamily 2504.07922}}.

\bibitem{Ekhammar:2024kzp}
S.~Ekhammar, N.~Gromov and B.~Stefa\'nski, \emph{{Demystifying the Massless Sector in AdS\_3 Quantum Spectral Curve}},  \href{https://arxiv.org/abs/2412.11915}{{\ttfamily 2412.11915}}.

\bibitem{witten2024instantonslargen4algebra}
E.~Witten, \emph{Instantons and the large n=4 algebra},  2024.

\bibitem{Chernyak:2020lgw}
D.~Chernyak, S.~Leurent and D.~Volin, \emph{{Completeness of Wronskian Bethe Equations for Rational ${\mathfrak {\mathfrak {gl}}_{{{\mathsf {m}}}|{{\mathsf {n}}}}}$ Spin Chains}}, \href{https://doi.org/10.1007/s00220-021-04275-9}{\emph{Commun. Math. Phys.} {\bfseries 391} (2022) 969} [\href{https://arxiv.org/abs/2004.02865}{{\ttfamily 2004.02865}}].

\bibitem{Ryan:2020rfk}
P.~Ryan and D.~Volin, \emph{{Separation of Variables for Rational $\mathfrak {gl}(\mathsf {n})$ Spin Chains in Any Compact Representation, via Fusion, Embedding Morphism and B\"acklund Flow}}, \href{https://doi.org/10.1007/s00220-021-03990-7}{\emph{Commun. Math. Phys.} {\bfseries 383} (2021) 311} [\href{https://arxiv.org/abs/2002.12341}{{\ttfamily 2002.12341}}].

\bibitem{Kazakov:2015efa}
V.~Kazakov, S.~Leurent and D.~Volin, \emph{{T-system on T-hook: Grassmannian Solution and Twisted Quantum Spectral Curve}}, \href{https://doi.org/10.1007/JHEP12(2016)044}{\emph{JHEP} {\bfseries 12} (2016) 044} [\href{https://arxiv.org/abs/1510.02100}{{\ttfamily 1510.02100}}].

\bibitem{Marboe_2017}
C.~Marboe and D.~Volin, \emph{Fast analytic solver of rational bethe equations}, \href{https://doi.org/10.1088/1751-8121/aa6b88}{\emph{Journal of Physics A: Mathematical and Theoretical} {\bfseries 50} (2017) 204002}.

\bibitem{Frassek:2023tka}
R.~Frassek and A.~Tsymbaliuk, \emph{{Orthosymplectic superoscillator Lax matrices}}, \href{https://doi.org/10.1007/s11005-024-01789-w}{\emph{Lett. Math. Phys.} {\bfseries 114} (2024) 49} [\href{https://arxiv.org/abs/2309.14199}{{\ttfamily 2309.14199}}].

\bibitem{Ferrando:2020vzk}
G.~Ferrando, R.~Frassek and V.~Kazakov, \emph{{QQ-system and Weyl-type transfer matrices in integrable SO(2r) spin chains}}, \href{https://doi.org/10.1007/JHEP02(2021)193}{\emph{JHEP} {\bfseries 02} (2021) 193} [\href{https://arxiv.org/abs/2008.04336}{{\ttfamily 2008.04336}}].

\bibitem{ekhammar2021extendedsystemsbaxterqfunctions}
S.~Ekhammar, H.~Shu and D.~Volin, \emph{Extended systems of baxter q-functions and fused flags i: simply-laced case},  2021.

\bibitem{Tsuboi:2023sfs}
Z.~Tsuboi, \emph{{Folding QQ-relations and transfer matrix eigenvalues: Towards a unified approach to Bethe ansatz for super spin chains}}, \href{https://doi.org/10.1016/j.nuclphysb.2024.116607}{\emph{Nucl. Phys. B} {\bfseries 1005} (2024) 116607} [\href{https://arxiv.org/abs/2309.16660}{{\ttfamily 2309.16660}}].

\bibitem{Abbott:2012dd}
M.~C. Abbott, \emph{Comment on strings in {$\AdS_3 \times \Sphere^3 \times \Sphere^3 \times \Sphere^1$} at one loop}, \href{https://doi.org/10.1007/JHEP02(2013)102}{\emph{JHEP} {\bfseries 1302} (2013) 102} [\href{https://arxiv.org/abs/1211.5587}{{\ttfamily 1211.5587}}].

\bibitem{Bombardelli_2010}
D.~Bombardelli, D.~Fioravanti and R.~Tateo, \emph{Tba and y-system for planar ads4/cft3}, \href{https://doi.org/10.1016/j.nuclphysb.2010.04.005}{\emph{Nuclear Physics B} {\bfseries 834} (2010) 543–561}.

\bibitem{Brizio:2024nso}
N.~Brizio, A.~Cavagli{\`a}, R.~Tateo and V.~Tripodi, \emph{{Regge trajectories and bridges between them in integrable AdS/CFT}}, \href{https://doi.org/10.1007/JHEP07(2025)122}{\emph{JHEP} {\bfseries 07} (2025) 122} [\href{https://arxiv.org/abs/2410.08927}{{\ttfamily 2410.08927}}].

\bibitem{Kazakov:2004qf}
V.~A. Kazakov, A.~Marshakov, J.~A. Minahan and K.~Zarembo, \emph{Classical/quantum integrability in {AdS/CFT}}, {\emph{JHEP} {\bfseries 5} (2004) 24} [\href{https://arxiv.org/abs/hep-th/0402207}{{\ttfamily hep-th/0402207}}].

\bibitem{SchaferNameki:2010jy}
S.~Sch{\"a}fer-Nameki, \emph{Review of {AdS/CFT} integrability, {C}hapter {II.4}: The spectral curve}, \href{https://doi.org/10.1007/s11005-011-0525-6}{\emph{Lett. Math. Phys.} {\bfseries 99} (2010) 169} [\href{https://arxiv.org/abs/1012.3989}{{\ttfamily 1012.3989}}].

\bibitem{Gromov:2008bz}
N.~Gromov and P.~Vieira, \emph{The {AdS${}_{4}$/CFT${}_{3}$} algebraic curve}, \href{https://doi.org/10.1088/1126-6708/2009/02/040}{\emph{JHEP} {\bfseries 0902} (2009) 040} [\href{https://arxiv.org/abs/0807.0437}{{\ttfamily 0807.0437}}].

\bibitem{Ekhammar:2021myw}
S.~Ekhammar and D.~Volin, \emph{{Bethe algebra using pure spinors}}, \href{https://doi.org/10.1007/s11005-024-01894-w}{\emph{Lett. Math. Phys.} {\bfseries 115} (2025) 20} [\href{https://arxiv.org/abs/2104.04539}{{\ttfamily 2104.04539}}].

\bibitem{Frolov:2021fmj}
S.~Frolov and A.~Sfondrini, \emph{New dressing factors for {$\AdS_3/\CFT_2$}}, \href{https://doi.org/10.1007/JHEP04(2022)162}{\emph{JHEP} {\bfseries 04} (2022) 162} [\href{https://arxiv.org/abs/2112.08896}{{\ttfamily 2112.08896}}].

\bibitem{Borsato:2016xns}
R.~Borsato, O.~Ohlsson~Sax, A.~Sfondrini, B.~Stefa{\'n}ski, jr. and A.~Torrielli, \emph{On the dressing factors, {B}ethe equations and {Y}angian symmetry of strings on {$\AdS_3 \times \Sphere^3 \times \Torus^4$}}, \href{https://doi.org/10.1088/1751-8121/50/2/024004}{\emph{J. Phys.} {\bfseries A50} (2017) 024004} [\href{https://arxiv.org/abs/1607.00914}{{\ttfamily 1607.00914}}].

\bibitem{Borsato:2013hoa}
R.~Borsato, O.~Ohlsson~Sax, A.~Sfondrini, B.~Stefa{\'n}ski, jr. and A.~Torrielli, \emph{Dressing phases of {AdS${}_{3}$/CFT${}_{2}$}}, \href{https://doi.org/10.1103/PhysRevD.88.066004}{\emph{Phys. Rev.} {\bfseries D88} (2013) 066004} [\href{https://arxiv.org/abs/1306.2512}{{\ttfamily 1306.2512}}].

\bibitem{Frolov:2025ozz}
S.~Frolov, D.~Polvara and A.~Sfondrini, \emph{{Exchange relations and crossing}},  \href{https://arxiv.org/abs/2506.04096}{{\ttfamily 2506.04096}}.

\bibitem{Ekhammar:2024neh}
S.~Ekhammar, N.~Gromov and M.~Preti, \emph{{Long Range Asymptotic Baxter-Bethe Ansatz for N=4 BFKL}},  \href{https://arxiv.org/abs/2406.18639}{{\ttfamily 2406.18639}}.

\bibitem{Ekhammar:2025vig}
S.~Ekhammar, N.~Gromov and M.~Preti, \emph{{Regge Trajectories of N=4 SYM Part I: General Asymptotic Baxter-Bethe Ansatz}},  \href{https://arxiv.org/abs/2507.15983}{{\ttfamily 2507.15983}}.

\bibitem{Beisert:2006ez}
N.~Beisert, B.~Eden and M.~Staudacher, \emph{Transcendentality and crossing}, \href{https://doi.org/10.1088/1742-5468/2007/01/P01021}{\emph{J. Stat. Mech.} {\bfseries 0701} (2007) P01021} [\href{https://arxiv.org/abs/hep-th/0610251}{{\ttfamily hep-th/0610251}}].

\bibitem{Ekhammar:2024rfj}
S.~Ekhammar, N.~Gromov and P.~Ryan, \emph{{New approach to strongly coupled $ \mathcal{N} $ = 4 SYM via integrability}}, \href{https://doi.org/10.1007/JHEP12(2024)165}{\emph{JHEP} {\bfseries 12} (2024) 165} [\href{https://arxiv.org/abs/2406.02698}{{\ttfamily 2406.02698}}].

\bibitem{Alday:2023mvu}
L.~F. Alday and T.~Hansen, \emph{{The AdS Virasoro-Shapiro amplitude}}, \href{https://doi.org/10.1007/JHEP10(2023)023}{\emph{JHEP} {\bfseries 10} (2023) 023} [\href{https://arxiv.org/abs/2306.12786}{{\ttfamily 2306.12786}}].

\bibitem{Chester:2024wnb}
S.~M. Chester and D.-l. Zhong, \emph{{AdS3{\texttimes}S3 Virasoro-Shapiro Amplitude with Ramond-Ramond Flux}}, \href{https://doi.org/10.1103/PhysRevLett.134.151602}{\emph{Phys. Rev. Lett.} {\bfseries 134} (2025) 151602} [\href{https://arxiv.org/abs/2412.06429}{{\ttfamily 2412.06429}}].

\bibitem{Jiang:2025oar}
H.~Jiang and D.-l. Zhong, \emph{{$AdS_3 \times S^3$ Virasoro-Shapiro amplitude with KK modes}},  \href{https://arxiv.org/abs/2508.06039}{{\ttfamily 2508.06039}}.

\bibitem{Eberhardt:2019niq}
L.~Eberhardt and M.~R. Gaberdiel, \emph{{Strings on $\text{AdS}_3 \times \text{S}^3 \times \text{S}^3 \times \text{S}^1$}}, \href{https://doi.org/10.1007/JHEP06(2019)035}{\emph{JHEP} {\bfseries 06} (2019) 035} [\href{https://arxiv.org/abs/1904.01585}{{\ttfamily 1904.01585}}].

\bibitem{sriprachyakul2024spacetimedilatonrmads3times}
V.~Sriprachyakul, \emph{Spacetime dilaton in ${\rm ads}_3\times x$ holography},  2024.

\bibitem{sriprachyakul2024superstringsnearconformalboundary}
V.~Sriprachyakul, \emph{Superstrings near the conformal boundary of $\rm ads_3$},  2024.

\bibitem{Gaberdiel:2024dva}
M.~R. Gaberdiel and V.~Sriprachyakul, \emph{{Tensionless strings on AdS$_{3}${\texttimes} S$^{3}${\texttimes} S$^{3}${\texttimes} S$^{1}$}}, \href{https://doi.org/10.1007/JHEP05(2025)003}{\emph{JHEP} {\bfseries 05} (2025) 003} [\href{https://arxiv.org/abs/2411.16848}{{\ttfamily 2411.16848}}].

\bibitem{Frolov:2024pkz}
S.~Frolov, D.~Polvara and A.~Sfondrini, \emph{{Dressing factors for mixed-flux AdS3{\texttimes}S3{\texttimes}T4 superstrings}}, \href{https://doi.org/10.1103/PhysRevD.111.L081901}{\emph{Phys. Rev. D} {\bfseries 111} (2025) L081901} [\href{https://arxiv.org/abs/2402.11732}{{\ttfamily 2402.11732}}].

\bibitem{Frolov:2025uwz}
S.~Frolov, D.~Polvara and A.~Sfondrini, \emph{{Massive dressing factors for mixed-flux AdS$_{3}$/CFT$_{2}$}}, \href{https://doi.org/10.1007/JHEP07(2025)171}{\emph{JHEP} {\bfseries 07} (2025) 171} [\href{https://arxiv.org/abs/2501.05995}{{\ttfamily 2501.05995}}].

\bibitem{Frolov:2025tda}
S.~Frolov, D.~Polvara and A.~Sfondrini, \emph{{Dressing Factors and Mirror Thermodynamic Bethe Ansatz for mixed-flux AdS3/CFT2}},  \href{https://arxiv.org/abs/2507.12191}{{\ttfamily 2507.12191}}.

\bibitem{Gaberdiel:2023lco}
M.~R. Gaberdiel, R.~Gopakumar and B.~Nairz, \emph{{Beyond the tensionless limit: integrability in the symmetric orbifold}}, \href{https://doi.org/10.1007/JHEP06(2024)030}{\emph{JHEP} {\bfseries 06} (2024) 030} [\href{https://arxiv.org/abs/2312.13288}{{\ttfamily 2312.13288}}].

\bibitem{Gaberdiel:2024dfw}
M.~R. Gaberdiel, D.~Kempel and B.~Nairz, \emph{{AdS$_{3}${\texttimes}S$^{3}$ magnons in the symmetric orbifold}}, \href{https://doi.org/10.1007/JHEP05(2025)229}{\emph{JHEP} {\bfseries 05} (2025) 229} [\href{https://arxiv.org/abs/2412.02741}{{\ttfamily 2412.02741}}].

\bibitem{Gaberdiel:2024nge}
M.~R. Gaberdiel, F.~Lichtner and B.~Nairz, \emph{{Anomalous dimensions in the symmetric orbifold}}, \href{https://doi.org/10.1007/JHEP05(2025)084}{\emph{JHEP} {\bfseries 05} (2025) 084} [\href{https://arxiv.org/abs/2411.17612}{{\ttfamily 2411.17612}}].

\bibitem{Primi2025AdS3S3S3S1}
N.~Primi, ``\href{https://www.youtube.com/watch?v=QGCNScxEfPk}{A proposal for the AdS$_3$ S$_3$ S$_3$ S$_1$ Quantum Spectral Curve}.'' Gong Show Presentation at IGST 2025, 2025.

\bibitem{Torrielli:2025lgu}
A.~Torrielli, \emph{{AdS3 integrability, Sine-Gordon and fractional supersymmetry}},  \href{https://arxiv.org/abs/2508.12316}{{\ttfamily 2508.12316}}.

\end{thebibliography}\endgroup
%\input{main.bbl}
%V18:52
\end{document}